\documentclass[draftcls,12pt,onecolumn]{IEEEtran}  % draft mode (1 column)
\usepackage[pdftex]{graphicx,color}
\usepackage{psfrag}
\usepackage{subfigure}
\usepackage{cite}

\graphicspath{{./}}

\usepackage{array,multirow}
\usepackage{graphicx}  % Written by David Carlisle and Sebastian Rahtz
\usepackage{psfrag}    % Written by Craig Barratt, Michael C. Grant,
\usepackage{subfigure} % Written by Steven Douglas Cochran
\usepackage{amsmath}   % From the American Mathematical Society
\usepackage{amssymb}   % A popular package that provides many helpful commands
\usepackage{eucal}     % Calligraphic letters

\newtheorem{lemma}{Lemma}
\newtheorem{proposition}{Proposition}
\newtheorem{observation}{Observation}

\newtheorem{corollary}{Corollary}

\newtheorem{theorem}{Theorem}
\newtheorem{remark}{Remark}

\newcommand{\qed}{\nobreak \ifvmode \relax \else
      \ifdim\lastskip<1.5em \hskip-\lastskip
      \hskip1.5em plus0em minus0.5em \fi \nobreak
      \vrule height0.5em width0.5em depth0.25em\fi}

\newcommand{\dv}{\mathbf} % determenistic vector
\newcommand{\mc}{\mathcal} % determenistic vector

\DeclareMathAlphabet{\eurm}{U}{eur}{m}{n}
\DeclareMathAlphabet{\mathbsf}{OT1}{cmss}{bx}{n}% bold sans serif
\DeclareMathAlphabet{\mathssf}{OT1}{cmss}{m}{sl}% slanted sans serif
\DeclareMathAlphabet{\mathcsf}{OT1}{cmss}{sbc}{n}% condensed sans serif
\DeclareSymbolFont{bsfletters}{OT1}{cmss}{bx}{n}  
\DeclareSymbolFont{ssfletters}{OT1}{cmss}{m}{n}
\DeclareMathSymbol{\bsfGamma}{0}{bsfletters}{'000}
\DeclareMathSymbol{\ssfGamma}{0}{ssfletters}{'000}
\DeclareMathSymbol{\bsfDelta}{0}{bsfletters}{'001}
\DeclareMathSymbol{\ssfDelta}{0}{ssfletters}{'001}
\DeclareMathSymbol{\bsfTheta}{0}{bsfletters}{'002}
\DeclareMathSymbol{\ssfTheta}{0}{ssfletters}{'002}
\DeclareMathSymbol{\bsfLambda}{0}{bsfletters}{'003}
\DeclareMathSymbol{\ssfLambda}{0}{ssfletters}{'003}
\DeclareMathSymbol{\bsfXi}{0}{bsfletters}{'004}
\DeclareMathSymbol{\ssfXi}{0}{ssfletters}{'004}
\DeclareMathSymbol{\bsfPi}{0}{bsfletters}{'005}
\DeclareMathSymbol{\ssfPi}{0}{ssfletters}{'005}
\DeclareMathSymbol{\bsfSigma}{0}{bsfletters}{'006}
\DeclareMathSymbol{\ssfSigma}{0}{ssfletters}{'006}
\DeclareMathSymbol{\bsfUpsilon}{0}{bsfletters}{'007}
\DeclareMathSymbol{\ssfUpsilon}{0}{ssfletters}{'007}
\DeclareMathSymbol{\bsfPhi}{0}{bsfletters}{'010}
\DeclareMathSymbol{\ssfPhi}{0}{ssfletters}{'010}
\DeclareMathSymbol{\bsfPsi}{0}{bsfletters}{'011}
\DeclareMathSymbol{\ssfPsi}{0}{ssfletters}{'011}
\DeclareMathSymbol{\bsfOmega}{0}{bsfletters}{'012}
\DeclareMathSymbol{\ssfOmega}{0}{ssfletters}{'012}

\newcommand{\calN}{{\mathcal{N}}}

\newcommand{\calS}{{\mathcal{S}}}
\newcommand{\calU}{{\mathcal{U}}}

\newcommand{\calX}{{\mathcal{X}}}
\newcommand{\calY}{{\mathcal{Y}}}

\begin{document}

% paper title
\title{Cooperative Relaying with State Available Non-Causally at the Relay}

%\markboth{IEEE Transactions on Information Theory (Submitted for Publication)}{Zaidi \MakeLowercase{\textit{et al.}}: Cooperative Relaying with State Available Non-Causally at the Relay}
\vspace{1.5cm}
%\pubid{\vspace{1.5cm}0000--0000/00\$00.00˜\copyright˜2008 IEEE}

\author{\vspace{0cm}
\authorblockN{ \small Abdellatif Zaidi \qquad Shiva Prasad Kotagiri \qquad J. Nicholas Laneman \qquad Luc Vandendorpe\thanks{Abdellatif Zaidi and Luc Vandendorpe are with \'{E}cole Polytechnique de Louvain, Universit\'e Catholique de Louvain, Louvain-la-Neuve 1348, Belgium. Email: \{abdellatif.zaidi,luc.vandendorpe\}@uclouvain.be}
\thanks{Shiva Prasad Kotagiri was with the Department of Electrical Engineering, University of Notre Dame,  Notre Dame, IN 46556, and is now with SERDES Technology Group, Xilinx Inc., San Jose, CA, USA-95124; Email: skotagir@gmail.com}
\thanks{J. Nicholas Laneman is with the Department of Electrical Engineering, University of Notre Dame, Notre Dame, IN 46556; Email: jnl@nd.edu} 
\thanks{The work of A. Zaidi and L. Vandendorpe has been supported in part by the EU framework program COOPCOM and the network of excellence NEWCOM++. The work of S. P. Kotagiri and J. N. Laneman has been supported in part by NSF Grants CNS06-26595 and CCF05-46618.}}}
\maketitle
% make the title area
\maketitle
\IEEEpeerreviewmaketitle
%\newpage
\vspace{-0.2cm}

\begin{abstract}
We consider a three-terminal state-dependent relay channel with the channel state noncausally available at only the relay. Such a model may be useful for designing cooperative wireless networks with some terminals equipped with cognition capabilities, i.e., the relay in our setup. In the discrete memoryless (DM) case, we establish lower and upper bounds on channel capacity. The lower bound is obtained by a coding scheme at the relay  that uses a combination of codeword splitting,  Gel'fand-Pinsker binning, and decode-and-forward relaying. The upper bound improves upon that obtained by assuming that the channel state is available at the source, the relay, and the destination. For the Gaussian case, we also derive lower and upper bounds on the capacity. The lower bound is obtained by a coding scheme at the relay that uses a combination of codeword splitting, generalized dirty paper coding, and decode-and-forward relaying; the upper bound is also better than that obtained by assuming that the channel state is available at the source, the relay, and the destination. In the case of degraded Gaussian channels, the lower bound meets with the upper bound for some special cases, and, so, the capacity is obtained for these cases.  Furthermore, in the Gaussian case, we also extend the results to the case in which the relay operates in a half-duplex mode.
\end{abstract}

\vspace{-0.5cm}

\IEEEpeerreviewmaketitle

\section{Introduction}\label{secI}
\begin{figure}[htpb]
\centering
\resizebox{0.8\linewidth}{!}{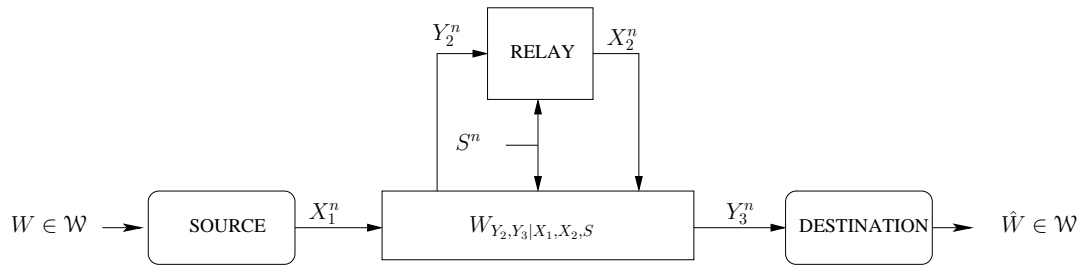}
\caption{Relay channel with state information $S^n$ available noncausally at only the relay.}
 \label{StateDependentDiscreteMemorylessRelayChannel}
 \end{figure}

We consider a three-terminal state-dependent relay channel (RC) in which, as shown in Figure~\ref{StateDependentDiscreteMemorylessRelayChannel}, the source wants to communicate a message $W$ to the destination through the state-dependent RC in $n$ uses of the channel, with the help of the relay. The channel outputs $Y_2$ and $Y_3$ for  the relay and the destination, respectively, are controlled by the channel input $X_1$, the relay input $X_2$ and the channel state $S$, through a given memoryless probability law $W_{Y_2,Y_3|X_1,X_2,S}$. The channel state $S$ is generated according to  a given memoryless probability law $Q_S$. It is assumed that the channel state is known, noncausally, to only the relay. The destination estimates the message sent by the source from the received channel output. In this paper we study the capacity of this communication system. We refer to the model under investigation as state-dependent RC with informed relay.

\subsection{Background}
Channels with random parameters or states have received  considerable attention due to a wide range of possible applications. Shannon initiated the study of single-user models with state available \textit{causally} at the encoder \cite{Sh58}. For the single-user discrete memoryless (DM) state-dependent models, Gel'fand and Pinsker derive the capacity for the setup in which the channel state is available \textit{noncausally} at the encoder \cite{GP80}. In this case, a random coding scheme based on binning, known as \textit{Gel'fand-Pinsker coding}, achieves the capacity \cite{GP80}. Costa considers an additive Gaussian channel with additive Gaussian state known at the encoder and shows that Gel'fand-Pinsker coding with a specific auxiliary random variable, widely known as \textit{dirty paper coding} (DPC), achieves the trivial upper bound obtained by assuming the channel state available also at the decoder \cite{C83}. Interestingly, DPC eliminates the effect of the additive channel state on the capacity, as if there were no channel state present in the model or the channel state were known to the decoder as well. It is worthnoting that since DPC achieves the trivial upper bound for this model there is no need to derive tighter upper bounds in this case. In \cite{HG83}, models with channel state available noncausally at the encoder are studied from the perspective of memories with defects. Practical coding realizations using concepts of lattices for the models with noncausal encoder state information are studied, e.g., in \cite{ZSE02,ESZ05}. For a review on the subject of state-dependent channels and related work, the reader may refer to \cite{KSM08}.

A growing body of work studies multi-user state-dependent models with noncausal encoder state information \cite{GP84,KSS04,KL04,CS05,S05,KL07,KL07a,SBSV07,SBSV07a,PKEZ07,ZV07b,ZV09b,ZKLV08a,ZKLV09a}. In the multi-user models, the channel state can be known to all, only some, or none of the users in the communication system. In the case of state-dependent DM models, the multiple access channel (MAC) with partial channel state at all the encoders and full channel state at the decoder is considered in \cite{CS05}, and the broadcast channel (BC) with state available at the encoder but not at the decoders is considered in \cite{S05, SS05}.

In the Gaussian case, the MAC with all encoders being informed, the BC with informed encoder, the physically degraded relay channel (RC) with informed source and informed relay, and the physically degraded relay broadcast channel (RBC) with informed source and informed relay are studied in \cite{GP84,KSS04,ZV07b}. In all these cases, it is shown that some variants of DPC achieve the respective capacity or capacity region. Also, since for all these models DPC achieves the trivial upper or outer bound obtained by assuming that the channel state is also available at the decoders, it is not necessary to derive non-trivial upper or outer bounds, i.e., bounds that are tighter than the cut-set bound. For all these models, the key assumption that makes the problem relatively easy is the availability of the channel state at {\it all} the encoders in the communication model, which allows these encoders to remove the effect of the channel state on their respective communication using variants of DPC. It is interesting to study state-dependent multi-user models in which {\it some, but not all,} encoders are informed of the channel state, because the uninformed encoders cannot apply DPC.

%%% FIXME:  Watch ordering of citations in the first \cite below.  The numbers appear out of sequence here relative to the above.

%-- FIXED 

The state-dependent MAC with some, but not all, encoders informed of the channel state is considered in \cite{KL04,KL07a,KL07,SBSV07,SBSV07a,ZKLV09a} and the state-dependent relay channel with informed source is considered in \cite{ZV07b,ZV09b}. For the Gaussian cases of these models, the informed encoder applies a slightly generalized DPC (GDPC) in which the channel input and the channel state are negatively correlated. It is interesting to note that in these models the uninformed encoders benefit from the GDPC applied by the informed encoders because the negative correlation can be viewed as partial state cancellation. The capacity region of the discrete memoryless state-dependent MAC with one informed encoder is characterized in \cite{KL07a,SBSV07a} for the case in which the messages sets are degraded and the informed encoder knows the message of the uninformed encoder. In \cite{SBSV07a}, the authors also study the Gaussian case and they characterize the capacity region by deriving a non-trivial outer bound that is strictly tighter than the cut-set outer bound. 

\iffalse
In \cite{ZKLV09a}, we study the model obtained by swapping the roles of the encoders in \cite{KL07a,SBSV07a}, i.e., a state-dependent MAC with one informed encoder and with the uninformed encoder knowing the message of the informed encoder. For this model, we establish inner and non-trivial outer bounds on the capacity region for both DM and Gaussian cases. Also, we show that the bounds for the Gaussian model meet in some special cases. 
\fi

For the study of communication models in which only some of the involved encoders are informed about the channel state, it is important to establish non-trivial upper or outer bounds. These bounds help characterize the rate loss due to not knowing the state at the uninformed encoders; and help assess the effectiveness of the coding schemes that are employed for the achievability results. In this paper, we study a state-dependent relay channel with the channel state known to only the relay. This model is conceptually different from the model considered previously in \cite{ZV07b,ZV09b} in which the channel state is noncausally known to only the source. 

\subsection{Motivation}
Channels whose  probabilistic input-output relationship depends on random parameters, or channel states, can model a large variety of scenarios. The assumption of noncausal channel state can hold \textit{naturally} or \textit{approximately}. Examples where the assumption of noncausal state holds naturally include information embedding \cite{MO03,CL02,SM04,ZPD05b,ZV09a,nM00}, certain storage applications such as computer memories with defective cells \cite{KT74} and certain broadcast scenarios such as multiple-input multiple-output (MIMO) broadcast channels \cite{CS03,VJG03,VT03} where DPC is a central ingredient in achieving the capacity region \cite{WSS06}. Examples where the assumption of noncausal state holds approximately include dispersive (ISI) channels \cite{ZSE02}, block fading in wireless environments \cite{BPS98}, network \cite{H-M02} and cooperative networks \cite{JFG04}.  

\begin{figure}[htpb]
\centering
\resizebox{0.8\linewidth}{!}{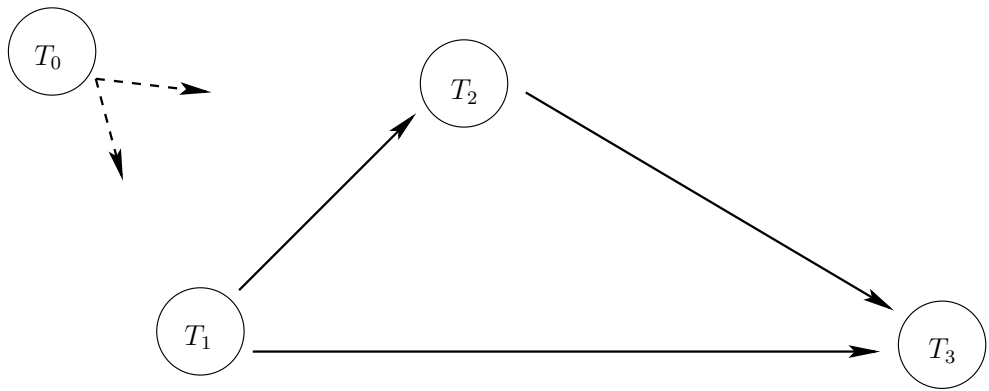}
\caption{Example wireless network with cognition capabilities. If the relay $T_2$ is cognizant of the competing source $T_0$, it can help the source $T_1$ cancel the effect of the interference from $T_0$.}
 \label{MotivatingExample1}
 \end{figure}

Yet, another example application is cooperation in the realm of cognition. Driven by the growing demand for frequency spectrum, cognitive radios, usually defined as smart radio devices that are capable of acquiring some knowledge about the channel state, are  introduced into communication systems in order to help non-cognitive radios in terms of spectral efficiency \cite{JM00}. In a wireless interference network in which some terminals compete and some others cooperate, equipping some specific terminals with cognition capabilities that allow them to learn the interference to high accuracy would help other non-cognitive terminals. These cognitive radios can exploit the knowledge of the interference or channel state to remove its effect on the transmission of their own messages and also that of the messages of the non-cognitive terminals as well. The study of fundamental performance limits of models with only a subset of the encoders being informed is relevant for a better understanding of communication systems that involve cognitive radios. For example, to increase system spectral efficiency, collaboration is investigated in the realm of cognition in \cite{DMT06a,DMT06b,JV06}. Also, the problem of collaborative signal transmission in the presence of some cognizant terminals is investigated for a MAC scenario in \cite{SBSV07a,KL07} and for an interference channel scenario in \cite{MGKS08,SVA-JS08,SE07a,SBSV08}. The setup we consider in this paper also models the building block for collaborative wireless networks in which only the relays, but neither sources nor destinations, are cognizant of the channel state. An example of such a scenario is shown in Figure~\ref{MotivatingExample1}.
 
\subsection{Main Contributions}
For the DM case, we derive lower and upper bounds on the capacity of the general state-dependent relay channel with informed relay. The lower bound is obtained by a coding scheme at the relay that uses a combination of codeword splitting, Gel'fand-Pinsker coding, and decode-and-forward (DF) relaying. For this model, designing a codebook at the relay is challenging since such a codebook should allow the source to generate codewords that are correlated with the channel input of the relay which exploits the available channel state. In this work, this is accomplished by codeword splitting at the relay. With codeword splitting, the channel input of the relay is generated from two codewords: the first of which is a function of the cooperative information (i.e, the information that is sent cooperatively by the source and the relay using a \textit{joint codebook}) and the channel state, and the second of which is a function of only the cooperative information. Since the source knows the cooperative information, it can generate its channel input in a way such that it is correlated with the latter codeword at the relay, which is a function of only the cooperative information.

%%% FIXME: Paragraph below feels more complicated than it needs to be. Basically trying to say that things are related to SBSV07a and KL07a, but there are some important differences.  At this stage of the paper, it's important to give the reader more of a sense of what those differences are. 

%-- FIXED

Our upper bound on the capacity is tighter than that obtained by assuming that the channel state is also available at the source and the destination. This upper bound is non-trivial and relates to the bounding technique developed in the context of multiple access channels with asymmetric channel state in \cite[Theorem 2]{SBSV07a}; however, we note that the present upper bound is proved using techniques that are different from those in \cite{SBSV07a}. On a related note, we mention that at a high level there is a connection between the multiple access transmission part in the RC with informed relay in this work and the models in \cite{SBSV07a,KL07}. However, there are also numerous conceptual differences that will be discussed whenever relevant. In particular, in contrast to \cite{SBSV07a,KL07},  here the uninformed encoder (the source) knows the message of the informed encoder (the relay). From this angle, the model in this paper connects more with the state-dependent MAC studied in \cite{ZKLV09a}.

Furthermore, we specialize the results to the case in which the channel is degraded. Also, we extend the lower bound for the DF relaying scheme to the case in which the source implements rate-splitting and the relay decodes-and-forwards only one part of the source message.

We apply the concepts developed in the DM case to the Gaussian case in which both the noise and the state are additive Gaussian random variables. In our analysis for the Gaussian RC, we first allow the relay to operate in a {\it full-duplex} mode in which it can transmit and receive simultaneously, and  then we constrain it to operate in a {\it half-duplex} mode in which it can either only transmit or only receive. 

In the case of full-duplex transmission, we derive lower and upper bounds on the capacity of the Gaussian relay channel with informed relay.  We obtain two lower bounds by using the concepts of codeword splitting, generalized DPC (GDPC) \cite{KL04,MS06}, and decode-and-forward relaying. Through codeword splitting, the channel input of the source is \textit{partially} coherent with the channel input of the relay. The first lower bound uses full DF at the relay and the second further enlarges it by allowing rate-splitting at the source. 

%%% FIXME: Feels too repetitive
\iffalse
If the channel input or the codeword generated by the relay is a function of the channel state and the cooperative information, it is very difficult for the source to generate codewords that are coherent with the codewords of the relay, because the source knows the cooperative information but not the channel states. But, at the same time, the relay should not completely ignore the known channel state by generating codewords that are function of only the cooperative information and, hence, allow the source to generate codewords that are coherent with the codewords of the relay. To overcome this difficulty, the relay generates the channel input as a superposition of one codeword which is a function of both the cooperative information and the channel state and another codeword which is a function of only the cooperative information and is independent of the channel state. The codeword that is a function of only the cooperative information is obtained by standard coding, and the codeword that is a function of both the channel state and the cooperative information is obtained through GDPC. Then, the source can generate codewords that are coherent with the codewords at the relay obtained by standard coding. This way,
\fi

We also point out the loss incurred by the availability of the channel state at only the relay in the upper bound. We show that the lower bound obtained with rate splitting at the source is in general close to the upper bound for general Gaussian channels. In the case of the degraded Gaussian channel, the two lower bounds meet and they meet with the upper bound for some special cases. 

%%% FIXME: Wonder if this is really useful?
In the case of half-duplex transmission, we derive lower and upper bounds for the capacity of the Gaussian relay channel with informed relay. In this case, we focus on relaying protocols in which the relay either fully or partially decodes the source message, re-encodes and sends it to the destination.

\subsection{Outline and Notation}\label{secI_subsecA}
An outline of the remainder of this paper is as follows. Section~\ref{secII} describes in mode detail the communication model that we consider in this work.  Section~\ref{secIII} provides lower and upper bounds on the capacity of the general discrete memoryless RC with informed relay. Section~\ref{secIV} provides lower and upper bounds on the capacity of the Gaussian RC with informed relay, and also contains some numerical results and discussions. Finally, Section \ref{secV} concludes the paper.

We use the following notations throughout the paper. Upper case letters are used to denote random variables, e.g., $X$; lower case letters are used to denote realizations of random variables, e.g., $x$; and calligraphic letters designate alphabets, i.e., $\mc X$. The probability distribution of a random variable $X$ is denoted by $P_X(x)$. Sometimes, for convenience, we write it as $P_X$.  We use the notation $\mathbb{E}_{X}[\cdot]$ to denote the expectation of random variable $X$. A probability distribution of a random variable $Y$ given $X$ is denoted by $P_{Y|X}$. The set of probability distributions defined on an alphabet $\mc X$ is denoted by $\mc P(\mc X)$. The cardinality of a set $\mc X$ is denoted by $|\mc X|$. The short-hand notation $X_i^j$ indicates a sequence of random variables $(X_i,X_{i+1},\cdots,X_j)$ and $x_i^j$ denotes a particular realization of a random sequence $X_i^j$. For convenience, the length $n$ vector $x^n$ will occasionally be denoted in boldface notation $\dv x$. Given random variables $X$, $Y$, $Z$, we denote the entropy of $X$ by $H(X)$, the mutual information between $X$ and $Y$ by $I(X;Y)$, and the conditional mutual information between $X$ and $Y$, conditioned on $Z$, by $I(X;Y|Z)$ \cite{CT91}. The Gaussian distribution with mean $\mu$ and variance $\sigma^2$ is denoted by $\mathcal{N}(\mu,\sigma^2)$. Finally, throughout the paper, logarithms are taken to base $2$, and the complement to unity of a scalar $u \in [0,1]$ is denoted by $\bar{u}$, i.e., $\bar{u}=1-u$.

\section{System Model and Definitions}\label{secII}
In this section, we formally present our communication model and the definitions related to it. As shown in Figure~\ref{StateDependentDiscreteMemorylessRelayChannel}, we consider a state-dependent relay channel  denoted by $W_{Y_2,Y_3|X_1,X_2,S}$ whose outputs $Y_2 \in \mc Y_2$ and $Y_3 \in \mc Y_3$ for the relay and the destination, respectively, are controlled by the channel inputs $X_1 \in \mc X_1$ from the source and $X_2 \in \mc X_2$ from the relay, along with a channel state $S \in \mc S$. It is assumed that the channel state $S_i$ at time instant $i$ is independently drawn from a given distribution $Q_S$ and the channel state $S^n$ is noncausally known at the relay. Also, each transmitted input block $x^n_1$ from the source and each transmitted input block $x^n_2$ from the relay are subject to additive normalized input constraints 
\begin{equation}
 \varphi^n_k(x^n_k) \triangleq \frac{1}{n} \sum_{i=1}^{n}\varphi_k(x_{k,i}) \leq \Gamma_k, \quad k=1,2
\end{equation}
where $\varphi_1\::\: \mc X_1 \rightarrow \mathbb{R}^{+}$ and $\varphi_2\::\: \mc X_2 \rightarrow \mathbb{R}^{+}$ are single-letter input cost functions for the source and the relay, respectively.

The source wants to transmit a message $W$ to the destination with the help of the relay, in $n$ channel uses. The message $W$ is assumed to be uniformly distributed over the set $\mc W=\{1,\cdots,M\}$. The information rate $R$ is defined as $\log M / n$ bits per transmission. 

An $(n,M,\Gamma_1,\Gamma_2)$ code for the state-dependent relay channel with informed relay consists of an encoding function at the source 
$$\phi_1^n:~~\{1,\cdots,M\} \rightarrow \mc X_1^n,$$ a sequence of encoding functions at the relay
$$\phi_{2,i}:~~ \calY_{2,1}^{i-1} \times \calS^n \rightarrow \calX_{2},$$ 
for $i=1,2,\ldots,n,$ and a decoding function at the destination 
$$\psi^n: ~~ \mc Y_3^n \rightarrow \{1,\cdots,M\}$$
such that $$\frac{1}{n}\sum_{i=1}^{n}\varphi_1(\phi^n_1(w)_i) \leq \Gamma_1$$
and $$\frac{1}{n}\sum_{i=1}^{n}\varphi_2(\phi_{2,i}(y^{i-1}_2,s^n)) \leq \Gamma_2$$
for $w \in \{1,\cdots,M\}$.

From an $(n,M,\Gamma_1,\Gamma_2)$ code, the sequences $X_{1}^n$ and $X_{2}^n$ from the source and the relay, respectively, are transmitted across a state-dependent relay channel modeled as a memoryless conditional probability distribution $W_{Y_2,Y_3|X_1,X_2,S}$. The joint probability mass function on ${\mc W}{\times}{\mc S^n}{\times}{\mc X^n_1}{\times}{\mc X^n_2}{\times}{\mc Y^n_2}{\times}{\mc Y^n_3}$ is given by
\begin{align}
P(w,s^n,x^n_1,x^n_2,y^n_2,y^n_3) &= P(w)\prod_{i=1}^{n}Q_S(s_i)P(x_{1,i}|w)P(x_{2,i}|s^n,y^{i-1}_2)\nonumber\\
&\qquad {\cdot}W_{Y_2,Y_3|X_1,X_2,S}(y_{2,i},y_{3,i}|x_{1,i},x_{2,i},s_i).
\end{align}
\iffalse
\begin{align}
&p(y_{2,i},y_{3,i}|x^i_1,x^i_2,s^n,y^{i-1}_2,y^{i-1}_3)=W_{Y_2,Y_3|X_1,X_2,S}(y_{2,i},y_{3,i}|x_{1,i},x_{2,i},s_i)
\end{align}
for all $i=1,\hdots,n$.
\fi

The channel is said to be \textit{physically degraded} if the conditional distribution 
$W_{Y_2,Y_3|X_1,X_2,S}$ factorizes as 
\begin{equation}
W_{Y_2,Y_3|X_1,X_2,S}=W_{Y_2|X_1,X_2,S}W_{Y_3|Y_2,X_2,S}.
\label{DistributionDegradedChannel}
\end{equation}

The destination estimates the message sent by the source from the channel output $Y_{3}^n$. The average probability of error is defined as $P_e^n = \mathrm{Pr}[\psi^n(Y_{3}^n)\neq W].$

An $(\epsilon,n,R,\Gamma_1,\Gamma_2)$ code for the state-dependent RC with informed relay is an $(n,\left\lceil 2^{nR} \right\rceil,\Gamma_1,\Gamma_2)$ code $(\phi_1^n,\phi_2^n,\psi^n)$ having average probability of error $P_e^n$ not exceeding $\epsilon$.

Given a pair $\dv \Gamma=(\Gamma_1,\Gamma_2)$, a rate $R$ is said to be $\dv \Gamma$-achievable if there exists a sequence of $(\epsilon_n,n,R,\Gamma_1,\Gamma_2)$-codes with $\lim_{n \rightarrow \infty} \epsilon_n=0$. The capacity $C(\dv \Gamma)$ of the state-dependent RC with informed relay is the supremum of the set of $\dv \Gamma$-achievable rates.

\section{The Discrete Memoryless RC with Informed Relay}\label{secIII} 
In this section, we assume that all the alphabets in the model, $\calS$, $\calX_1$, $\calX_2$,  $\calY_2$ and $\calY_3$, are discrete and finite. 

\subsection{Lower Bound on Capacity}
The following theorem provides a lower bound on the capacity of the state-dependent DM RC with informed relay.
\begin{theorem}\label{TheoremAchievabeRateNonCausalCaseDiscreteMemorylessChannel}
Let $\dv \Gamma=(\Gamma_1,\Gamma_2)$ be given. The capacity $C(\dv \Gamma)$ of the state-dependent DM RC with informed relay satisfies $C(\dv \Gamma) \ge R^{\text{lo}}(\dv \Gamma)$, where
\begin{align}
R^{\text{lo}}(\dv \Gamma)\:=\: \max \min \Big\{&I(X_1;Y_2|S,U_1,X_2), \nonumber\\
&I(X_1,U_1,U_2;Y_3)-I(U_2;S|U_1)\Big\},
\label{AchievabeRateNonCausalCaseDiscreteMemorylessChannel}
\end{align}
with the maximization over all probability distributions of the form 
\begin{align}
&P_{S,U_1,U_2,X_1,X_2,Y_2,Y_3}=\nonumber\\
&\hspace{0.8cm}Q_SP_{U_1}P_{X_1|U_1}P_{U_2|U_1,S}P_{X_2|U_1,U_2,S}W_{Y_2,Y_3|X_1,X_2,S}
\label{MeasureForAchievabeRateNonCausalCaseDiscreteMemorylessChannel}
\end{align}
and satisfying $\mathbb{E}[\varphi_i(X_i)] \leq \Gamma_i$, $i=1,2$, and $U_1 \in \calU_1$, $U_2 \in \calU_2$ are auxiliary random variables with
\begin{subequations}
\begin{align}
\label{BoundsOnCardinalityOfAuxiliaryRandonVariableU1ForAchievabeRateNonCausalCaseDiscreteMemorylessChannel}
&|\mc U_1| \leq |\mc S||\mc X_1||\mc X_2|+1\\
&|\mc U_2| \leq \Big(|\mc S||\mc X_1||\mc X_2|+1\Big)|\mc S||\mc X_1||\mc X_2|,
\label{BoundsOnCardinalityOfAuxiliaryRandonVariableU2ForAchievabeRateNonCausalCaseDiscreteMemorylessChannel}
\end{align}
\label{BoundsOnCardinalityOfAuxiliaryRandonVariablesForAchievabeRateNonCausalCaseDiscreteMemorylessChannel}
\end{subequations}
respectively.
\end{theorem}
%%% FIXME: I can't quite decide if the technique is really "codeword splitting" or "superposition" of GP onto DF.  I think it would help to have a dependence diagram of all the random variables involved.
%-- I think the denomination 'codeword splitting' is better suited for our code construction as it tells about the necessity of having one part of the relay's codeword not exploiting the known state, so that this part can be utilized to get a coherent transmission with the source. I find referring to our code construction as 'superposition of GP onto DF' not enough precise, as it does not really convey the above idea.

%-- A dependence diagram is added in Figure 3.

\begin{figure}[htpb]
\centering
\resizebox{0.5\linewidth}{!}{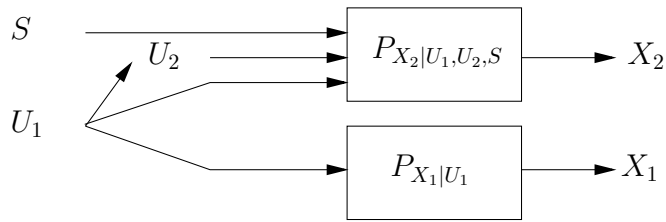}
\caption{Dependence diagram of the random variables for the lower bound in Theorem~\ref{TheoremAchievabeRateNonCausalCaseDiscreteMemorylessChannel}.}
\label{RandomVariablesDependenceDiagram}
\end{figure}

\begin{remark}\label{remark1}
The lower bound \eqref{AchievabeRateNonCausalCaseDiscreteMemorylessChannel} is based upon a technique at the relay we call \textit{codeword splitting}, combining decode-and-forward (DF) relaying \cite[Theorem 1]{CG79} with 
Gel'fand-Pinsker coding \cite{GP80}. In conventional DF strategies, the source knows the relay input, allowing the source and relay to utilize a joint codebook to transmit cooperative information.  However, in our model there is a tension between the utility of a joint codebook for relaying and the utility of the relay's making use of the channel state, which is unknown to the source.
\iffalse
what cooperative information the relay transmits. Consequently, the source generates the codeword that achieves some \textit{coherence} gain as in multi-antenna transmission, by having the channel input of the source correlated with that of the relay. In our model, at one hand, if the relay generates its channel input as a function of both the cooperative information and the channel state, then it is not possible for the source to know the relay input, because the source does not know the channel state. So, in this case, the source cannot generate codewords that are coherent with the relay input. On the other hand, the relay should not completely ignore the known channel state which it can use to remove its effect on the communication.
\fi
To resolve this tension, we generate two codebooks at the relay. In one codebook, the codewords are generated using a random variable $U_1$ that is independent of the channel state $S$. The relay chooses the appropriate codeword from this codebook using only the cooperative information. In the other codebook, the codewords are generated using a random variable $U_2$ that is correlated with the channel state $S$ and the variable $U_1$ through $P_{U_2|U_1,S}$. The relay chooses the appropriate codeword from this codebook using both the cooperative information and the channel state, in order to combat the effect of the channel state on the communication. Finally, the relay generates the channel input $X_2^n$ from $(U_1^n,U_2^n)$ using the conditional probability law $P_{X_2|U_1,U_2,S}$. The source knows $U_1^n$ as this is a function of only the cooperative information, and, given $U_1^n$, it generates the random codeword $X_1^n$ according to the conditional probability law $P_{X_1|U_1}$. Thus, the channel inputs of the source and the relay are correlated through $U_1^n$. A dependence diagram of the random variables that are involved in the coding scheme is shown in Figure~\ref{RandomVariablesDependenceDiagram}.
\end{remark}

%%% FIXME: Would a block diagram highlighting the differences be more appropriate?

%-- I think the remark as is conveys well enough the differences. 

\begin{remark}\label{remark2}
The term $[I(X_1,U_1,U_2;Y_3)-I(U_2;S|U_1)]$ in  \eqref{AchievabeRateNonCausalCaseDiscreteMemorylessChannel} can be interpreted as an achievable sum rate over a state-dependent MAC with one informed encoder and degraded messages, i.e., one common and one individual message.  In our model, the informed encoder sends only the common message, i.e., the cooperative information of DF relaying, and the uninformed encoder sends both the common and individual messages.  By contrast, \cite{SBSV07a,KL07} derive the capacity region for the reverse situation in which the informed encoder sends both the common and individual messages, and the uninformed encoder sends only the common message.  This swapping of roles makes coding at the relay more involved than in \cite{SBSV07a,KL07} for the state-dependent MAC and the \cite{ZV07b,ZV09b} for the related state-dependent RC with informed source. As we mentioned earlier, a MAC model that has closer connection with the model in this paper is investigated in \cite{ZKLV09a}. This model is obtained by swapping the roles of the encoders in \cite{KL07,SBSV07a}. 
\end{remark}

\noindent \textbf{Outline of Proof of Theorem~\ref{TheoremAchievabeRateNonCausalCaseDiscreteMemorylessChannel}:}\\
First we generate a random codebook that we use to obtain the lower bound in Theorem~\ref{TheoremAchievabeRateNonCausalCaseDiscreteMemorylessChannel}. Next, we outline the encoding and decoding procedures at the source and the relay. The coding scheme is based on a combination of codeword splitting, regular-encoding backward decoding for DF \cite{W82}, and a variation of Gel'fand-Pinsker binning. A formal proof with complete error analysis is given in Appendix~\ref{appendixTheorem1}. In the formal proof we also show that the input constraints are satisfied.

\textbf{Codebook generation:} Fix a measure $P_{S,U_1,U_2,X_1,X_2,Y_2,Y_3}$ satisfying \eqref{MeasureForAchievabeRateNonCausalCaseDiscreteMemorylessChannel} and $\mathbb{E}[\varphi_i(X_i)] \leq \Gamma_i$, $i=1,2$. Fix $\epsilon > 0$ and denote
\begin{subequations}
\begin{align}
J &=  2^{n(I(U_2;S|U_1)+2\epsilon)}\\
M &=  2^{n(R-4\epsilon)}.
\end{align}
\label{ValuesForBinningVariablesInTheorem1}
\end{subequations}

\begin{enumerate}
\item[1.] We generate $M$ independent and identically distributed (i.i.d.) codewords $\{\dv u_1(w')\}$ indexed by $w'=1,\hdots,M$, each with i.i.d. components drawn according to $P_{U_1}$.
\item[2.] For each codeword $\dv u_1(w')$, we generate $M$ i.i.d. codewords $\{\dv x_1(w',w)\}$ at the source indexed by $w=1,\hdots,M$, and $J$ auxiliary codewords $\{\dv u_2(w',j)\}$ at the relay indexed by $j=1,\hdots,J$. The codewords $\dv x_1(w',w)$ and  $\dv u_2(w',j)$ are with i.i.d. components  given $\dv u_1(w')$ drawn according to $P_{X_1|U_1}$ and $P_{U_2|U_1}$, respectively.
\end{enumerate}

%%% FIXME: I *still* think it is important to have a time index for the
%%% state vectors \mathbf{s} in the proofs.  We have to make it clear that
%%% for encoding message i, the relay knows the state in block i for the
%%% purpose of decoding the fresh information from the source, and knows
%%% and uses the state in block i+1 for the purpose of Gel'fand-Pinsker
%%% coding the cooperative information.

%-- FIXED

\textbf{Outline of the coding scheme:} We outline the coding scheme in the following. The message $W$ to be sent from the source node is divided into $B$ blocks $w_1,w_2,\hdots,w_B$ of $nR$ bits each. For convenience we let $w_{B+1}=1$. The transmission is performed in $B+1$ blocks. We denote by $\dv s[i]$ the channel state in block $i$, $i=1,\hdots,B+1$.

\noindent Continuing with the strategy, in the first block, the source transmits $\dv x_1(1,w_1)$. The relay searches for the smallest $j \in \{1,\cdots,J\}$ such that $\dv u_1(1)$, ${\dv u_2}(1,j)$ and $\dv s[1]$ are jointly typical (the properties of strongly typical sequences guarantee that there exists one such $j$). Denote this $j$ by $j^{\star}=j(\dv s[1], 1)$. Then, the relay transmits a vector $\dv x_2(1)$ with i.i.d. components given $(\dv u_1(1), \dv u_2(1,j^{\star}),\dv s[1])$ drawn according to the marginal $P_{X_2|U_1,U_2,S}$ induced by the distribution \eqref{MeasureForAchievabeRateNonCausalCaseDiscreteMemorylessChannel}. 

%By distortion typical we mean that $\dv u_1(1)$, ${\dv u_2}(1,j)$ and $\dv s[1]$ are jointly typical and $\varphi_2(\dv x_2(1)) \leq \Gamma_2$. (the achievability theorem from conventional rate-distortion theory guarantee that there exists one such $j$). 

The decoder at the relay uses joint typicality. It declares that message $\hat{w}_1$ is sent if there is a unique $\hat{w}_1$ such that $\dv x_1(1,\hat{w}_1)$ is jointly typical with $(\dv y_2[1],\dv s[1])$ given $\dv u_1(1)$, $\dv u_2(1,j^{\star})$ and $\dv x_2(1)$, where $\dv y_2[1]$ denotes the information received at the relay in block $1$. One can show that the relay can decode reliably as long as $n$ is large and
\begin{equation}
R < I(X_1;Y_2|S,U_1,X_2).
\label{Condition1OnAchievabeRateNonCausalCaseDiscreteMemorylessChannel}
\end{equation}

\noindent So, suppose the relay correctly obtains $w_1$. In the second block, the source transmits $\dv x_1(w_1,w_2)$ and the relay transmits a vector $\dv x_2(w_1)$ with i.i.d. components given $\dv u_1(w_1)$, $\dv u_2(w_1,j(\dv s[2],w_1))$, $\dv s[2]$ drawn according to the marginal $P_{X_2|U_1,U_2,S}$; the sequence $\dv u_2(w_1,j(\dv s[2],w_1))$ is chosen such that $j(\dv s[2],w_1)$ is the smallest $j \in \{1,\cdots,J\}$ satisfying $\dv u_1(w_1)$, ${\dv u_2}(w_1,j)$ and $\dv s[2]$ are jointly typical. Upon observation of $\dv y_2[2]$, the decoder at the relay declares that $\hat{w}_2$ is sent if there is a unique $\hat{w}_2$ such that $\dv x_1(w_1,\hat{w}_2)$ is jointly typical with $(\dv y_2[2],\dv s[2])$ given $\dv u_1(w_1)$, $\dv u_2(w_1,j(\dv s[2],w_1))$ and $\dv x_2(w_1)$. Again, it can decode reliably as long as $n$ is large and \eqref{Condition1OnAchievabeRateNonCausalCaseDiscreteMemorylessChannel} is true. At the relay, one continues in this way until block $B+1$.

Consider now the destination, and let $\dv y_3[i]$ be the received information at the destination in block $i$. Suppose these information are collected until the last block of transmission is completed. The destination can then perform Willem's \textit{backward decoding} \cite{W82}, by first decoding $w_B$ from $\dv y_3[B+1]$. Note that $\dv y_3[B+1]$ depends on $\dv x_1(w_B,1)$, $\dv u_1(w_B)$ and $\dv u_2(w_B,j(\dv s[B+1], w_B))$, which in turn depends only on $w_B$. The decoder at the destination uses joint typicality. It declares that $\hat{w}_B$ is sent if there is a unique $\hat{w}_B$ such that $\dv x_1(\hat{w}_B,1)$, $\dv u_1(\hat{w}_B)$, $\dv u_2(\hat{w}_B,j_B)$, $\dv y_3[B+1]$ are jointly typical, for some index $j_B \in \{1,\hdots,J\}$. One can show that the destination can decode reliably as long as $n$ is large and
\begin{equation}
R < I(X_1,U_1,U_2;Y_3)-I(U_2;S|U_1).
\label{Condition2OnAchievabeRateNonCausalCaseDiscreteMemorylessChannel}
\end{equation}

\noindent So, suppose the destination correctly obtains $w_B$. Next, the destination decodes $w_{B-1}$ from $\dv y_3[B]$, which depends on $\dv x_1(w_{B-1},w_B)$, $\dv u_1(w_{B-1})$ and $\dv u_2(w_{B-1},j(\dv s[B], w_{B-1}))$. Since the destination knows $w_{B}$, it can again decode reliably as long as $n$ is large and \eqref{Condition2OnAchievabeRateNonCausalCaseDiscreteMemorylessChannel} is true. At the destination, one continues in this fashion until all message blocks have been decoded. The average rate over the $B+1$ blocks is $RB/(B+1)$ bits per use, and by making $B$ large one can get the rate as close to $R$ as desired.

%%% FIXME: Is there a way to signal the end of remarks?  Otherwise, it is
%%% not clear where this remark really ends and the next paragraph of the
%%% discussion begins.
%-- I added a \qed to signal the end of the remarks.
\begin{remark}\label{remark3}
 In the case of classic RC without state, one can consider three different decode-and-forward strategies: irregular encoding successive decoding \cite{CG79}, regular encoding sliding-window decoding \cite{C82} and regular encoding backward decoding \cite{W82}. It is well known that these three strategies achieve the same rate in this case \cite{KGG05}.  In the state-dependent case with informed relay, one can show that backward decoding achieves rates higher than those of sliding-window decoding. More precisely, sliding window decoding at the destination at the end of block $i$ is as follows (we use the notation in the proof of Theorem \ref{TheoremAchievabeRateNonCausalCaseDiscreteMemorylessChannel}). The destination knows $w_{i-2}$ and also the correct index $j(\dv s[i-1],w_{i-2})$, and decodes $w_{i-1}$ based on the information received in the two adjacent blocks $i-1$ and $i$. It declares that the message $\hat{w}_{i-1}$ is sent if there is a unique pair $(\hat{w}_{i-1},\hat{j}_{i-1})$ such that the vectors $\dv u_1(w_{i-2})$, $\dv u_2(w_{i-2},j(\dv s[i-1],w_{i-2}))$, $\dv x_1(w_{i-2},\hat{w}_{i-1})$, $\dv y_3[i-1]$ are jointly typical, \textit{and} the vectors $\dv u_1(\hat{w}_{i-1})$, $\dv u_2(\hat{w}_{i-1},\hat{j}_{i-1})$, $\dv y_3[i]$ are jointly typical. Thus, the destination obtains the message $w_{i-1}$ if
\begin{align}
&R < I(X_1,U_1,U_2;Y_3)-I(U_2;S|U_1)\\ 
&I(U_2;Y_3|U_1)-I(U_2;S|U_1) > 0.
\label{AddionalConditionForWindowDecoding}
\end{align}
Hence, with window decoding also, the achievable rate is obtained by maximizing the RHS of \eqref{AchievabeRateNonCausalCaseDiscreteMemorylessChannel}. However, unlike the above backward decoding scheme, the maximization is over a set of distributions of the form \eqref{MeasureForAchievabeRateNonCausalCaseDiscreteMemorylessChannel} that satisfy the constraint \eqref{AddionalConditionForWindowDecoding}. Because of the additional constraint, this set is smaller than the one used in Theorem~\ref{TheoremAchievabeRateNonCausalCaseDiscreteMemorylessChannel}. Informally speaking, the additional constraint \eqref{AddionalConditionForWindowDecoding} guarantees that, in the decoding of the vectors $\dv u_1$ and $\dv u_2$, the destination can actually decode the vector $\dv u_2$ \textit{fully}, i.e., it can determine not only the bin index (i.e., the message $w_{i-1}$) but also the correct sequence in the bin (i.e., the index $j(\dv s[i], w_{i-1})$).\qed 
\end{remark}

The achievable rate in \eqref{AchievabeRateNonCausalCaseDiscreteMemorylessChannel} requires the relay to \textit{fully} decode the message sent by the source, and this can be rather a severe constraint. We can generalize Theorem~\ref{TheoremAchievabeRateNonCausalCaseDiscreteMemorylessChannel} by allowing the relay to decode the source message \textit{only partially} \cite{GA82}. This can be done by implementing rate-splitting at the source \cite{H-MZ05} and introducing a new random variable $U$ that represents the information decoded by the relay. The following corollary gives the resulting rate.

\begin{corollary}\label{CorollaryPartialDecodeAndForwardNonCausalCaseDiscreteMemorylessChannel}
The capacity $C(\dv \Gamma)$ of the state-dependent DM RC with informed relay satisfies $C(\dv \Gamma) \ge R'^{\text{lo}}(\dv \Gamma)$, where
\begin{align}
 R'^{\text{lo}}(\dv \Gamma)\: = \:\: \max & \min \Big\{I(U;Y_2|S,U_1,X_2)+I(X_1;Y_3|U,U_1,U_2)\nonumber\\
&+\min\{0,I(U_2;Y_3|U,U_1)-I(U_2;S|U_1)\}, I(X_1,U_1,U_2;Y_3)-I(U_2;S|U_1)\Big\},
%&I(U;Y_2|S,U_1)+I(X_1;Y_3|U,U_1,U_2)+\min\{0,I(U_2;Y_3|U,U_1)-I(U_2;S|U_1)\}\Big\},
\label{AchievabeRatePartialDecodeAndForwardNonCausalCaseDiscreteMemorylessChannel}
\end{align}
with the maximization over all probability distributions of the form
\begin{align}
&P_{S,U_1,U_2,U,X_1,X_2,Y_2,Y_3}=\nonumber\\
&\hspace{0.5cm}Q_SP_{U_1}P_{U|U_1}P_{X_1|U_1,U}P_{U_2|U_1,S}P_{X_2|U_1,U_2,S}W_{Y_2,Y_3|X_1,X_2,S}
\label{MeasureForAchievabeRatePartialDecodeAndForwardNonCausalCaseDiscreteMemorylessChannel}
\end{align}
and satisfying $\mathbb{E}[\varphi_i(X_i)] \leq \Gamma_i$, $i=1,2$, and $U_1 \in \calU_1$ , $U_2 \in \calU_2$, $U \in \calU$ are auxiliary random variables with
\begin{subequations}
\begin{align}
\label{BoundsOnCardinalityOfAuxiliaryRandonVariableU1ForAchievabeRatePartialDecodeAndForwardNonCausalCaseDiscreteMemorylessChannel}
&|\mc U_1| \leq |\mc S||\mc X_1||\mc X_2|+2\\ % INSTEAD OF +1
\label{BoundsOnCardinalityOfAuxiliaryRandonVariableU2ForAchievabeRatePartialDecodeAndForwardNonCausalCaseDiscreteMemorylessChannel}
&|\mc U_2| \leq \Big(|\mc S||\mc X_1||\mc X_2|+2\Big)|\mc S||\mc X_1||\mc X_2|+2\\ % INSTEAD OF +1 / + 0
\label{BoundsOnCardinalityOfAuxiliaryRandonVariableUForAchievabeRatePartialDecodeAndForwardNonCausalCaseDiscreteMemorylessChannel}
&|\mc U| \leq \Big(|\mc S||\mc X_1||\mc X_2|+2\Big)|\mc S||\mc X_1||\mc X_2|+2, % INSTEAD OF +1 / +0
\end{align}
\label{BoundsOnCardinalityOfAuxiliaryRandonVariablesForAchievabeRatePartialDecodeAndForwardNonCausalCaseDiscreteMemorylessChannel}
\end{subequations}
respectively.
\end{corollary}

The proof of Corollary \ref{CorollaryPartialDecodeAndForwardNonCausalCaseDiscreteMemorylessChannel} is similar to that of Theorem~\ref{TheoremAchievabeRateNonCausalCaseDiscreteMemorylessChannel} and, hence, only an outline of it is given in Appendix \ref{appendixCorollary1}. For instance, the particular choice $U=X_1$ in Corollary~ \ref{CorollaryPartialDecodeAndForwardNonCausalCaseDiscreteMemorylessChannel} gives the lower bound in Theorem~\ref{TheoremAchievabeRateNonCausalCaseDiscreteMemorylessChannel}.

An informal interpretation of the rate \eqref{AchievabeRatePartialDecodeAndForwardNonCausalCaseDiscreteMemorylessChannel} for the case in which $[I(U_2;Y_3|U,U_1)-I(U_2;S|U_1)] >0$ is as follows. Since $I(U;Y_3|U_1,U_2,X_1)=0$ for the distribution considered in \eqref{MeasureForAchievabeRatePartialDecodeAndForwardNonCausalCaseDiscreteMemorylessChannel}, the second term of the minimization in \eqref{AchievabeRatePartialDecodeAndForwardNonCausalCaseDiscreteMemorylessChannel} can be written as
\begin{equation*} 
I(U,U_1,U_2;Y_3)-I(U_2;S|U_1)+I(X_1;Y_3|U,U_1,U_2).
\end{equation*} 
The rate \eqref{AchievabeRatePartialDecodeAndForwardNonCausalCaseDiscreteMemorylessChannel} can then be interpreted as the rate achievable if the message $W$ transmitted by the source is split into two independent parts, one of which is transmitted through the relay, say at rate $R_r$, and the other is transmitted directly to the destination without the help of the relay, say at rate $R_d$. The total rate is $R=R_r+R_d$. In \eqref{AchievabeRatePartialDecodeAndForwardNonCausalCaseDiscreteMemorylessChannel} the auxiliary variable $U$ stands for the information decoded by the relay and plays the role of $X_1$ in Theorem \ref{TheoremAchievabeRateNonCausalCaseDiscreteMemorylessChannel}. Thus, it follows from \eqref{AchievabeRateNonCausalCaseDiscreteMemorylessChannel} that the message transmitted through the relay can be decoded correctly at the destination if rate $R_r$ satisfies
\begin{align}
R_r \: < \:\: \min \Big\{&I(U;Y_2|S,U_1,X_2), I(U,U_1,U_2;Y_3)-I(U_2;S|U_1)\Big\}.
\label{RateRrAchievabeRateRrPartialDecodeAndForwardNonCausalCaseDiscreteMemorylessChannel}
\end{align}
It can also be easily argued (see Appendix \ref{appendixCorollary1}) that the additional information, which is sent on top of the information transmitted through the relay, can be decoded correctly at the destination if rate $R_d$ satisfies
\begin{align}
 &R_d \: < \:\: I(X_1;Y_3|U,U_1,U_2).
\label{RateRdAchievabeRatePartialDecodeAndForwardNonCausalCaseDiscreteMemorylessChannel}
\end{align}
This shows that message $W$ can be sent at the rate \eqref{AchievabeRatePartialDecodeAndForwardNonCausalCaseDiscreteMemorylessChannel}.
%%% FIXME: For some reason, I crossed off the sentences below in reading
%%% the hardcopy.  Not sure why now.
\iffalse
We note that choosing $U=X_1$ in \eqref{AchievabeRatePartialDecodeAndForwardNonCausalCaseDiscreteMemorylessChannel} gives the lower bound in Theorem~\ref{TheoremAchievabeRateNonCausalCaseDiscreteMemorylessChannel}, as we mentioned previously. Further, this choice of $U$, i.e., $U=X_1$, is relevant in the special case in which the channel is physically degraded as the relay can decode \textit{all} the information intended to the destination in this case.  
\fi

%%% FIXME: I think the point of the discussion below is that CF is beyond
%%% our scope.  We should just say that, and not try to justify things.
%-- FIXED
%%% That being said, we should consider whether or not it is important to
%%% drop some of the discussion about general / degraded and full- / half-
%%% duplex and work in some discussion of CF.  Probably much more involved
%%% technically to do the latter, and the paper is already rather long.

\begin{remark}
The relay can employ other relaying schemes to assist the source, such as estimate-and-forward \cite{CG79}, amplify-and-forward \cite{SG00,GV02,LTW04} or combinations of these schemes. However, none of these schemes achieves capacity even if the channel is state-independent.  Hence, though some of these schemes may perform well in terms of achievable rates for some particular channels, we do not focus on these schemes in this paper.
\end{remark}

\iffalse
We close this section by noting that the relay can employ other relaying schemes to assist the source, such as estimate-and-forward \cite{CG79}, amplify-and-forward \cite{SG00,GV02,LTW04} or combinations of these schemes. However, in general, none of these schemes truly extracts the potential benefits of cooperation even if the channel is state-independent; and which scheme gives the largest rate depends on the particular channel of interest. Hence, though some of these schemes may perform well in terms of achievable rates for some particular channels, we do not focus on these schemes in this paper. Instead, we derive an upper bound on channel capacity that we will use to investigate the tightness of the established lower bounds.
\fi
\subsection{Upper Bound on Capacity}
The following theorem provides an upper bound on the capacity of the state-dependent DM RC with informed relay.

\begin{theorem}\label{TheoremOuterBoundNonCausalCaseDiscreteMemorylessChannel}
Let $\dv \Gamma=(\Gamma_1,\Gamma_2)$ be given. The capacity $C(\dv \Gamma)$ of the state-dependent DM RC with informed relay satisfies $C(\dv \Gamma) \le R^{\text{up}}(\dv \Gamma)$, where
\begin{align}
 R^{\text{up}}(\dv \Gamma) \:= \:\: \max \min \Big\{&I(X_1;Y_2,Y_3|S,X_2), \nonumber\\
&I(X_1,X_2;Y_3|S)-I(X_1;S|Y_3)\Big\}
\label{OuterBoundNonCausalCaseDiscreteMemorylessChannel}
\end{align}
with the maximization over all probability distributions of the form
\begin{align}
&P_{S,X_1,X_2,Y_2,Y_3}=Q_SP_{X_1}P_{X_2|X_1,S}W_{Y_2,Y_3|X_1,X_2,S}
\label{MeasureForOuterBoundNonCausalCaseDiscreteMemorylessChannel}
\end{align}
and satisfying $\mathbb{E}[\varphi_i(X_i)] \leq \Gamma_i$, $i=1,2$.
\end{theorem}
The proof of Theorem \ref{TheoremOuterBoundNonCausalCaseDiscreteMemorylessChannel} appears in Appendix \ref{appendixTheorem2}. 

In the second term of the minimum in \eqref{OuterBoundNonCausalCaseDiscreteMemorylessChannel}, $I(X_1;S|Y_3)$ can be interpreted as the rate penalty caused by the source's not knowing the channel state. This rate loss makes the above upper bound tighter than the trivial upper bound obtained by assuming that the channel state is also available at the source and the destination, i.e., the cut set upper bound
\begin{align}
&R^{\text{up}}_{\text{triv}}(\dv \Gamma) \:=\:\: \max \min \Big\{I(X_1;Y_2,Y_3|S,X_2),I(X_1,X_2;Y_3|S)\Big\}
\label{TrivialOuterBoundNonCausalCaseDiscreteMemorylessChannel}
\end{align}
with the maximization over all distributions of the form
\begin{align}
&P_{S,X_1,X_2,Y_2,Y_3}=Q_SP_{X_1|S}P_{X_2|X_1,S}W_{Y_2,Y_3|X_1,X_2,S}
\label{MeasureForTrivialOuterBoundNonCausalCaseDiscreteMemorylessChannel}
\end{align}
and satisfying $\mathbb{E}[\varphi_i(X_i)] \leq \Gamma_i$, $i=1,2$.

If the channel is physically degraded, the upper bound in Theorem \ref{TheoremOuterBoundNonCausalCaseDiscreteMemorylessChannel} reduces to the one in the following corollary.  
\begin{corollary}\label{CorollaryOuterBoundDegradedRelayChannelNonCausalCaseDiscreteMemorylessChannel}
The capacity of the state-dependent physically degraded RC with informed relay satisfies $C_{\text{D}}(\dv \Gamma) \leq R^{\text{up}}_{\text{D}}(\dv \Gamma)$, where
\begin{align}
R^{\text{up}}_{\text{D}}(\dv \Gamma) \:=\:\: \max \min \Big\{&I(X_1;Y_2|S,X_2), \nonumber\\
&I(X_1,X_2;Y_3|S)-I(X_1;S|Y_3)\Big\}
\label{OuterBoundDegradedChannelNonCausalCaseDiscreteMemorylessChannel}
\end{align}
with the maximization over all probability distributions of the form
\begin{align}
P_{S,X_1,X_2,Y_2,Y_3}&=Q_SP_{X_1}P_{X_2|X_1,S}W_{Y_2|X_1,X_2,S}W_{Y_3|Y_2,X_2,S}
\label{MeasureForOuterBoundDegradedChannelNonCausalCaseDiscreteMemorylessChannel}
\end{align}
and satisfying $\mathbb{E}[\varphi_i(X_i)] \leq \Gamma_i$, $i=1,2$.
\end{corollary}

Similar to the general case in Theorem~\ref{TheoremOuterBoundNonCausalCaseDiscreteMemorylessChannel}, the upper bound in Corollary~\ref{CorollaryOuterBoundDegradedRelayChannelNonCausalCaseDiscreteMemorylessChannel} is tighter than the trivial upper bound in \eqref{TrivialOuterBoundNonCausalCaseDiscreteMemorylessChannel} for the degraded case.

\section{The Gaussian RC with Informed Relay}\label{secIV}
In this section, we consider a state-dependent Gaussian RC in which both the channel state and the noise are additive and Gaussian. We also assume that the additive channel state is noncausally known to only the relay. First, we consider full-duplex transmission at the relay, i.e., the relay transmits and receives at the same time, and we derive lower and upper bounds on channel capacity for this case. Then we extend these results to the half-duplex mode in which the relay is constrained to operate in a time-division (TD) manner.

%The results obtained in Section \ref{secIII} for the DM case can be extended to memoryless channels with discrete time and continuous alphabets using standard techniques \cite[Chapter 7]{G68}. We use the bounds in Theorem \ref{TheoremAchievabeRateNonCausalCaseDiscreteMemorylessChannel} and Theorem \ref{TheoremOuterBoundNonCausalCaseDiscreteMemorylessChannel} to compute bounds on the capacity for the Gaussian case.

\subsection{Full-Duplex Channel Model}\label{secIV_subsecA}
For the full-duplex state-dependent Gaussian RC, the channel outputs $Y_{2,i}$ and $Y_{3,i}$ at time instant $i$ for the relay and the destination, respectively, are related to the channel input $X_{1,i}$ from the source and $X_{2,i}$ from the relay, and the channel state $S_i$ by
%%% FIXME: Just thinking out loud here, but what about scenarios in which
%%% the state affects one, but not both, links?  Might be worth thinking
%%% about in an applications context and publishing in Globecom / ICC....
\begin{subequations}
\begin{align}
\label{ReceivedAtRelayFullDuplexRegimeGaussianRCWithState}
Y_{2,i}&=X_{1,i}+S_i+Z_{2,i}\\
Y_{3,i}&=X_{1,i}+X_{2,i}+S_i+Z_{3,i},
\label{ReceivedAtDestinationFullDuplexRegimeGaussianRCWithState}
\end{align}
\label{ChannelModelForFullDuplexRegimeGaussianRCWithState}
\end{subequations}
where $S_i$ is a zero mean Gaussian random variable with variance $Q$, $Z_{2,i}$ is zero mean Gaussian with variance $N_2$, and $Z_{3,i}$ is zero mean Gaussian with variance $N_3$. The random variables $S_i$, $Z_{2,i}$ and $Z_{3,i}$ at time instant $i \in \{1,2,\ldots,n\}$ are mutually independent,  and are independent of $(S_j, Z_{2,j}, Z_{3,j})$ for $j \neq i$. The random variables $Z_{2,i}$ and $Z_{3,i}$ are also independent of the channel inputs $(X_1^n,X_2^n)$.

For the full-duplex degraded additive Gaussian RC, the channel outputs $Y_{2,i}$ and $Y_{3,i}$ for the relay and the destination, respectively, are related to the channel inputs $X_{1,i}$ and $X_{2,i}$ and the state $S_i$ by
\begin{subequations}
\begin{align}
\label{ReceivedAtRelayFullDuplexRegimeDegradedGaussianRCWithState}
Y_{2,i}&=X_{1,i}+S_i+Z_{2,i}\\
Y_{3,i}&=X_{2,i}+Y_{2,i}+Z'_{3,i},
\label{ReceivedAtDestinationFullDuplexRegimeDegradedGaussianRCWithState}
\end{align}
\label{ChannelModelForFullDuplexRegimeDegradedGaussianRCWithState}
\end{subequations}
where $(Z'_{3,1},\cdots,Z'_{3,n})$ is a sequence of i.i.d. zero mean Gaussian random variables with variance $N'_3=N_3-N_2$ which is independent of $Z^n_2$. 

The channel inputs from the source and the relay should satisfy the following average power constraints,
\begin{equation}
\sum_{i=1}^{n}X_{1,i}^2 \leq nP_1, \qquad \sum_{i=1}^{n}X_{2,i}^2 \leq nP_2.
\label{IndividualPowerConstraintsFullDuplexRegime}
\end{equation}
As we indicated previously, we assume that the channel state $S^n$ is noncausally known at only the relay. The definition of a code for this channel is the same as that given in Section \ref{secII}, with the additional constraint that the channel inputs should satisfy the power constraints \eqref{IndividualPowerConstraintsFullDuplexRegime}.

\subsection{Lower Bounds on Capacity}\label{secIV_subsecB}
In this section, we derive lower bounds on the capacity of the state-dependent full-duplex Gaussian RC with informed relay. The results obtained in Section \ref{secIII} for the DM case can be extended to memoryless channels with discrete time and continuous alphabets using standard techniques \cite[Chapter 7]{G68}. 

%We use the bounds in Theorem \ref{TheoremAchievabeRateNonCausalCaseDiscreteMemorylessChannel} and Corollary~\refTheorem \ref{TheoremOuterBoundNonCausalCaseDiscreteMemorylessChannel} to compute bounds on the capacity for the Gaussian case. 

The following theorem provides a lower bound on the capacity of the state-dependent full-duplex Gaussian RC with informed relay.
\begin{theorem}\label{TheoremAchievabeRateNonCausalCaseGaussianChannelFullDuplexRegime}
The capacity $C_{\text{G}}$ of the state-dependent Gaussian RC with informed relay satisfies $C_{\text{G}} \ge R^{\text{lo}}_{\text{G}}$, where
\begin{align}
 R^{\text{lo}}_{\text{G}}\:=\:\max_{\rho'_{12}} \min \Bigg\{&\frac{1}{2}\log(1+\frac{P_1(1-\rho'^2_{12})}{N_2}),\nonumber\\
&\max_{\theta, \rho'_{2s}}\:\:\frac{1}{2}\log\Big(1+\frac{P_1+\bar{\theta}P_2+2\rho'_{12}\sqrt{\bar{\theta}P_1P_2}}{{\theta}P_2+Q+N_3+2\rho'_{2s}\sqrt{{\theta}P_2Q}}\Big)+\frac{1}{2}\log(1+\frac{{\theta}P_2(1-\rho'^2_{2s})}{N_3})\Bigg\},
\label{AchievabeRateNonCausalCaseGaussianChannelFullDuplexRegime}
\end{align}
with the maximization over parameters $\rho'_{12} \in [0,1]$, $\theta \in [0,1]$, and $\rho'_{2s} \in [-1,0]$.
\end{theorem}
\noindent \textbf{Proof:} A formal proof of Theorem \ref{TheoremAchievabeRateNonCausalCaseGaussianChannelFullDuplexRegime} is given in Appendix \ref{appendixTheorem3}.

\noindent \textbf{Outline of Proof:}
\begin{itemize}
%%% FIXME: I feel this remark is too repetitive with corresponding one for
%%% the DM case.  Think we should say something like "By extension, Remark %%% 1 applies Gaussian case, with the following differences.", and then
%%% only highlight the differences and key citations for the Gaussian case.
%%% Hope that makes sense.
%-- I rephrased parts of the remark. Hope it is clearer now.
\item We compute the lower bound \eqref{AchievabeRateNonCausalCaseDiscreteMemorylessChannel} for an appropriate choice of the input distribution that will be specified in the sequel. By extension, Remark~\ref{remark1} also applies for the Gaussian case. More specifically, we should consider two important features in the design of an efficient coding scheme at the relay: obtaining correlation or coherence between the channel inputs from the source and the relay, and   exploiting the channel state to remove the effect of the state on the communication. As we already mentioned, it is not obvious to accomplish these features because the channel state is not available at the source. Proceeding like for the code construction in the DM case, we split the relay input $X_2^n$ into two parts, namely $U_1^n$ and $\tilde{X}_2^n$. Furthermore, here we set $U_1^n$ and $\tilde{X}_2^n$ to be \textit{independent}. The first part, $U_1^n$, is a function of only the cooperative information, and is generated using standard coding. Since the source knows the cooperative information at the relay, it can generate its codeword $X_1^n$ in such a way that it is coherent with $U_1^n$, by allowing correlation between $X_1^n$ and $U_1^n$. The second part, $\tilde{X}_2^n$, which is independent of the source input $X_1^n$, is a function of both the cooperative information and the channel state $S^n$ at the relay, and is generated using a GDPC similar to that in \cite{KL07a, SBSV07a, ZV07b}. 

\item More formally, we decompose the relay input random variable $X_2$ as
\begin{equation}
X_2=U_1+\tilde{X}_2,
\label{RelayInputAchievableRateFullDuplexGaussianRC}
\end{equation}
where: $U_1$ is zero mean Gaussian with variance $\bar{\theta}P_2$, is independent of both $\tilde{X}_2$ and $S$, and is correlated with $X_1$ with $\mathbb{E}[U_1X_1]=\rho'_{12}\sqrt{\bar{\theta}P_1P_2}$, for some $\theta \in [0,1]$, $\rho'_{12} \in [-1,1]$ ; and $\tilde{X}_2$ is zero mean Gaussian with variance ${\theta}P_2$, is independent of $X_1$, and is correlated with the channel state $S$ with $\mathbb{E}[\tilde{X}_2S]=\rho'_{2s}\sqrt{{\theta}P_2Q}$, for some $\rho'_{2s} \in [-1,1]$. Expressed in terms of the covariances $\sigma_{12}=\mathbb{E}[X_1X_2]=\mathbb{E}[X_1U_1]$ and $\sigma_{2s}=\mathbb{E}[X_2S]=\mathbb{E}[\tilde{X}_2S]$, the parameters $\rho'_{12}$, $\rho'_{2s}$ are given by 
\begin{equation}
\rho'_{12}=\frac{\sigma_{12}}{\sqrt{\bar{\theta}P_1P_2}},\quad \rho'_{2s}=\frac{\sigma_{2s}}{\sqrt{{\theta}P_2Q}}.
\label{CorrelationCoefficientsFullDuplexGaussianRC}
\end{equation}

For the GDPC, we  choose the auxiliary random variable $U_2$ as 
\begin{equation}
U_2=\tilde{X}_2+\alpha_{\text{opt}}S
\label{GeneralizedDPCatRelayFullDuplexGaussianRC}
\end{equation}
with 
\begin{align}
\alpha_{\text{opt}}&=\frac{{\theta}P_2(1-\rho'^2_{2s})-\rho'_{2s}\sqrt{\frac{{\theta}P_2}{Q}}N_3}{{\theta}P_2(1-\rho'^2_{2s})+N_3}.
\label{OptimalCostaParameterGeneralizedDPCatRelayFullDuplexGaussianRC}
\end{align}
\end{itemize}

Similarly to in the DM case, we can generalize Theorem~\ref{TheoremAchievabeRateNonCausalCaseGaussianChannelFullDuplexRegime} by allowing the relay to decode the source message only partially, through rate-splitting at the source. The following corollary gives the resulting rate.

\begin{corollary}\label{CorollaryAchievabeRatePartialDecodeAndForwardNonCausalCaseGaussianChannel}
The capacity $C_{\text{G}}$ of the state-dependent Gaussian RC with informed relay satisfies $C_{\text{G}} \ge R'^{\text{lo}}_{\text{G}}$, where
\begin{equation}
 R'^{\text{lo}}_{\text{G}}\:=\:\max \min \{ T_1, T_2, T_3\}
\label{AchievabeRate__PartialDF__GaussianCase__FullDuplexRegime}
\end{equation}
with
\begin{align}
\label{AchievabeRate__PartialDF__GaussianCase__FullDuplexRegime_T1}
T_1 &= \frac{1}{2}\log\Big(1+\frac{\bar{\gamma}P_1(1-\rho'^2_{12})}{N_2+{\gamma}P_1}\Big) +\frac{1}{2}\log\Big(1+\frac{{\gamma}P_1}{N_3+\Phi(\alpha',\theta,\rho'_{2s})}\Big)\\
\label{AchievabeRate__PartialDF__GaussianCase__FullDuplexRegime_T2}
T_2 &= \frac{1}{2}\log\Big(1+\frac{\bar{\gamma}P_1(1-\rho'^2_{12})}{N_2+{\gamma}P_1}\Big)+\frac{1}{2}\log\Big(\frac{P'_2(P'_2+Q'+{\gamma}P_1+N_3)}{P'_2Q'(1-\alpha')^2+N_3(P'_2+\alpha'^2Q')}\Big)\\
T_3 &=\frac{1}{2}\log\Big(1+\frac{P_1+\bar{\theta}P_2+2\rho'_{12}\sqrt{\bar{\theta}\bar{\gamma}P_1P_2}}{{\theta}P_2+Q+N_3+2\rho'_{2s}\sqrt{{\theta}P_2Q}}\Big) + \frac{1}{2}\log\Big(\frac{P'_2(P'_2+Q'+N_3)}{P'_2Q'(1-\alpha')^2+N_3(P'_2+\alpha'^2Q')}\Big);
\label{AchievabeRate__PartialDF__GaussianCase__FullDuplexRegime_T3}
\end{align}
$P'_2:={\theta}P_2(1-\rho'^2_{2s})$, $Q':=(\sqrt{Q}+\rho'_{2s}\sqrt{{\theta}P_2})^2$, $\Phi(\alpha',\theta,\rho'_{2s}):=\frac{P'_2Q'(1-\alpha')^2}{P'_2+\alpha'^2Q'}$; and the maximization is over parameters  $\gamma \in [0,1]$,  $\theta \in [0,1]$, $\rho'_{12} \in [0,1]$, $\rho'_{2s} \in [-1,0]$, and $\alpha' \in \mathbb{R}$ such that the second logarithm terms in $T_2$ and $T_3$ are defined.\\
\end{corollary}

\noindent \textbf{Outline of Proof:} An informal proof of Corollary~\ref{CorollaryAchievabeRatePartialDecodeAndForwardNonCausalCaseGaussianChannel} is as follows. We decompose the message $W$ to be sent from the source into two independent parts $W_r$ and $W_d$. The message $W_r$ will be sent through the relay, at rate $R_r$; and the message $W_d$ will be sent directly to the destination, at rate $R_d$. The total rate is $R'_{\text{G}}=R_r+R_d$. The input $X^n_1$ from the source is divided accordingly into two independent parts, i.e., $X^n_1=U^n+\tilde{X}^n_1$, where $U^n$ carries message $W_r$ and has power constraint $n\bar{\gamma}P_1$ and $\tilde{X}^n_1$ carries message $W_d$ and has power constraint $n{\gamma}P_1$, for some $\gamma \in [0,1]$. The relay decodes and forwards only the part $U^n$, and its input sequence is obtained in a manner which is similar to that in the coding scheme for Theorem~\ref{TheoremAchievabeRateNonCausalCaseGaussianChannelFullDuplexRegime} (with $U^n$ playing the role of $X^n_1$ therein).  

The rest of the proof follows by computing the lower bound in Corollary~\ref{CorollaryPartialDecodeAndForwardNonCausalCaseDiscreteMemorylessChannel} using an input distribution and techniques that are essentially similar to those in the proof of Theorem~\ref{TheoremAchievabeRateNonCausalCaseGaussianChannelFullDuplexRegime}. An outline of the important steps is given in Appendix~\ref{appendixCorollary3}.

\iffalse
The formal proof of Corollary~\ref{TheoremAchievabeRatePartialDecodeAndForwardNonCausalCaseGaussianChannel} is similar to that of Theorem~\ref{TheoremAchievabeRateNonCausalCaseGaussianChannelFullDuplexRegime}, and, hence, we only outline it in the following. 

More specifically, we decompose the source $X_1$ and the relay input $X_2$ as
\begin{align}
X_1 &= U+\tilde{X}_1\\
X_2 &= U_1+\tilde{X}_2,
\end{align}
where $U$ and $\tilde{X}_1$ are independent zero mean Gaussian random variables with variances $\bar{\gamma}P_1$ and ${\gamma}P_1$, respectively, for some $\gamma \in [0,1]$; and $U_1$ and $\tilde{X}_2$ are independent zero mean Gaussian random variables with variances $\bar{\theta}P_2$ and ${\theta}P_2$, respectively, for some $\theta \in [0,1]$. Furthermore, $\tilde{X}_1$ is independent of all other variables; $U$ and $U_1$ are correlated, with $\mathbb{E}[UU_1]=\rho'_{12}\sqrt{\bar{\theta}\bar{\gamma}P_1P_2}$ for some $\rho'_{12} \in [0,1]$, and are both independent of $S$; $\tilde{X}_2$ is independent of $U$ and is correlated with $S$, with $\mathbb{E}[\tilde{X}_2S]=\rho'_{2s}\sqrt{{\theta}P_2Q}$, for some $\rho'_{2s} \in [-1,0]$. 

The rest of the proof follows by computing the lower bound in Corollary~\ref{CorollaryPartialDecodeAndForwardNonCausalCaseDiscreteMemorylessChannel}, with the auxiliary random variable $U_2$ chosen as
\begin{equation} 
U_2 = \tilde{X}_2 +[\alpha'(1+\rho'_{2s}\sqrt{\frac{{\theta}P_2}{Q}})-\rho'_{2s}\sqrt{\frac{{\theta}P_2}{Q}}]S,
\end{equation}
for some $\alpha' \in \mathbb{R}$.
\fi

\subsection{Upper Bound on Capacity}\label{secIV_subsecC}

The following theorem provides an upper bound on the capacity of the state-dependent full-duplex general Gaussian RC with informed relay.

\begin{theorem}\label{TheoremOuterBoundNonCausalCaseGaussianChannelFullDuplexRegime}
The capacity $C_{\text{G}}$ of the state-dependent general Gaussian RC with informed relay satisfies $C_{\text{G}} \le  R^{\text{up}}_{\text{G}}$, where
\begin{align}
 R^{\text{up}}_{\text{G}} \:= &\max \min \Bigg\{\frac{1}{2}\log\Big(1+P_1(1-\frac{\rho^2_{12}}{1-\rho^2_{2s}})(\frac{1}{N_2}+\frac{1}{N_3})\Big),\nonumber\\
&\frac{1}{2}\log\Big(1+\frac{(\sqrt{P_1}+\rho_{12}\sqrt{P_2})^2}{P_2(1-\rho^2_{12}-\rho^2_{2s})+(\sqrt{Q}+\rho_{2s}\sqrt{P_2})^2+N_3}\Big)+\frac{1}{2}\log(1+\frac{P_2(1-\rho^2_{12}-\rho^2_{2s})}{N_3})\Bigg\},
\label{OuterBoundNonCausalCaseGaussianChannelFullDuplexRegime}
\end{align}
 with the maximization over parameters $\rho_{12} \in [0,1]$ and $\rho_{2s} \in [-1,0]$ such that
\begin{equation}
\rho^2_{12}+\rho^2_{2s} \leq 1.
\label{MaximizationRangeOuterBoundNonCausalCaseGaussianChannelFullDuplexRegime}
\end{equation}
\end{theorem}

\noindent \textbf{Proof:} The proof of Theorem \ref{TheoremOuterBoundNonCausalCaseGaussianChannelFullDuplexRegime} is given in Appendix \ref{appendixTheorem4}. In the proof, we evaluate\footnote{In Theorem \ref{TheoremOuterBoundNonCausalCaseGaussianChannelFullDuplexRegime}, if the maximizing $\rho_{2s}$ in \eqref{OuterBoundNonCausalCaseGaussianChannelFullDuplexRegime} has absolute value equal to unity then \eqref{MaximizationRangeOuterBoundNonCausalCaseGaussianChannelFullDuplexRegime} implies that $\rho_{12}$ is zero. In this case, and also in the rest of this paper, we use the convention that $\frac{0}{0}=0$.} the upper bound \eqref{OuterBoundNonCausalCaseDiscreteMemorylessChannel} using an appropriate joint distribution of $S,X_1,X_2,Y_2,Y_3$. 

Following straightforwardly the proof of Theorem~\ref{TheoremOuterBoundNonCausalCaseGaussianChannelFullDuplexRegime} in Appendix \ref{appendixTheorem4}, it can be easily shown that the capacity of the state-dependent degraded Gaussian RC is upper-bounded as in the following corollary.
\vspace{0.2cm}

%%% FIXME: Don't use "replaced by" here; write out the full result.
%-- FIXED
\begin{corollary}\label{CorollaryUpperBoundDegradedGaussianChannelFullDuplexRegime}
The capacity $C_{\text{DG}}$ of the state-dependent degraded Gaussian RC with informed relay satisfies $C_{\text{DG}} \leq R^{\text{up}}_{\text{DG}}$, where 
\begin{align}
 R^{\text{up}}_{\text{DG}} \:= &\max \min \Big\{\frac{1}{2}\log\Big(1+\frac{P_1(1-\rho^2_{12}-\rho^2_{2s})}{N_2(1-\rho^2_{2s})}\Big),\nonumber\\
&\frac{1}{2}\log\Big(1+\frac{(\sqrt{P_1}+\rho_{12}\sqrt{P_2})^2}{P_2(1-\rho^2_{12}-\rho^2_{2s})+(\sqrt{Q}+\rho_{2s}\sqrt{P_2})^2+N_3}\Big)+\frac{1}{2}\log(1+\frac{P_2(1-\rho^2_{12}-\rho^2_{2s})}{N_3})\Big\},
\label{OuterBoundNonCausalCaseDegradedGaussianChannelFullDuplexRegime}
\end{align}
 with the maximization over parameters $\rho_{12} \in [0,1]$ and $\rho_{2s} \in [-1,0]$ such that
\begin{equation}
\rho^2_{12}+\rho^2_{2s} \leq 1.
\end{equation}
\end{corollary}

\subsection{Analysis of some Special Cases}\label{secIV_subsecD}

We note that comparing the above lower and upper bounds  analytically can be tedious in the general case. In what follows, we identify a few cases in which the lower bound in Theorem~\ref{TheoremAchievabeRateNonCausalCaseGaussianChannelFullDuplexRegime} and the upper bound in  Corollary~\ref{CorollaryUpperBoundDegradedGaussianChannelFullDuplexRegime} meet for degraded Gaussian channels, and some extreme cases for which the lower bound in  Corollary~\ref{CorollaryAchievabeRatePartialDecodeAndForwardNonCausalCaseGaussianChannel} and the upper bound in Theorem~\ref{TheoremOuterBoundNonCausalCaseGaussianChannelFullDuplexRegime} meet for general, i.e., not necessarily degraded, Gaussian channels; and so we obtain the capacity expression for these cases. 

In the following corollary we recast the lower bound \eqref{AchievabeRateNonCausalCaseGaussianChannelFullDuplexRegime} into an equivalent form  by substituting $\varrho_{12}=\rho'_{12}\sqrt{\bar{\theta}}$ and $\varrho_{2s}=\rho'_{2s}\sqrt{\theta}$. Also, we recast the upper bound given in Corollary~\ref{CorollaryUpperBoundDegradedGaussianChannelFullDuplexRegime} into an equivalent form by substituting $\kappa=\rho_{12}/\sqrt{1-\rho^2_{2s}}$ and $\rho=\rho_{2s}$. 

\begin{corollary}\label{CorollaryEquivalentFormOuterBoundNonCausalCaseDegradedGaussianChannelFullDuplexRegime}
For the Gaussian RC, the lower bound \eqref{AchievabeRateNonCausalCaseGaussianChannelFullDuplexRegime} in Theorem \ref{TheoremAchievabeRateNonCausalCaseGaussianChannelFullDuplexRegime} can be written as
\begin{align}
 R^{\text{lo}}_{\text{G}}=&\max \min \Bigg\{\frac{1}{2}\log(1+\frac{P_1(1-\varrho^2_{12}-\theta)}{N_2(1-\theta)}),\nonumber\\
&\frac{1}{2}\log\Big(1+\frac{(\sqrt{P_1}+\varrho_{12}\sqrt{P_2})^2+(\bar{\theta}-\varrho^2_{12})P_2}{P_2(1-\bar{\theta}-\varrho^2_{2s})+(\sqrt{Q}+\varrho_{2s}\sqrt{P_2})^2+N_3}\Big)+\frac{1}{2}\log(1+\frac{P_2(1-\bar{\theta}-\varrho^2_{2s})}{N_3})\Bigg\},
\label{EquivalentFormForAchievabeRateNonCausalCaseGaussianChannelFullDuplexRegime}
\end{align}
where the maximization is over parameters $\theta \in [0,1]$, $\varrho_{12} \in [0,1]$, $\varrho_{2s} \in [-1,0]$ such that
\begin{equation}
\varrho^2_{12}+\varrho^2_{2s} \leq 1.
\label{MaximizationRangeAchievableRateNonCausalCaseGaussianChannelFullDuplexRegime}
\end{equation}
For the physically degraded case, the upper bound in Corollary \ref{CorollaryUpperBoundDegradedGaussianChannelFullDuplexRegime} can be written as
\begin{align}
 R^{\text{up}}_{\text{DG}} \:=\:\: \max_{\kappa} \min \Bigg\{&\frac{1}{2}\log\Big(1+\frac{P_1(1-\kappa^2)}{N_2}\Big),\nonumber\\
&\max_{\rho} \frac{1}{2}\log(1+\frac{P_2(1-\kappa^2(1-\rho^2)-\rho^2)}{N_3})\nonumber\\
&+\frac{1}{2}\log\Big(1+\frac{P_1+\kappa^2(1-\rho^2)P_2+2\kappa\sqrt{1-\rho^2}\sqrt{P_1P_2}}{P_2(1-\kappa^2(1-\rho^2))+Q+2\rho\sqrt{P_2Q}+N_3}\Big)\Bigg\},
\label{EquivalentFormOuterBoundNonCausalCaseDegradedGaussianChannelFullDuplexRegime}
\end{align}
where the maximization is over parameters $\kappa \in [0,1]$ and $\rho \in [-1,0]$.\\ 
\end{corollary}

By investigating the bounds in Theorem~\ref{TheoremAchievabeRateNonCausalCaseGaussianChannelFullDuplexRegime} and Corollary~\ref{CorollaryUpperBoundDegradedGaussianChannelFullDuplexRegime} and the equivalent expressions of these bounds in Corollary~\ref{CorollaryEquivalentFormOuterBoundNonCausalCaseDegradedGaussianChannelFullDuplexRegime}, it can be shown that the lower bound for the degraded case is tight for certain values of $P_1$, $P_2$, $Q$, $N_2$, $N_3$. The following observation states some cases for which the lower bound is tight. 

\begin{observation}\label{CapacityForSpecialCases}
For the physically degraded Gaussian RC, we have:
%%% FIXME: Is this really interesting, or just a reasonable sanity check?
%%% Noise at the relay limits performance, relay knows state, so can cancel
%%% its effect.  We should be careful about inflating the observation to
%%% a proposition.
%-- Changed into 'Observation'. Originally inflated to proposition because of the converse (but the converse has been dropped out after discussion with Shiva, for reasons of simplification). The converse in now stated in the text after the observation.

$1)$ If $P_1$, $P_2$, $Q$, $N_2$, $N_3$ satisfy 
\begin{align}
N_2 \geq \max_{\zeta \in [-1,0]} \frac{P_1N_3(P_2+Q+N_3+2{\zeta}\sqrt{P_2Q})}{P_1N_3+P_2(1-\zeta^2)(P_1+P_2+Q+N_3+2{\zeta}\sqrt{P_2Q})},
\label{SnrRangeForChannelCapacityLowSnrAtRelay}
\end{align}
then channel capacity is given by
\begin{equation}
C_{\text{DG}} = \frac{1}{2}\log(1+\frac{P_1}{N_2}),
\label{ChannelCapacityLowSnrAtRelay}
\end{equation}
which is the same as the interference-free capacity, i.e., the capacity if the channel state were not present in the model, or were also known to the source.

$2)$ If the maximizing $\rho_{12}$ and $\rho_{2s}$ in the upper bound in Corollary \ref{CorollaryUpperBoundDegradedGaussianChannelFullDuplexRegime} are such that condition \eqref{MaximizationRangeOuterBoundNonCausalCaseGaussianChannelFullDuplexRegime} is met with equality, i.e., $\rho^2_{12}+\rho^2_{2s}=1$, then the lower bound \eqref{EquivalentFormForAchievabeRateNonCausalCaseGaussianChannelFullDuplexRegime} is tight and gives the capacity.
\end{observation}
\noindent \textbf{Proof:} The proof of observation \ref{CapacityForSpecialCases} appears in  Appendix~\ref{appendixProposition1}.

\begin{remark}
The condition in \eqref{SnrRangeForChannelCapacityLowSnrAtRelay} specifies a range of values $(P_1,P_2,Q,N_2,N_3)$ for which the lower bound for the degraded Gaussian case is tight. In this case, the capacity  is the same as that of the degraded Gaussian RC with informed relay and informed source or interference-free capacity. Thus, the first statement in Observation~\ref{CapacityForSpecialCases} also provides a \textit{sufficient} condition for the rate loss incurred by not knowing the interference at the source as well to be zero. At a high level, condition \eqref{SnrRangeForChannelCapacityLowSnrAtRelay} means that there is no rate loss due to the asymmetry when capacity is constrained by the broadcast part in the model, i.e, transmission from the source to the relay and the destination. By investigating the upper bound \eqref{EquivalentFormOuterBoundNonCausalCaseDegradedGaussianChannelFullDuplexRegime} and comparing it with the interference-free capacity, it can be shown that this condition is also \textit{necessary}. That is, the interference-free capacity is attained \textit{only} if \eqref{SnrRangeForChannelCapacityLowSnrAtRelay} is fulfilled. If the capacity of our model is constrained by the sum rate of the cooperative MAC part, i.e., the cooperative transmission from the source and the relay to the destination, the asymmetry resulting from not knowing the interference at the source as well causes an inevitable rate loss, i.e., the term $I(X_1;S|Y_3)$ in Corollary \ref{CorollaryOuterBoundDegradedRelayChannelNonCausalCaseDiscreteMemorylessChannel}.
\end{remark}

\iffalse
having the channel state at only the relay (relative to the interference-free capacity) is zero if the capacity is constrained by the broadcast part in the model, i.e, transmission from the source to the relay and the destination. Investigating the upper bound \eqref{EquivalentFormOuterBoundNonCausalCaseDegradedGaussianChannelFullDuplexRegime}, it can be easily shown that the rate loss caused by not knowing the interference at the source as well is zero \textit{only} if condition in \eqref{SnrRangeForChannelCapacityLowSnrAtRelay} is fulfilled. If the capacity of our model is constrained by the sum rate of the cooperative MAC part, i.e., the cooperative transmission from the source and the relay to the destination, the asymmetry resulting from not knowing the interference at the source as well causes a rate loss, i.e., the term $I(X_1;S|Y_3)$ in Corollary \ref{CorollaryOuterBoundDegradedRelayChannelNonCausalCaseDiscreteMemorylessChannel}. 
\fi

\noindent \textbf{Extreme Cases} \\
We now summarize the behavior of the above bounds in some extreme cases.
\begin{enumerate}
\item[1)] \textit{Arbitrarily strong channel state:} In the asymptotic case $Q \rightarrow \infty$, the lower bound in Theorem~\ref{TheoremAchievabeRateNonCausalCaseGaussianChannelFullDuplexRegime} and the upper bound in Corollary~\ref{CorollaryEquivalentFormOuterBoundNonCausalCaseDegradedGaussianChannelFullDuplexRegime} meet, thus yielding the capacity for degraded Gaussian RC  
\begin{align}
& C_{\text{DG}}(Q=\infty) \: =\:\: \min \Big\{\frac{1}{2}\log(1+\frac{P_1}{N_2}),\:\frac{1}{2}\log(1+\frac{P_2}{N_3})\Big\}.
\label{CapacityVeryStrongSideInformationDegradedGaussianChannelFullDuplexRegime}
\end{align}

Equation \eqref{CapacityVeryStrongSideInformationDegradedGaussianChannelFullDuplexRegime} suggests that traditional multi-hop transmission achieves the capacity in this case. A two-hop scheme allows to completely cancel the effect of the channel state by subtracting it out upon reception at the relay, and by applying standard DPC for transmission from the relay to the destination.

For arbitrarily strong channel state and general, i.e., not necessarily degraded, Gaussian RC, the lower bound in Corollary~\ref{CorollaryAchievabeRatePartialDecodeAndForwardNonCausalCaseGaussianChannel} and the upper bound in Theorem~\ref{TheoremOuterBoundNonCausalCaseGaussianChannelFullDuplexRegime} meet if $P_2N_2 \leq P_1N_3$ or $P_2+N_3 \leq P_1$, and capacity in these cases is given by 
\begin{align}
& C_{\text{G}}(Q=\infty) \: =\:\: \frac{1}{2}\log(1+\frac{P_2}{N_3}).
\label{CapacityVeryStrongSideInformationGeneralGaussianChannelFullDuplexRegime}
\end{align}
It is interesting to note that if $P_2+N_3 \leq P_1$ the lower bound in Corollary~\ref{CorollaryAchievabeRatePartialDecodeAndForwardNonCausalCaseGaussianChannel} is maximized for $\alpha'=P_2/(P_2+N_3)$ and $\gamma=1$, meaning that direct transmission from the source to the relay is, not only possible, but also optimal in this case. The relay transmits independent information and decoding this information and subtracting it out at the destination, in a sense, clears the channel for the direct transmission.

\item[2)] \textit{Deaf helper problem:} In the case in which the relay is unable to \textit{hear} the source (e.g., due to a very noisy or broken link source-to-relay) and $Q \rightarrow \infty$,  the lower bound in Corollary~\ref{CorollaryAchievabeRatePartialDecodeAndForwardNonCausalCaseGaussianChannel} and the upper bound in Theorem~\ref{TheoremOuterBoundNonCausalCaseGaussianChannelFullDuplexRegime} meet if $N_3 \leq |P_1-P_2|$, giving 
\begin{equation} 
C_{\text{G}}(Q=\infty,N_2=\infty)\: =\:\frac{1}{2}\log(1+\frac{\min\{P_1,P_2\}}{N_3}).
\label{RateDeafHelperProblem}
\end{equation}
If $Q,N_2 \longrightarrow \infty$ and $N_3 > |P_1-P_2|$ the bounds do not meet. However, the lower bound is "within one bit" from the upper bound if $P_1+N_3>P_2$, and it reaches it asymptotically in the power at the relay if $P_2+N_3>P_1$ and $P_2 \gg N_3$, i.e.,
\begin{align}
 R'^{\text{lo}}_{\text{G}}(Q=\infty,N_2=\infty) &= R^{\text{up}}_{\text{G}}(Q=\infty,N_2=\infty)-o(1)\nonumber\\
&= \frac{1}{2}\log(1+\frac{P_1}{N_3})-o(1)
\end{align}
where $o(1) \longrightarrow 0$ as $P_2 \longrightarrow \infty$.

\iffalse
The rate \eqref{RateDeafHelperProblem} can also be achieved as follows. At time $i$, the source sends a Gaussian codeword $X_{1,i}$ which is independent of the state $S_i$. Independently, the relay generates its input $X_{2,i}$ using a \textit{dummy} DPC as $X_2=U_2-S$, where $X_2 \sim \mc N(0,P_2)$ is independent of $S$ and $U_2$ is Costa's auxiliary random variable. Upon reception of $Y_{3,i}=X_{1,i}+X_{2,i}+S_i+Z_{3,i}$ at the destination, the decoder first decodes the codeword $U_{2,i}$ fully, i.e., not only the bin index but also the correct sequence in the bin. This can be done reliably as long as $I(U_2;Y_3)-(U_2;S) >0$. Then, the decoder at the destination subtracts out $U_{2,i}$ from $Y_{3,i}$ to obtain $\tilde{Y}_{3,i}=X_{1,i}+Z_{3,i}$ from which it decodes the source's message using standard decoding, at rate \eqref{RateDeafHelperProblem}. A related scenario for a helper over a state-dependent Gaussian MAC is studied in \cite{PKEZ07}.
\fi
\item[3)] For $Q=0$, the lower bound in Corollary~\ref{CorollaryAchievabeRatePartialDecodeAndForwardNonCausalCaseGaussianChannel} reduces to the rate achievable using a partial decode-and-forward scheme in an interference-free relay channel, i.e.,
\begin{align}
R_{\text{G}}(Q=0) = \max_{0 \leq \rho'_{12}, \gamma \leq 1} \min \{&\frac{1}{2}\log(1+\frac{\bar{\gamma}P_1(1-\rho'^2_{12})}{N_2+{\gamma}P_1}),\nonumber\\
&\frac{1}{2}\log(1+\frac{\bar{\gamma}P_1+P_2+2\rho'_{12}\sqrt{\bar{\gamma}P_1P_2}}{N_3+{\gamma}P_1})\}+\frac{1}{2}\log(1+\frac{{\gamma}P_1}{N_3}),
\end{align}
and the upper bound in Theorem~\ref{TheoremOuterBoundNonCausalCaseGaussianChannelFullDuplexRegime} reduces to the cut-set upper bound. Furthermore, if the channel is degraded these bounds meet and give the capacity of standard degraded Gaussian RC \cite[Theorem 5]{CG79}.

\iffalse
i.e., no channel state at all in the model, capacity for the degraded Gaussian case is given by
\begin{align}
C_{\text{DG}} \: =\:\: \max_{0\leq \beta \leq 1} \min \Big\{&\frac{1}{2}\log\Big(1+\frac{P_1(1-\beta^2)}{N_2}\Big), \frac{1}{2}\log\Big(1+\frac{P_1+P_2+2\beta\sqrt{P_1P_2}}{N_3}\Big)\Big\},
\label{CapacityNoSideInformationDegradedGaussianChannelFullDuplexRegime}
\end{align}
which is the same as the capacity of the standard degraded Gaussian channel \cite[Theorem 5]{CG79}.
This can be directly obtained by putting $Q=0$ in \eqref{AchievabeRateNonCausalCaseGaussianChannelFullDuplexRegime} and \eqref{EquivalentFormOuterBoundNonCausalCaseDegradedGaussianChannelFullDuplexRegime}. In this case, the maximizing parameters are $\theta=0$, $\rho'_{2s}=0$ for \eqref{AchievabeRateNonCausalCaseGaussianChannelFullDuplexRegime} and $\rho=0$ for \eqref{EquivalentFormOuterBoundNonCausalCaseDegradedGaussianChannelFullDuplexRegime}.
\fi
\item[4)] If $P_2=0$, capacity for general Gaussian RC is given by
\begin{align}
&C_{\text{G}}(P_2=0)\: =\:\: \frac{1}{2}\log(1+\frac{P_1}{Q+N_3}).
\label{CapacityZeroPowerAtRelayDegradedGaussianChannelFullDuplexRegime}
\end{align}
%In this case, the informed relay cannot help the source, and the interference is simply treated as an unknown noise.  
\end{enumerate}

\subsection{Numerical Examples and Discussion}\label{secIV_subsecE}
%%% FIXME: There is a lot of symmetry in the examples.  I wonder if the
%%% reviewers might complain about that.
In this section we discuss some numerical examples, for both the degraded Gaussian case and the general Gaussian case. We consider two numerical examples, a) $P_1=P_2=Q=10$ dB, $N_3=20$ dB; and b)$P_1=P_2=Q=N_3=10$ dB. 

\begin{figure}[!htpb]

        \begin{minipage}[t]{\linewidth}
	\vspace{-0.5cm}        
	\begin{center}
        \subfigure[]
        {
        \includegraphics[width=0.7\linewidth,height=0.4\linewidth]{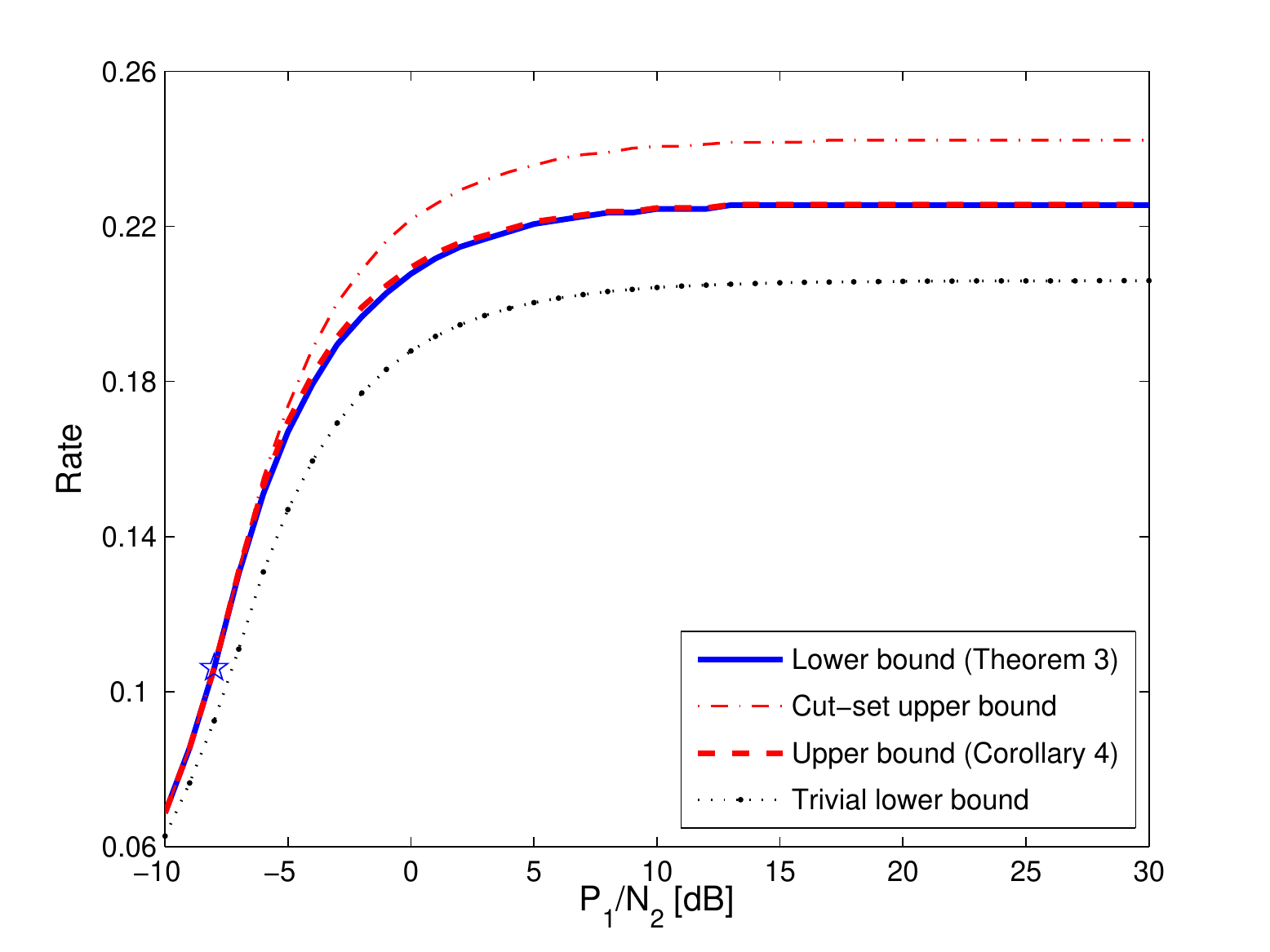}
        \label{Subfig1Fig1IllustrativeExamples}
        }
        \subfigure[]
        {
        \includegraphics[width=0.7\linewidth,height=0.4\linewidth]{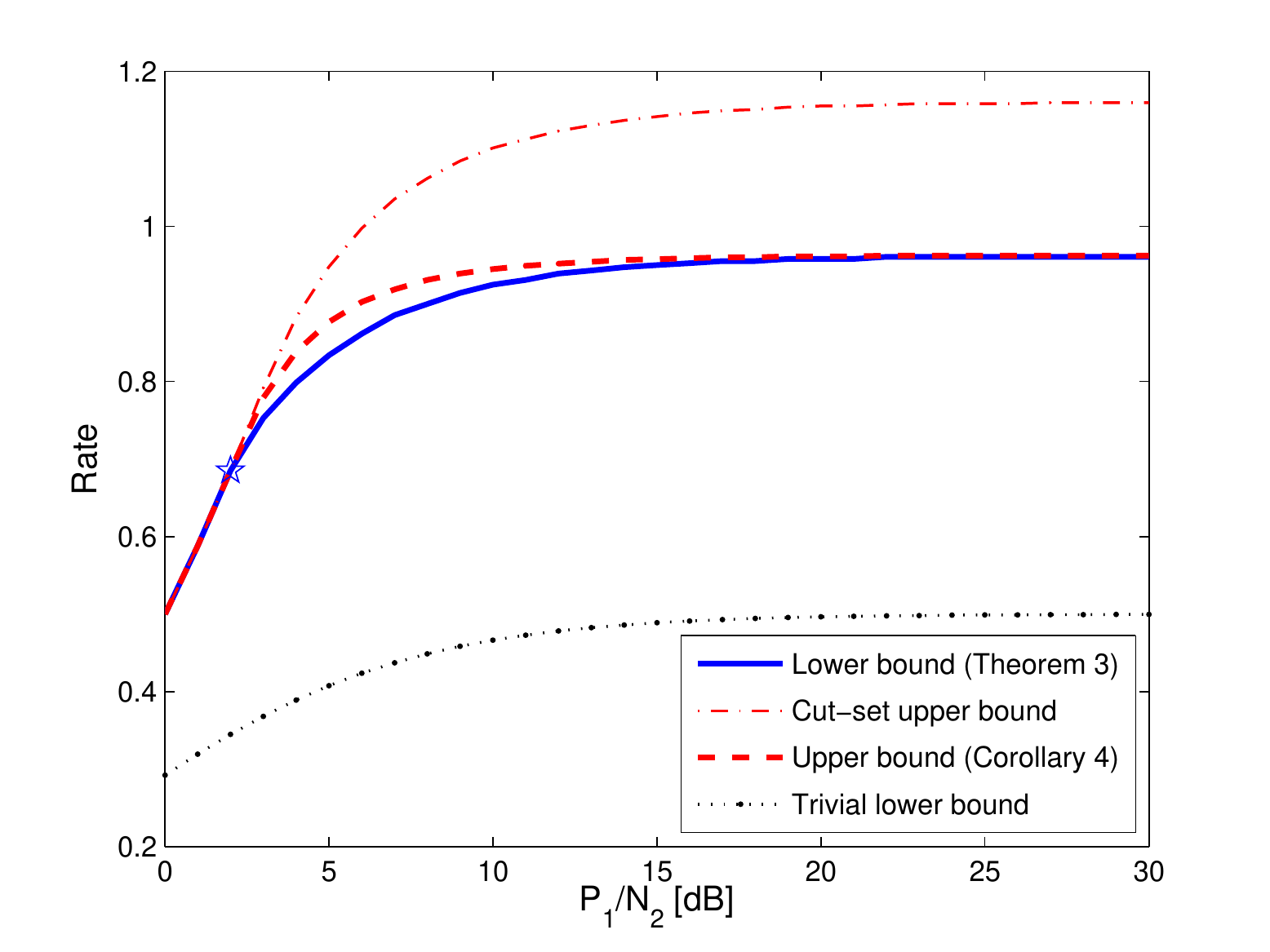}
        \label{Subfig2Fig1IllustrativeExamples}
        }
	\vspace{-0cm}        
        \caption{Lower and upper bounds on the capacity of the state-dependent degraded Gaussian RC with informed relay versus the SNR in the link source-to-relay, for two examples of numerical values (a) $P_1=P_2=Q=10$ dB, $N_3=20$ dB, and (b) $P_1=P_2=Q=N_3=10$ dB.}
        \label{Fig1IllustrativeExamples}
        \end{center}
        \end{minipage}
\end{figure}

\begin{figure}[!htpb]
	\vspace{-2cm}        
	\begin{minipage}[t]{\linewidth}
	\begin{center}
	\includegraphics[width=0.6\linewidth]{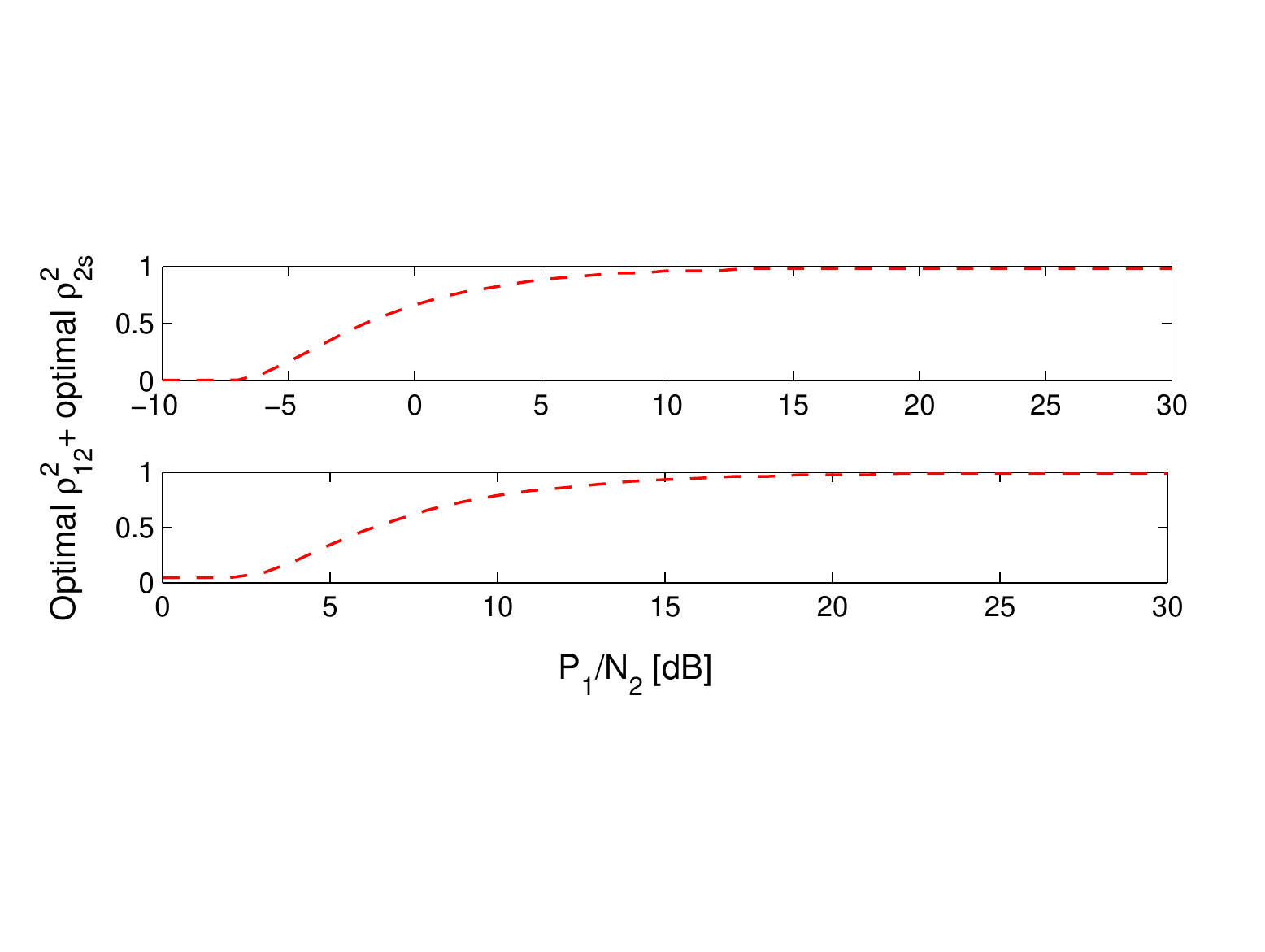}
	\vspace{-2cm}        
	\caption{The sum $\rho^2_{12}+\rho^2_{2s}$ in the constraint \eqref{MaximizationRangeOuterBoundNonCausalCaseGaussianChannelFullDuplexRegime}. Optimal $\rho_{12}$ and $\rho_{2s}$ are the maximizing for the upper bound for the numerical examples considered in Figure~\ref{Fig1IllustrativeExamples}. The upper subfigure is for the upper bound curve in Figure~\ref{Subfig1Fig1IllustrativeExamples}, and the lower subfigure is for the upper bound curve in Figure~\ref{Subfig2Fig1IllustrativeExamples}.}
	\label{Fig2IllustrativeExamples}
        \end{center}
        \end{minipage}
\end{figure}

Figure~\ref{Fig1IllustrativeExamples} illustrates the lower bound \eqref{EquivalentFormForAchievabeRateNonCausalCaseGaussianChannelFullDuplexRegime} and the upper bound \eqref{EquivalentFormOuterBoundNonCausalCaseDegradedGaussianChannelFullDuplexRegime} as functions of the signal-to-noise-ratio (SNR) at the relay, i.e., $\text{SNR}=P_1/N_2$ (in decibels), for a degraded channel\footnote{Note that for the full-duplex degraded Gaussian RC, the rate in Corollary~\ref{CorollaryAchievabeRatePartialDecodeAndForwardNonCausalCaseGaussianChannel} reduces to that in Theorem~\ref{TheoremAchievabeRateNonCausalCaseGaussianChannelFullDuplexRegime}.}. Also shown for comparison are the cut-set upper bound \eqref{TrivialOuterBoundNonCausalCaseDiscreteMemorylessChannel} computed for the degraded Gaussian case and the trivial lower bound obtained by considering the channel state as an unknown noise and implementing full-DF at the relay \cite[Theorem 5]{CG79}.

%%% FIXME: Think its important to suggest that we are being efficient on
%%% the MAC side, but not the BC side, of the relay channel.  That right?
%-- This comment on the curves of Figure 6 is added: "The gap beween the lower bound and the upper bound which is visible at low $\text{SNR}$ is due to the fact that DF relaying is not effective at this range of $\text{SNR}$ and also to that our upper bounding technique is efficient on the MAC side, but not on the BC side of the relay channel."

%-- What makes us inefficient on the BC side is the fact that we are revealing the state to the destination as well in the term I(X_1;Y_2,Y_3|S,X_2) in the upper bound. In the degraded case, Y_3 disappears from this term, and, hence, the lower and upper bound meet in the degraded Gaussian case.
%-- To be efficient on the BC side as well, we need an non-trivial outer bound on the capacity of the corresponding model for BC with one informed receiver (i.e, one which is tighter than the cut-set bound).

%%% Btw, have we explored the corresponding BC models?  These might be
%%% interesting.

The curves show that the lower bound and the upper bound do not meet for all SNR regimes. However, as it is visible from the depicted numerical examples, the gap between the two bounds is small for the degraded case. Furthermore, the curves in Figure~\ref{Fig1IllustrativeExamples} also illustrate the results in observation \ref{CapacityForSpecialCases}, by showing that the lower bound and the upper bound meet for the cases identified in Observation~\ref{CapacityForSpecialCases}. We note that the pentagram marker visible in Figure ~\ref{Fig1IllustrativeExamples} indicates capacity when the noise at the relay is equal to the RHS of \eqref{SnrRangeForChannelCapacityLowSnrAtRelay}; and this illustrates the first case  for which the lower bound and the upper bound meet in Proposition~\ref{CapacityForSpecialCases}. Also, Figure~\ref{Fig2IllustrativeExamples} depicts the variation of $\rho^2_{12}+\rho^2_{2s}$, where $\rho_{12}$ and $\rho_{2s}$ are the maximizing for the upper bound, as a function of the SNR for the two numerical examples considered in Figure~\ref{Fig1IllustrativeExamples}; and this illustrates the second case for which the lower and upper bounds meet in Proposition \ref{CapacityForSpecialCases}.

\iffalse
\begin{figure}[!htpb]
	\vspace{-2cm}        
	\begin{minipage}[t]{\linewidth}
	\begin{center}
	\includegraphics[width=0.6\linewidth]{SimFigOptimalOuterBoundCorrelationCoefficientsDegradedRC.pdf}
	\vspace{-2cm}        
	\caption{The sum $\rho^2_{12}+\rho^2_{2s}$ in the constraint \eqref{MaximizationRangeOuterBoundNonCausalCaseGaussianChannelFullDuplexRegime}. Optimal $\rho_{12}$ and $\rho_{2s}$ are the maximizing for the upper bound for the numerical examples considered in Figure~\ref{Fig1IllustrativeExamples}. The upper subfigure is for the upper bound curve in Figure~\ref{Subfig1Fig1IllustrativeExamples}, and the lower subfigure is for the upper bound curve in Figure~\ref{Subfig2Fig1IllustrativeExamples}.}
	\label{Fig2IllustrativeExamples}
        \end{center}
        \end{minipage}
\end{figure}
\fi

\begin{figure}[!htpb]

        \begin{minipage}[t]{\linewidth}
        %%\vspace{-1cm}
        \begin{center}
        \includegraphics[width=0.7\linewidth]{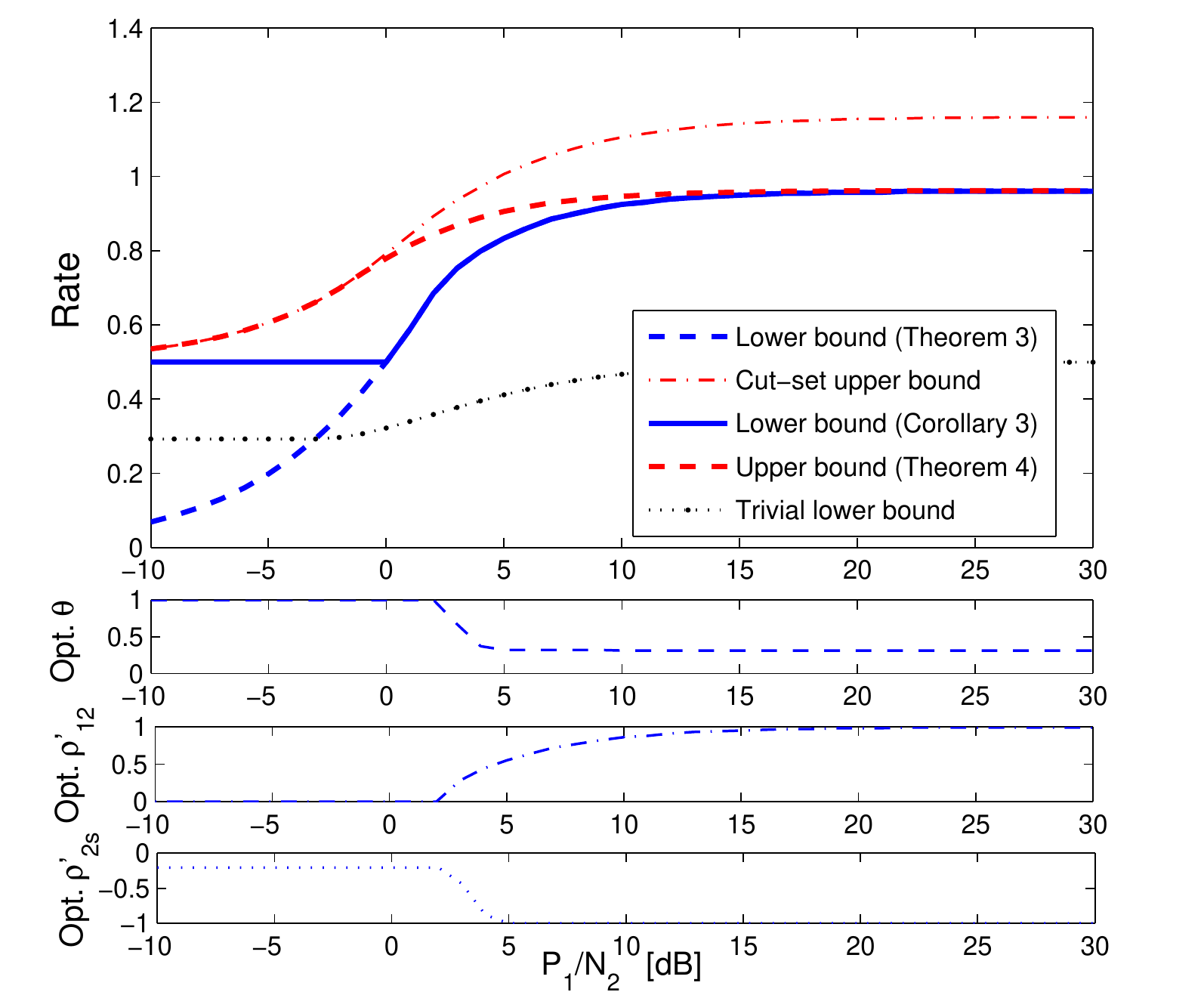}
        \vspace{-0.5cm}
       \caption{Lower and upper bounds on the capacity of the state-dependent general Gaussian RC with informed relay and the maximizing $\theta, \rho'_{12}, \rho'_{2s}$ in \eqref{AchievabeRateNonCausalCaseGaussianChannelFullDuplexRegime} as functions of the SNR at the relay. Numerical values are $P_1=P_2=Q=N_3=10$ dB.}
        \label{Fig5IllustrativeExamples}
        \end{center}
        \end{minipage}

\end{figure}

Figure~\ref{Fig5IllustrativeExamples} shows similar curves for the general Gaussian channel. The curves show that the lower bound \eqref{AchievabeRate__PartialDF__GaussianCase__FullDuplexRegime} is close to the upper bound \eqref{OuterBoundNonCausalCaseGaussianChannelFullDuplexRegime} at large $\text{SNR}$, i.e., when capacity of the channel is determined by the sum rate of the MAC formed by transmission from the uninformed source and the informed relay to the destination. At small $\text{SNR}$, the lower bound given in Corollary~\ref{CorollaryAchievabeRatePartialDecodeAndForwardNonCausalCaseGaussianChannel} improves upon that in Theorem~\ref{TheoremAchievabeRateNonCausalCaseGaussianChannelFullDuplexRegime} due to rate-splitting.  

Furthermore, Figure~\ref{Fig5IllustrativeExamples} also shows the variation of the maximizing $\theta$, $\rho'_{12}$, $\rho'_{2s}$ in \eqref{AchievabeRateNonCausalCaseGaussianChannelFullDuplexRegime} as function of the $\text{SNR}$ at the relay. This shows how the informed relay allocates its power among combating the interference for the source (related to the value of $\rho'_{2s}$) and sending signals that are coherent with the transmission from the source (related to the values of $\theta$ and $\rho'_{12}$).

\begin{remark}
In standard, i.e., state-independent, Gaussian relay channels, partial DF simply reduces to direct transmission if the link source-to-relay is too noisy, i.e, at low $\text{SNR}$. For the studied model, however, it is insightful to observe that the relay can still help the source even at very small $\text{SNR}$. This can be seen by observing that the lower bound \eqref{AchievabeRate__PartialDF__GaussianCase__FullDuplexRegime} is better than the trivial lower bound even at this range of $\text{SNR}$ (the trivial lower bound in Figure~\ref{Fig5IllustrativeExamples} is obtained by treating the channel state as additional noise and implementing partial DF). This observation has some connection with the aforementioned \textit{deaf helper problem} (see Case 2, Section "Extreme Cases"), and it can be interpreted as follows. The relay does not hear the source and generates its input  $X_{2,i}$ using a \textit{dummy} DPC as $X_2=U_2-S$, where $X_2 \sim \mc N(0,P_2)$ is independent of $S$ and $U_2$ is Costa's auxiliary random variable. Upon reception of $Y_{3,i}=X_{1,i}+X_{2,i}+S_i+Z_{3,i}$ at the destination, the decoder first decodes the codeword $U_{2,i}$ fully, i.e., not only the bin index but also the correct sequence in the bin. This can be done reliably as long as $I(U_2;Y_3)-(U_2;S) >0$. Then, the decoder at the destination subtracts out $U_{2,i}$ from $Y_{3,i}$ to obtain $\tilde{Y}_{3,i}=X_{1,i}+Z_{3,i}$ from which it decodes the source's message using standard decoding, at full rate $0.5\log(1+P_1/N_3)$. A related scenario for a helper over a state-dependent Gaussian MAC is studied in \cite{PKEZ07}.
\end{remark} 

\begin{remark}
The gap between the lower bound and the upper bound which is visible at low $\text{SNR}$ is due to that DF relaying (even partial) is not effective at small $\text{SNR}$ and also to that our upper bounding technique is efficient on the MAC side but not on the BC side of the relay channel. 
\end{remark}

\begin{figure}[!htpb]
        \begin{minipage}[t]{\linewidth}
        \vspace{-1cm}
        \begin{center}
	\subfigure[Degraded Gaussian RC. $P_2=N_2=10$, $N_3=20$ dB]
	{
        \includegraphics[width=0.7\linewidth,height=0.5\linewidth]{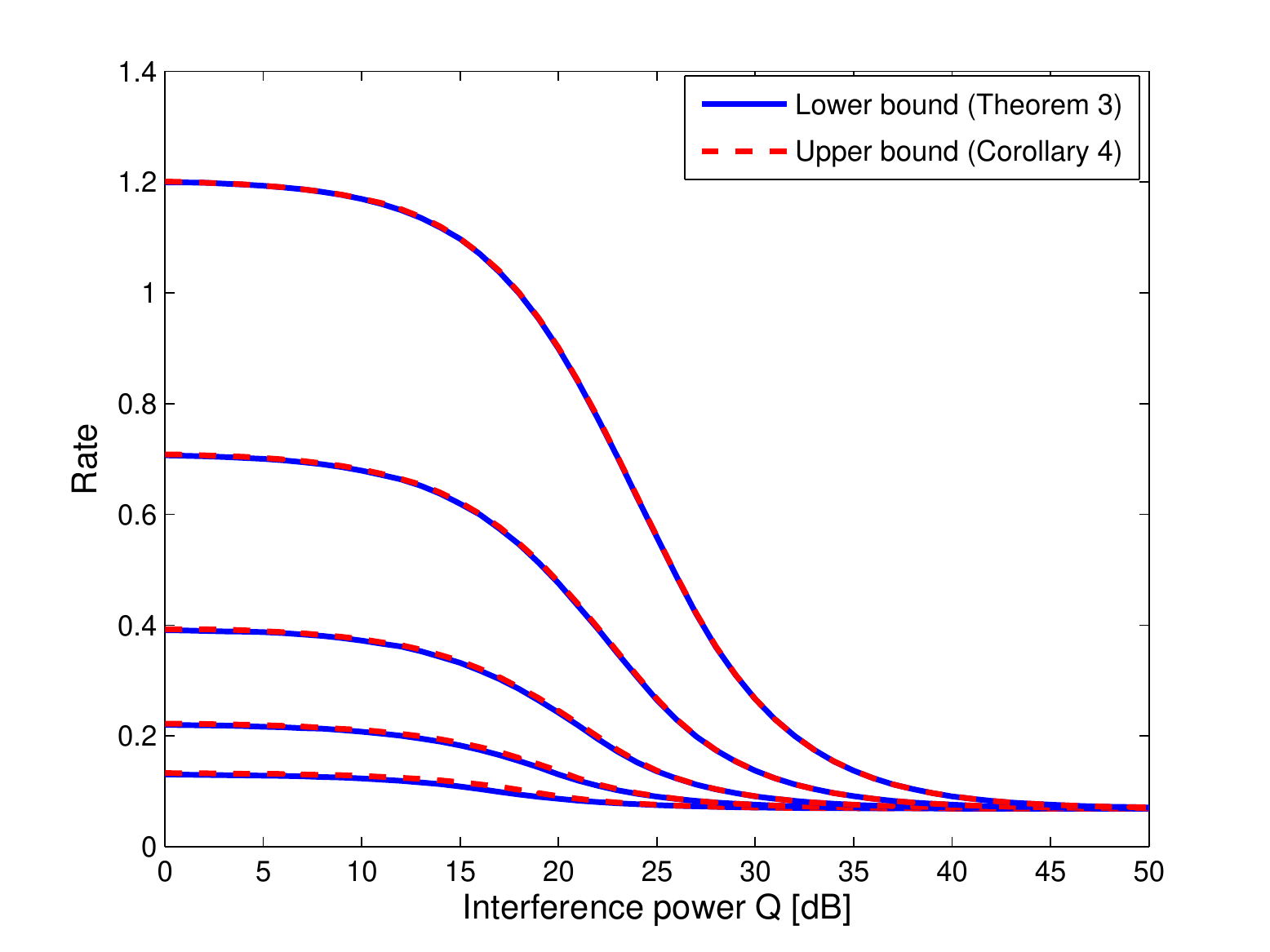}
	}
	\subfigure[General Gaussian RC. $P_2=N_3=10$, $N_2=20$ dB]
        {
        \includegraphics[width=0.7\linewidth,height=0.5\linewidth]{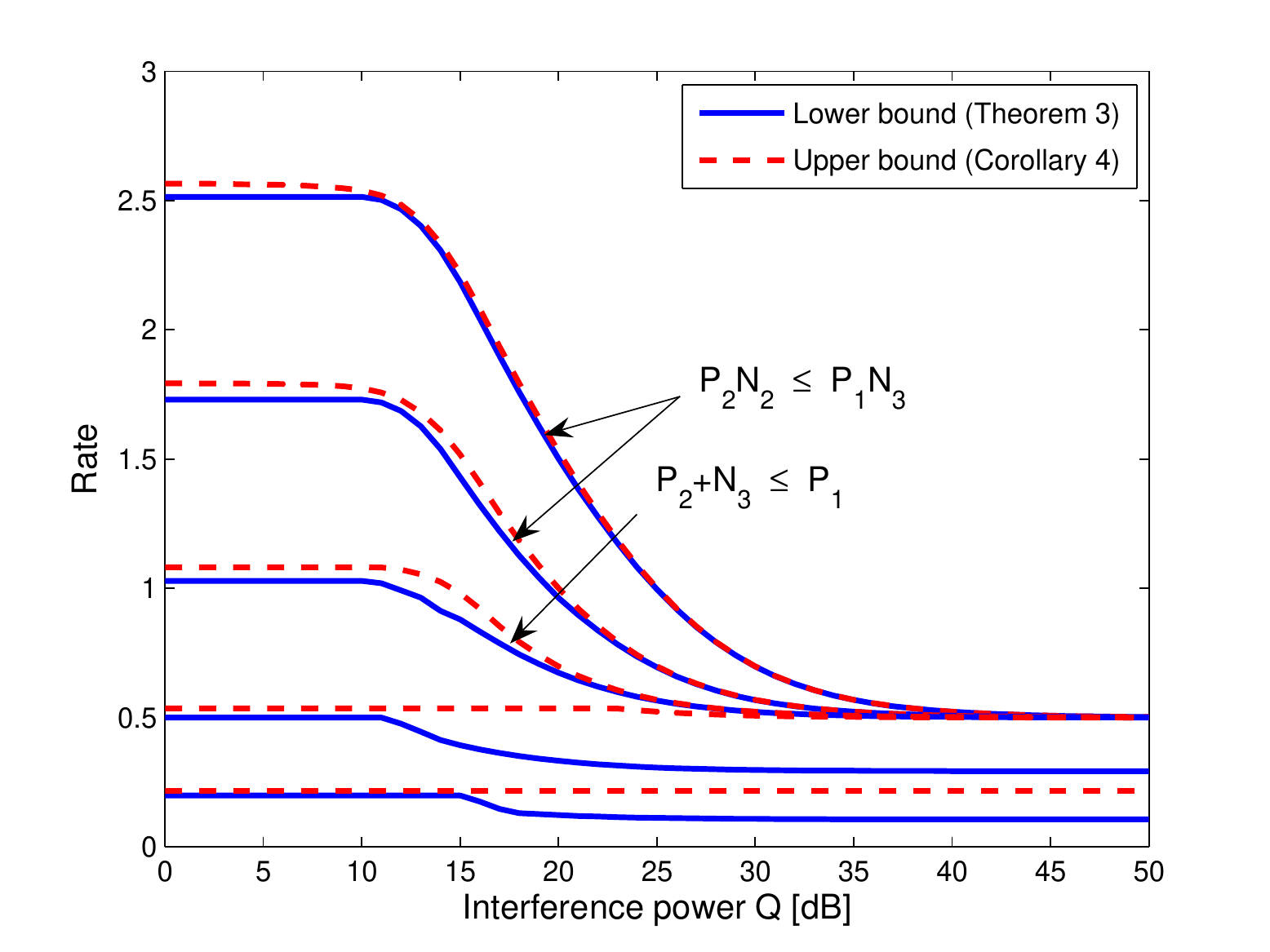}
        }
        \caption{Bounds on channel capacity as function of the interference power $Q$. The curves correspond to different choices of power at the source: from bottom to top $P_1=5,10,15,20,25$ dB.}
        \label{Fig4IllustrativeExamples}
        \end{center}
        \end{minipage}

 \end{figure}

In Figure~\ref{Fig4IllustrativeExamples}, the lower and upper bounds are plotted as function of the interference power $Q$, for fixed value of the power at the relay and several choices of the power at the source. The curves are depicted for two examples of noise configuration: $N_2 < N_3$ ($N_2=10$ dB and $N_3=20$ dB), and $N_2 > N_3$ ($N_2=20$ dB and $N_3=10$ dB). The curves illustrate the discussion in the above extreme cases analysis. For instance, for both noise configurations, that the rate achievable for very large values of $Q$ is strictly positive illustrates that transmission from the uninformed source to the uninformed destination is possible even in presence of an infinitely strong interference. Furthermore, the lower and upper bounds meet for the cases identified in the "Extreme Cases" Section, for both degraded Gaussian and General Gaussian channels.

%%% FIXME: We have to ask ourselves if it is really worth including this
%%% aspect, or if it just makes the paper too long.
\subsection{Half-Duplex Channel Model}\label{secIV_subsecF}
In this section, we extend the results of Section \ref{secIV_subsecA} to the case of half-duplex relaying, i.e., the relay  can either transmit only or receive only. We consider a state-dependent Gaussian RC with informed relay, and we assume that the relay operates in a time-division (TD) relaying mode. In the TD mode, for a given time window, the relay is in the receive mode for a fraction of the given time and in the transmit mode for the remaining fraction of this time. Since the message from the source is transmitted to the destination in $n$ channel uses, in the remaining of this section, we refer to the time indices from \footnote{For a scalar $x$, $\lfloor x \rfloor$ stands for the largest integer small than or equal to $x$.} $1$ to $\lfloor \nu n \rfloor$ as the {\it relay-receive period} and the time indices from $\lfloor \nu n \rfloor + 1$ to $n$ as the {\it relay-transmit period}, for some $\nu \in [0,1]$. Furthermore, to generalize the model, we assume that the channel state $S^{(1)}$ is zero mean Gaussian with variance $Q^{(1)}$ during the relay-receive period, and the channel state $S^{(2)}$  is zero mean Gaussian with variance $Q^{(2)}$ during the relay-transmit period. The channel output $Y_{2,i}$ at time instant $i$ at the relay is given by
$$Y_{2,i}=X_{1,i}^{(1)}+S_i^{(1)}+Z_{2,i},$$
during the relay-receive period, and is zero with probability one during the relay-transmit period. The channel output at time-instant $i$ at the destination is given by
\begin{subequations}
\begin{align}
Y_{3,i}^{(1)} &  =X_{1,i}^{(1)}+S_i^{(1)}+Z_{3,i}\qquad \text{during the relay-receive period}  \\
Y_{3,i}^{(2)}&=X_{1,i}^{(2)}+X_{2,i}+S_i^{(2)}+Z_{3,i}\qquad \text{during the relay-transmit period.}
\end{align}
\label{ReceivedForGaussianRCWithStateTimeDivisionRegime}
\end{subequations}
Furthermore, the source has average power constraint $P_1^{(1)}$ during the relay-receive period and average power constraint $P_1^{(2)}$ during the relay-transmit period ; the relay has average power constraint $P_2$.

For fixed values of $\nu$, $P_1^{(1)}$, $P_1^{(2)}$ and $P_2$, we have the following upper and lower bounds on the capacity of the state-dependent half-duplex Gaussian RC with informed relay.

\begin{proposition}\label{PropositionOuterBoundNonCausalCaseGaussianChannelTimeDivisionRelaying}
The capacity of the state-dependent TD Gaussian RC with informed relay is upper-bounded by
\begin{align}
& R^{\text{up}}_{\text{G}}(\text{TD}) \:=\:\: \max \; \min \{R^{\text{up}}_1,\:R^{\text{up}}_2\}
\label{OuterBoundNonCausalCaseGaussianChannelTimeDivisionRelaying}
\end{align}
with 
\begin{subequations}
\begin{align}
\label{OuterBound1NonCausalCaseGaussianChannelTimeDivisionRelaying}
R^{\text{up}}_1 &= \frac{\nu}{2}\log\Big(1+P_1^{(1)}(\frac{1}{N_2}+\frac{1}{N_3})\Big) + \frac{\bar{\nu}}{2}\log\Big(1+\frac{P_1^{(2)}(1-\rho^2_{12}-\rho^2_{2s})}{N_3(1-\rho^2_{2s})}\Big),\\
R^{\text{up}}_2 &= \bar{\nu}\Psi(P_1^{(2)},P_2,Q^{(2)},\rho_{12},\rho_{2s})+\frac{\nu}{2}\log\Big(1+\frac{P_1^{(1)}}{N_3+Q^{(1)}}\Big),
\label{OuterBound2NonCausalCaseGaussianChannelTimeDivisionRelaying}
\end{align}
\end{subequations}
where $\Psi(P_1,P_2,Q,\rho_{12},\rho_{2s})$ is defined as the second term of the minimization in \eqref{OuterBoundNonCausalCaseGaussianChannelFullDuplexRegime}, and the  maximization is over parameters $\rho_{12} \in [0,1]$ and $\rho_{2s} \in [-1,0]$ such that $\rho^2_{12}+\rho^2_{2s} \leq 1$.
\end{proposition}
\vspace{0.2cm}
\iffalse
For the state-dependent TD degraded Gaussian RC, an upper bound on the capacity is given by \eqref{OuterBoundNonCausalCaseGaussianChannelTimeDivisionRelaying} in which the first term on the RHS of \eqref{OuterBound1NonCausalCaseGaussianChannelTimeDivisionRelaying} is replaced by $\frac{\nu}{2}\log(1+P_1^{(1)}/N_2)$. 
\fi
\vspace{0.2cm}

\begin{proposition}\label{PropositionAchievabeRateNonCausalCaseGaussianChannelTimeDivisionRelaying}
The capacity of the state-dependent TD Gaussian RC with informed relay is lower-bounded by 
\begin{align}
& R^{\text{lo}}_{\text{G}}(\text{TD})\:=\:\max \; \min \{R^{\text{lo}}_1,\:R^{\text{lo}}_2,\:R^{\text{lo}}_3\}
\label{AchievableRateNonCausalCaseGaussianChannelTimeDivisionRelaying}
\end{align}
with
\begin{subequations}
\begin{align}
\label{AchievableRate1NonCausalCaseGaussianChannelTimeDivisionRelaying}
R^{\text{lo}}_1 &= \frac{\nu}{2}\log\Big(1+\frac{P_1^{(1)}}{N_2}\Big)+ \frac{\bar{\nu}}{2}\log\Big(1+\frac{(1-\rho'^2_{12})P_1^{(2)}}{N_3+\Phi(\alpha',\theta,\rho'_{2s})}\Big)\\
\label{AchievableRate2NonCausalCaseGaussianChannelTimeDivisionRelaying}
R^{\text{lo}}_2 &=  \frac{\nu}{2}\log\Big(1+\frac{P_1^{(1)}}{N_2}\Big)+\frac{\bar{\nu}}{2}\log\Big(\frac{P'_2(P'_2+Q'^{(2)}+(1-\rho'^2_{12})P_1^{(2)}+N_3)}{P'_2Q'^{(2)}(1-\alpha')^2+N_3(P'_2+\alpha'^2Q'^{(2)})}\Big).\\
R^{\text{lo}}_3 &= \frac{\nu}{2}\log\Big(1+\frac{P_1^{(1)}}{N_3+Q^{(1)}}\Big)\nonumber\\
& +\frac{\bar{\nu}}{2}\log\Big(1+\frac{P_1^{(2)}+\bar{\theta}P_2+2\rho'_{12}\sqrt{\bar{\theta}P_1^{(2)}P_2}}{{\theta}P_2+Q^{(2)}+2\rho'_{2s}\sqrt{{\theta}P_2Q^{(2)}}+N_3}\Big)\nonumber\\
& +\frac{\bar{\nu}}{2}\log\Big(\frac{P'_2(P'_2+Q'^{(2)}+N_3)}{P'_2Q'^{(2)}(1-\alpha')^2+N_3(P'_2+\alpha'^2Q'^{(2)})}\Big),
\label{AchievableRate3NonCausalCaseGaussianChannelTimeDivisionRelaying}
\end{align}
\end{subequations}
where, maximization is over parameters $\theta \in [0,1]$, $\rho'_{12} \in [0,1]$, $\rho'_{2s} \in [-1,0]$ and $\alpha' \in \mathbb{R}$ such that the last logarithm terms on the RHSs of \eqref{AchievableRate2NonCausalCaseGaussianChannelTimeDivisionRelaying} and \eqref{AchievableRate3NonCausalCaseGaussianChannelTimeDivisionRelaying} are defined; 
\begin{align}
\label{IntermediateFunction1ForAchievableRate1NonCausalCaseGaussianChannelTimeDivisionRelaying}
\Phi(\alpha',\theta,\rho'_{2s}) &:= \frac{P'_2Q'^{(2)}(1-\alpha')^2}{P'_2+\alpha'^2Q'^{(2)}}
%\Theta(t,\rho'_{12},\theta,\rho'_{2s}) &= \frac{1}{2}\log\Big(\frac{P'_2[P'_2+Q'^{(2)}+N_3+P_1^{(2)}(1-\rho'^2_{12})]}{P'_2Q'^{(2)}(1-t)^2+[N_3+P_1^{(2)}(1-\rho'^2_{12})][P'_2+t^2Q'^{(2)}]}\Big),
%\label{IntermediateFunction2ForAchievableRate1NonCausalCaseGaussianChannelTimeDivisionRelaying}
\end{align}
and $P'_2:={\theta}P_2(1-\rho'^2_{2s})$, \:\: $Q'^{(2)}:=(\sqrt{Q^{(2)}}+\rho'_{2s}\sqrt{{\theta}P_2})^2$.\\
\end{proposition}

The proofs of Proposition \ref{PropositionOuterBoundNonCausalCaseGaussianChannelTimeDivisionRelaying} and Proposition \ref{PropositionAchievabeRateNonCausalCaseGaussianChannelTimeDivisionRelaying} appear in Appendix \ref{appendixTimeDivisionRelaying}.

\begin{remark}
The coding scheme employed for the proof of Proposition \ref{PropositionAchievabeRateNonCausalCaseGaussianChannelTimeDivisionRelaying} pre-assigns the time slots for the relay's receiving and transmitting modes. All the nodes then know ahead of time when the relay receives and when it transmits. This is relevant for nodes synchronization but suboptimal in general for information rate. Instead, one can let the source and the relay choose the relay's mode and, so, in a sense, transmit additional information to the destination through that choice. This idea is introduced in \cite{K04} in the context of wireline and wireless networks without state and is called \textit{mode coding} therein; see also \cite[Section 4.3]{KMY06}. More specifically, let $M$ denote a random variable that takes on values $1$ ("receive") and $2$ ("transmit") with probabilities $\nu$ and $\bar{\nu}$, respectively. Also, let us redefine the channel so as to include the relay's operating mode as $W_{Y_2,Y_3|X_1,X_2,S,M}$; set $X'_1=(X_1,M)$, $X'_2=(X_2,M)$, $U'=(U,M)$, $U'_1=(U_1,M)$, $U'_2=(U_2,M)$ and choose $U=X_1^{(1)}$ if $M=1$ and $U=0$ if $M=2$. Then, using $(X'_1,X'_2,U',U'_1,U'_2)$ in place of $(X_1,X_2,U,U_1,U_2)$ in \eqref{AchievabeRatePartialDecodeAndForwardNonCausalCaseDiscreteMemorylessChannel}, it can be shown that this yields a rate which is obtained by maximizing the minimum among $R^{\text{lo}}_1$, $R^{\text{lo}}_2$ and $R^{\text{lo}}_3+I(M;Y_3)$, i.e., larger than \eqref{AchievableRateNonCausalCaseGaussianChannelTimeDivisionRelaying}. However, as mentioned in \cite[Section 4.3]{KMY06}, the improvement is no larger than $1$ bit per block and, also, harnessing it in practice requires some challenges in general.
\end{remark}

\section{Conclusion}\label{secV}

In this paper, we consider a state-dependent relay channel with the channel state available noncausally at only the relay, i.e., neither at the source nor at the destination. We refer to this communication model as \textit{state-dependent RC with informed relay}. This setup may model the basic building block for node cooperation over wireless networks in which some of the terminals may be equipped with cognition capabilities that enable estimating to high accuracy the states of the channel.

We investigate this problem in the discrete memoryless (DM) case and in the Gaussian case, and we derive bounds on the channel capacity.  For both cases,  the upper bounds are tighter than those obtained by assuming that the channel state is also available at the source and the destination, and they help characterizing the rate loss due to the asymmetry, i.e., having the channel state available at the relay but not the source. Key to the development of the lower bounds is a coding scheme that splits the codeword at the informed relay into two parts: one part depends only on the cooperative information, not on the known channel state, and is used to enable coherent transmission from the source and the relay to the destination; another part is a function of both the cooperative information and the known channel state, and is used to combat the effects of the channel state on the communication through a generalized Gel'fand-Pinsker binning scheme. In the Gaussian case, we consider average power constraints at the source and the relay and power allocation at the relay among the two parts of the code, allowing for a tradeoff between the coherence gain obtained through the coherent transmission and the mitigation of the channel state.

Specializing the results to the case in which the channel is physically degraded, we show that the developed lower and upper bounds meet in some cases, thus characterizing the channel capacity. For the general Gaussian case, the bounds are in general close, but they meet only in some extreme cases.

%Finally, we note that some of the concepts developed in this paper can be applied to a state-dependent multiple access channel (MAC) with degraded message sets in which the uninformed encoder knows the message to be sent by the informed encoder \cite{ZKLV09a}.
%%% FIXME: Cite a document in preparation?
%-- FIXED

%%% FIXME: JNL stopped here on 2008/10/09

\appendix

Throughout this section we denote the set of strongly jointly $\epsilon$-typical sequences \cite[Chapter 14.2]{CT91} with respect to the distribution $P_{X,Y}$ as $T_{\epsilon}^n(P_{X,Y})$. 

\renewcommand{\theequation}{A-\arabic{equation}}
\setcounter{equation}{0}  % reset counter
\subsection{Proof of Theorem \ref{TheoremAchievabeRateNonCausalCaseDiscreteMemorylessChannel}}\label{appendixTheorem1}
Consider the random coding scheme that we outlined in Section \ref{secIII}. We now give a formal description of the coding scheme and analyse the average probability of error. 

\noindent As we outlined after Theorem \ref{TheoremAchievabeRateNonCausalCaseDiscreteMemorylessChannel} we transmit in $B+1$ blocks, each of length $n$. During each of the first $B$ blocks, the source encodes a message $w_i \in [1,2^{nR}]$ and sends it over the channel, where $i=1,\hdots,B$ denotes the index of the block. For fixed $n$, the average rate $R\frac{B}{B+1}$ over $B+1$ blocks approaches $R$ as $B \longrightarrow +\infty$.

\textbf{Encoding:}
Let $w_i$ be the new message to be sent from the source node at the beginning of block $i$, and $w_{i-1}$ be the message sent in the previous block $i-1$. At the beginning of block $i$, the relay has decoded the message $w_{i-1}$ correctly and the source sends $\dv x_1(w_{i-1},w_i)$. The relay searches for the smallest $j \in \{1,\cdots,J\}$ such that $\dv u_1(w_{i-1})$, ${\dv u_2}(w_{i-1},j)$ and $\dv s[i]$ are jointly typical. Denote this $j$ by $j^{\star}=j(\dv s[i], w_{i-1})$. If such $j^{\star}$ is not found, or if the observed state is not typical, an error is declared and $j^{\star}$ is set to $J$. Then, the relay transmits a vector $\dv x_2(w_{i-1})$ with i.i.d. components given $(\dv u_1(w_{i-1}), \dv u_2(w_{i-1},j^{\star}), \dv s[i])$ drawn according to the marginal $P_{X_2|U_1,U_2,S}$ induced by the distribution \eqref{MeasureForAchievabeRateNonCausalCaseDiscreteMemorylessChannel}. 

\noindent The encoder at the source declares an error if the chosen codeword exceeds the power constraint, that is, $\varphi^n_1(\dv x_1(w_{i-1},w_i))$ $ > \Gamma_1 +\gamma_1(\epsilon)$ for some $\gamma_1(\epsilon) > 0$. Similarly, the encoder at the relay declares an error if $\varphi^n_2(\dv x_2(w_{i-1})) > \Gamma_2 +\gamma_2(\epsilon)$, for some $\gamma_2(\epsilon) > 0$.

\noindent For convenience, we list the codewords at the source and the relay that are used for transmission in the first four blocks in Figure~\ref{ExampleTransmittedCodewords}.
\begin{figure}[htbp]
\begin{center}
\resizebox{0.7\linewidth}{!}{
\begin{tabular}{|*{5}{l|}}
 \hline
 & block 1 & block 2 & block 3 & block 4\\
\hline
\hline
Source codewords & $\dv x_1(1,w_1)$ & $\dv x_1(w_1,w_2)$ & $\dv x_1(w_2,w_3)$ & $\dv x_1(w_3,1)$\\
\hline
\multirow{3}{*}{Relay codewords} & $\dv u_1(1)$     & $\dv u_1(w_1)$     & $\dv u_1(w_2)$     & $\dv u_1(w_3)$\\
\cline{2-5}
 & $\dv u_2(1,j(\dv s[1],1))$     & $\dv u_2(w_1,j(\dv s[2],w_1))$     & $\dv u_2(w_2,j(\dv s[3],w_2))$     & $\dv u_2(w_3,j(\dv s[4],w_3))$\\
\cline{2-5}
& $\dv x_2(1)$ & $\dv x_2(w_1)$ & $\dv x_2(w_2)$ & $\dv x_2(w_3)$\\
\hline
\end{tabular}}
\end{center}
\caption{Regular encoding for DF for the state-dependent RC with informed relay. At the beginning of block $i$, the source transmits $\dv x_1(w_{i-1},w_i)$ and the relay transmits a codeword $\dv x_2(w_{i-1})$ with i.i.d. components given $(\dv u_1(w_{i-1}), \dv u_2(w_{i-1},j(\dv s[i],w_{i-1})), \dv s[i])$ drawn according to the marginal $P_{X_2|U_1,U_2,S}$.}
\label{ExampleTransmittedCodewords}
\end{figure}

\textbf{Decoding:} The decoding procedure at the relay is based on joint typicality. The decoding procedure at the destination is based on a combination of joint typicality and backward-decoding.

\begin{enumerate}
\item[1.] At the end of block $i$, the relay knows $w_{i-1}$ and declares that $\hat{w}_i$ is sent if there is a unique $\hat{w}_i$ such that $\dv x_1(w_{i-1},\hat{w}_i)$ and $(\dv y_2[i],\dv s[i])$ are jointly typical given $\dv u_1(w_{i-1})$, ${\dv u_2}(w_{i-1},j^{\star})$ and $\dv x_2(w_{i-1})$, where $\dv y_2[i]$ denotes the output of the channel at the relay in block $i$ and $j^{\star}=j(\dv s[i], w_{i-1})$ as mentioned earlier. One can show that the decoding error in this step is small for sufficiently large $n$ if
\begin{align}
R &< I(X_1;Y_2|S,U_1,X_2).
\label{Rate1AchievabeRateNonCausalCaseDiscreteMemorylessChannel}
\end{align}
\item[2.] At the end of the transmission, the destination has collected all the blocks of channel outputs $\dv y_3[1],\dv y_3[2],\hdots,\dv y_3[B+1]$, and can then perform Willem's backward-decoding by first decoding $w_B$ from $\dv y_3[B+1]$.

First, the destination declares that $\hat{w}_B$ is sent if there is a unique $\hat{w}_B$ such that $\dv u_1(\hat{w}_B)$, $\dv u_2(\hat{w}_B,j_B)$, $\dv x_1(\hat{w}_B,1)$, $\dv y_3[B+1]$ are jointly typical, for some $j_B \in \{1,\hdots,J\}$. One can show that the decoding error in this step is small for sufficiently large $n$ if
\begin{equation}
R < I(X_1,U_1,U_2;Y_3)-I(U_2;S|U_1).
\label{Rate2AchievabeRateNonCausalCaseDiscreteMemorylessChannel}
\end{equation}

Next, for $b$ ranging from $B$ to $2$, the destination knows $w_{b}$ and decodes $w_{b-1}$ based on the information received in block $b$. It declares that $\hat{w}_{b-1}$ is sent if there is a unique $\hat{w}_{b-1}$ such that $\dv u_1(\hat{w}_{b-1})$, $\dv u_2(\hat{w}_{b-1},j_{b-1})$, $\dv x_1(\hat{w}_{b-1},w_b)$, $\dv y_3[b]$ are jointly typical, for some $j_{b-1} \in \{1,\hdots,J\}$. One can show that the decoding error in this step is small for sufficiently large $n$ if \eqref{Rate2AchievabeRateNonCausalCaseDiscreteMemorylessChannel} is true.

\end{enumerate}

\textbf{Analysis of Probability of Error:}

Fix a  probability distribution $P_{S,U_1,U_2,X_1,X_2,Y_2,Y_3}$ satisfying \eqref{MeasureForAchievabeRateNonCausalCaseDiscreteMemorylessChannel} and  and $\mathbb{E}[\varphi_i(X_i)] < \Gamma_i$, $i=1,2$. Let $\dv s[i]$, $w_{i-1}$ and $w_{i}$ be the state sequence in block $i$, the message sent from the source node in block $i-1$ and the message sent in block $i$, respectively. As we already mentioned above, at the beginning of block $i$ the source transmits $\dv x_1(w_{i-1},w_i)$ and the relay transmits a vector $\dv x_2(w_{i-1})$ with i.i.d. components conditionally given $(\dv u_1(w_{i-1}),\dv u_2(w_{i-1},j^{\star}),\dv s[i])$, with $j^{\star}=j(\dv s[i],w_{i-1})$, drawn according to the marginal $P_{X_2|U_1,U_2,S}$.

The average probability of error is such that
\begin{align}
\text{Pr}(\text{Error}) & \leq \sum_{(\dv s,\dv u_1) \notin T_{\epsilon}^n(Q_SP_{U_1})}\text{Pr}(\dv s)\text{Pr}(\dv u_1)\nonumber\\
                        & + \sum_{(\dv s,\dv u_1) \in T_{\epsilon}^n(Q_SP_{U_1})}\text{Pr}(\dv s)\text{Pr}(\dv u_1)\text{Pr}(\text{error}|\dv s,\dv u_1).
\label{AverageProbabilityOfError}
\end{align}
The first term, $\text{Pr}((\dv s,\dv u_1) \notin T_{\epsilon}^n(Q_SP_{U_1}))$, on the RHS of \eqref{AverageProbabilityOfError} goes to zero as $n \rightarrow \infty$, by the asymptotic equipartition property (AEP) \cite[p. 384 ]{CT91}. Thus, it is sufficient to upper bound the second term on the RHS of \eqref{AverageProbabilityOfError}.

We now examine the probabilities of the error events associated with the encoding and decoding procedures. The error event is contained in the union of the error events given below, where the events $E_{1i}$, $E_{2i}$ and $E_{3i}$ correspond to the encoding step at block $i$; the events $E_{4i}$ and $E_{5i}$ correspond to decoding at the relay at block $i$; the events $E_{6B}$ and $E_{7B}$ correspond to decoding at the destination at block $B+1$, and for $b$ ranging from $B$ to $2$, the events $E_{8(b-1)}$ and  $E_{9(b-1)}$ correspond to decoding at the destination at block $b$.

\begin{itemize}

\item Let $E_{1i}$ be the event that there is no sequence $\dv u_2(w_{i-1},j)$ jointly typical with $\dv s[i]$ given $\dv u_1(w_{i-1})$, i.e.,
\begin{equation*}
E_{1i} = \Big\{\nexists \:j \in \{1,\hdots,J\}\:\text{s.t.}\: \Big(\dv u_1(w_{i-1}),{\dv u_2}(w_{i-1},j),\dv s[i]\Big) \in T_{\epsilon}^n(P_{U_1,U_2,S})\Big\}.
\end{equation*}
To bound the probability of the event $E_{1i}$, we use a standard argument \cite{GP80}. More specifically, for $\dv u_2(w_{i-1},j)$ and $\dv s[i]$ generated independently given $\dv u_1(w_{i-1})$, with i.i.d. components drawn according to $P_{U_2|U_1}$ and $Q_S$, respectively, the probability that $\dv u_2(w_{i-1},j)$ is jointly typical with $\dv s[i]$ given $\dv u_1(w_{i-1})$ is greater than $(1-\epsilon)2^{-n(I(U_2;S|U_1)+\epsilon)}$ for sufficiently large $n$. There is a total of $J$ such $\dv u_2$'s in each bin. The probability of the event $E_{1i}$, the probability that there is no such $\dv u_2$, is therefore bounded as
\begin{equation}
\text{Pr}(E_{1i}) \leq [1-(1-\epsilon)2^{-n(I(U_2;S|U_1)+\epsilon)}]^J.
\label{BoundingProbabilityErrorEventE1}
\end{equation}
Taking the logarithm on both sides of \eqref{BoundingProbabilityErrorEventE1} and substituting $J$ using \eqref{ValuesForBinningVariablesInTheorem1} we obtain $\ln(\text{Pr}(E_{1i})) \leq -(1-\epsilon)2^{n\epsilon}$. Thus, $\text{Pr}(E_{1i}) \rightarrow 0 \quad \text{as}\quad  n \rightarrow \infty$.

%------------------------ EVENTS RELATED TO INPUT CONSTRAINTS ---------------------

\item Let $E_{2i}$ be the event that the chosen codeword at the source, $\dv x_1(w_{i-1,w_i})$, exceeds the power constraint $\Gamma_1$ by $\gamma_1(\epsilon)$,
\begin{align}
E_{2i} &= \Big\{\varphi^n_1(\dv x_1(w_{i-1,w_i})) > \Gamma_1 + \gamma_1(\epsilon)\Big\}.
\end{align}

By the weak law of large numbers, we have
\begin{align}
\text{Pr}(E_{2i}) &= \text{Pr}\Big(\frac{1}{n}\sum_{i=1}^{n} \varphi_1(x_{1,i}(w)) > \Gamma_1 +\gamma_1(\epsilon) \Big)\nonumber\\
& < \epsilon
\end{align}
for $n$ large enough and $\mathbb{E}[\varphi_1(X_1)] < \Gamma_1$.

\item Let $E_{3i}$ be the event that the chosen codeword at the relay, $\dv x_2(w_{i-1})$, exceeds the power constraint $\Gamma_2$ by $\gamma_2(\epsilon)$,
\begin{align}
E_{3i} &= \Big\{\varphi^n_2(\dv x_2(w_{i-1})) > \Gamma_2 + \gamma_2(\epsilon)\Big\}.
\end{align}
Using arguments similar to those for the event $E_{2i}$, we get $\text{Pr}(E_{3i}|E_{1i}^c) < \epsilon$ for $n$ large enough and $\mathbb{E}[\varphi_2(X_2)] < \Gamma_2$, where $E_{1i}^c$ denotes the event complement of $E_{1i}$.
%----------------------------------------------------------------------------------

\item Let $E_{4i}$ be the event that $\dv x_1(w_{i-1},w_i)$, $\dv y_2[i]$, $\dv s[i]$ are not jointly typical given $\dv u_1(w_{i-1})$, $\dv u_2(w_{i-1},j^{\star})$ and $\dv x_2(w_{i-1})$, i.e.,
\begin{equation*}
E_{4i} = \Big\{\Big(\dv u_1(w_{i-1}),\dv u_2(w_{i-1},j^{\star}),\dv x_1(w_{i-1},w_i), \dv x_2(w_{i-1}), \dv y_2[i],\dv s[i]\Big) \notin T_{\epsilon}^n(P_{U_1,U_2,X_1,X_2,Y_2,S})\Big\}.
\end{equation*}
\noindent Conditioned on $E_{1i}^c$, $E_{2i}^c$, $E_{3i}^c$, we have that $(\dv s[i],\dv u_1(w_{i-1}))$ is jointly typical with $\dv u_2(w_{i-1},j^{\star})$ and with the source input $\dv x_1(w_{i-1},w_i)$ and the relay input $\dv x_2(w_{i-1})$, i.e.,
\begin{align}
&\Big(\dv s[i],\dv u_1(w_{i-1}),\dv u_2(w_{i-1},j^{\star}),\dv x_1(w_{i-1},w_i),\dv x_2(w_{i-1})\Big) \in T_{\epsilon}^n(Q_SP_{U_1}P_{X_1|U_1}P_{U_2,X_2|S,U_1,X_1}).
\label{ErrorEventE2E4}
\end{align}
For $\dv s[i]$, $\dv u_1(w_{i-1})$, $\dv u_2(w_{i-1},j^{\star})$, $\dv x_1(w_{i-1},w_i)$ and $\dv x_2(w_{i-1})$ jointly typical, we have $\text{Pr}(E_{4i}|\cap_{k=1}^{3} E_{ki}^c)$ $\longrightarrow 0$ as $n \longrightarrow \infty$, by the Markov Lemma \cite[p. 436]{CT91}.

\item Let $E_{5i}$ be the event that $\dv x_1(w_{i-1},w'_i)$, $\dv y_2[i]$, $\dv s[i]$ are jointly typical given $\dv u_1(w_{i-1})$, $\dv u_2(w_{i-1},j^{\star})$, $\dv x_2(w_{i-1})$ for some $w'_i \neq w_i$, i.e.,
\begin{align*}
E_{5i} = \Big\{\exists \; &w'_i \in \{1,\hdots,M\}\; \text{s.t.}\; w'_i\neq w_i, \\
&\Big(\dv u_1(w_{i-1}),\dv u_2(w_{i-1},j^{\star}),\dv x_1(w_{i-1},w'_i),\dv x_2(w_{i-1}),\dv y_2(i),\dv s\Big) \in T_{\epsilon}^n(P_{U_1,U_2,X_1,X_2,Y_2,S})\Big\}.
\end{align*}

Using the union bound and standard arguments on strongly typical sequences, the probability of the event $E_{5i}$ conditioned on $E_{1i}^c$, $E_{2i}^c$, $E_{3i}^c$, $E_{4i}^c$, can be easily bounded as
\begin{subequations}
\begin{align}
\text{Pr}(E_{5i}|E_{1i}^c,E_{2i}^c,E_{3i}^c,E_{4i}^c)  & \leq M2^{-n(I(X_1;Y_2,S|U_1,U_2,X_2)-\epsilon)}\\
                                     & =  2^{-n(I(X_1;Y_2|S,U_1,U_2,X_2)-R+3\epsilon)},
\label{ProbabilityErrorEvent3Step3ForAchievabilityInTheoreom1}
\end{align}
\label{ProbabilityErrorEvent3ForAchievabilityInTheoreom1}
\end{subequations}
where in \eqref{ProbabilityErrorEvent3Step3ForAchievabilityInTheoreom1} we used the fact that $I(X_1;S|U_1,U_2,X_2)=0$ under the joint distribution \eqref{MeasureForAchievabeRateNonCausalCaseDiscreteMemorylessChannel}. Thus, $\text{Pr}(E_{5i}|\cap_{k=1}^{4} E_{ki}^c) \longrightarrow 0$ as $n \longrightarrow \infty$ if $R < I(X_1;Y_2|S,U_1,U_2,X_2)$. This condition, can be rewritten equivalently as
\begin{align}
R &< I(X_1;Y_2|S,U_1,U_2,X_2)\nonumber\\
  &= H(Y_2|S,U_1,U_2,X_2)- H(Y_2|S,U_1,U_2,X_1,X_2)\nonumber\\
  &= H(Y_2|S,U_1,X_2)- H(Y_2|S,X_1,X_2)\nonumber\\
  &=I(X_1;Y_2|S,U_1,X_2)
\end{align}
where the second equality holds since the measure \eqref{MeasureForAchievabeRateNonCausalCaseDiscreteMemorylessChannel} satisfies $P_{Y_2|S,U_1,U_2,X_2}=P_{Y_2|S,U_1,X_2}$; and $Y_2$ and $(U_1,U_2)$ are conditionally independent given $(S,X_1,X_2)$.

\item For the decoding of message $w_B$ at the destination, let $E_{6B}$ be the event that $\dv u_1(w_B)$, $\dv u_2(w_B,j(\dv s[B+1],w_B))$, $\dv x_1(w_B,1)$, $\dv y_3[B+1]$ are not jointly typical, i.e.,
\begin{equation*}
E_{6B} = \Big\{\Big(\dv u_1(w_B), \dv u_2(w_B,j(\dv s[B+1],w_B)),\dv x_1(w_B,1), \dv y_3[B+1]\Big) \notin T_{\epsilon}^n(P_{U_1,U_2,X_1,Y_3})\Big\}.
\end{equation*}

For $\dv s[B+1]$, $\dv u_1(w_B)$, $\dv u_2(w_B,j(\dv s[B+1], w_B))$, $\dv x_1(w_B,1)$ and $\dv x_2(w_B)$ jointly typical as shown by \eqref{ErrorEventE2E4}, $\text{Pr}(E_{6B}|\cap_{k=1}^{5} E_{ki}^c) \longrightarrow 0$ as $n \longrightarrow \infty$, by the Markov Lemma.

\item For the decoding of message $w_B$ at the destination, let $E_{7B}$ be the event that $\dv u_1(w'_B)$, $\dv u_2(w'_B,j'_B)$, $\dv x_1(w'_B,1)$, $\dv y_3[B+1]$ are jointly typical for some $w'_B \neq w_B$ and some $j'_B \in \{1,\hdots,J\}$, i.e.,
\begin{align}
E_{7B} = \Big\{ \exists \; w'_B \in \{1,\hdots,&M\},  j'_B \in \{1,\hdots,J\} \; \text{s.t.} \; w'_B\neq w_B, \nonumber\\
&\Big(\dv u_1(w'_B), \dv u_2(w'_B,j'_B),\dv x_1(w'_B,1), \dv y_3[B+1]\Big) \in T_{\epsilon}^n(P_{U_1,U_2,X_1,Y_3})\Big\}.\nonumber
\end{align}
Conditioned on the events $E_{1i}^c$, $E_{2i}^c$, $E_{3i}^c$, $E_{4i}^c$, $E_{5i}^c$, $E_{6B}^c$, the probability of the event $E_{7B}$ can be bounded using the union bound, as
\begin{subequations}
\begin{align}
\text{Pr}(E_{7B}|\cap_{k=1}^{5} E_{ki}^c, E_{6B}^c) & \leq MJ2^{-n(I(X_1,U_1,U_2;Y_3)-\epsilon)}\\& = 2^{-n(I(X_1,U_1,U_2;Y_3)-I(U_2;S|U_1)-R+\epsilon)}.
\end{align}
\label{ProbabilityErrorEvent5ForAchievabilityInTheoreom1}
\end{subequations}
Thus $\text{Pr}(E_{7B}|\cap_{k=1}^{5} E_{ki}^c,E_{6B}^c) \longrightarrow 0$ as $n \longrightarrow \infty$ if $R < I(X_1,U_1,U_2;Y_3)-I(U_2;S|U_1)$.
\item For the decoding of message $w_{b-1}$ at the destination, $b=B,\hdots,2$, let $E_{8(b-1)}$ be the event that $\dv u_1(w_{b-1})$, $\dv u_2(w_{b-1},j(\dv s[b],w_{b-1}))$, $\dv x_1(w_{b-1},w_b)$, $\dv y_3[b]$ are not jointly typical, i.e.,
\begin{equation*}
E_{8(b-1)}= \Big\{\Big(\dv u_1(w_{b-1}), \dv u_2(w_{b-1},j(\dv s[b],w_{b-1})),\dv x_1(w_{b-1},w_b), \dv y_3[b]\Big) \notin T_{\epsilon}^n(P_{U_1,U_2,X_1,Y_3})\Big\}.
\end{equation*}
For $\dv s[b]$, $\dv u_1(w_{b-1})$, $\dv u_2(w_{b-1},j(\dv s[b], w_{b-1}))$, $\dv x_1(w_{b-1},w_b)$ and $\dv x_2(w_{b-1})$ jointly typical as shown by \eqref{ErrorEventE2E4}, $\text{Pr}(E_{8(b-1)}|\cap_{k=1}^{5} E_{ki}^c,E_{6B}^c,E_{7B}^c) \longrightarrow 0$ as $n \longrightarrow \infty$, by the Markov Lemma.

\item For the decoding of message $w_{b-1}$ at the destination, let $E_{9(b-1)}$ be the event that $\dv u_1(w'_{b-1})$, $\dv u_2(w'_{b-1},j'_{b-1})$, $\dv x_1(w'_{b-1},w_b)$, $\dv y_3[b]$ are jointly typical for some $w'_{b-1} \neq w_{b-1}$ and some $j'_{b-1} \in \{1,\hdots,J\}$, i.e.,
\begin{align}
E_{9(b-1)} = \Big\{ \exists \; w'_{b-1} \in & \{1,\hdots,M\},  j'_{b-1} \in \{1,\hdots,J\}, \; \text{s.t.} \; w'_{b-1}\neq w_{b-1}, \nonumber\\
&\Big(\dv u_1(w'_{b-1}), \dv u_2(w'_{b-1},j'_{b-1}),\dv x_1(w'_{b-1},w_b), \dv y_3[b]\Big) \in T_{\epsilon}^n(P_{U_1,U_2,X_1,Y_3})\Big\}.\nonumber
\end{align}
Proceeding like for the event $E_{7B}$, one can easily show that  $\text{Pr}(E_{9(b-1)}|\cap_{k=1}^{5} E_{ki}^c,E_{6B}^c,E_{7B}^c,E_{8(b-1)}^c)$ can be bounded similarly to in \eqref{ProbabilityErrorEvent5ForAchievabilityInTheoreom1}, and this shows that $ \text{Pr}(E_{9(b-1)}|\cap_{k=1}^{5} E_{ki}^c,E_{6B}^c,E_{7B}^c,E_{8(b-1)}^c)$ $\longrightarrow 0$ as $n \longrightarrow \infty$ if $R < I(X_1,U_1,U_2;Y_3)-I(U_2;S|U_1)$.

\end{itemize}

It remains to show that the rate \eqref{AchievabeRateNonCausalCaseDiscreteMemorylessChannel} is not altered if one restricts the random variables $U_1$ and $U_2$ to have their alphabet sizes limited as indicated in \eqref{BoundsOnCardinalityOfAuxiliaryRandonVariablesForAchievabeRateNonCausalCaseDiscreteMemorylessChannel}. This is done by invoking the support lemma \cite[p. 310]{CK81}. Fix a distribution $\mu$ of $(S,U_1,U_2,X_1,X_2,Y_2,Y_3)$ on $\mc P({\mc S}{\times}{\mc U_1}{\times}{\mc U_2}{\times}{\mc X_1}{\times}{\mc X_2}{\times}{\mc Y_2}{\times}{\mc Y_3})$ that has the form \eqref{MeasureForAchievabeRateNonCausalCaseDiscreteMemorylessChannel} and satisfies $\mathbb{E}[\varphi_i(X_i)] \leq \Gamma_i$, $i=1,2$.

\noindent To prove the bound \eqref{BoundsOnCardinalityOfAuxiliaryRandonVariableU1ForAchievabeRateNonCausalCaseDiscreteMemorylessChannel} on $|\mc U_1|$, note that we have 
\begin{subequations}
\begin{align}
\label{BoundsOnCardinalityForTheorem1Equation11}
I_{\mu}(X_1;Y_2|S,U_1,X_2) &= I_{\mu}(X_1;Y_2,S,X_2|U_1)\\
&= H_{\mu}(X_1|U_1)+H_{\mu}(Y_2,S,X_2|U_1)-H_{\mu}(X_1,X_2,Y_2,S|U_1),
\label{BoundsOnCardinalityForTheorem1Equation12}
\end{align}
\end{subequations}
where \eqref{BoundsOnCardinalityForTheorem1Equation11} follows since $(S,X_2) \leftrightarrow U_1 \leftrightarrow X_1$ under the distribution $\mu$. Also, we have 
\begin{subequations}
\begin{align}
I_{\mu}(X_1,U_1,U_2;Y_3)-I_{\mu}(U_2;S|U_1)&=I_{\mu}(U_1;Y_3)+I_{\mu}(X_1,U_2;Y_3|U_1)-I_{\mu}(U_2;S|U_1)\\
&=H_{\mu}(Y_3)-H_{\mu}(S)-H_{\mu}(U_2|U_1)+H_{\mu}(U_2,S|U_1)\nonumber\\
&+H_{\mu}(X_1,U_2|U_1)-H_{\mu}(X_1,U_2,Y_3|U_1).
\end{align}
\label{BoundsOnCardinalityForTheorem1Equation2}
\end{subequations}
Hence, it suffices to show that the following functionals of $\mu(S,U_1,U_2,X_1,X_2,Y_2,Y_3)$
\begin{subequations}
\begin{align}
\label{FunctionalRBoundsOnCardinalityForTheorem1}
r_{s,x,x'}(\mu) &= \mu(s,x,x') \quad \forall \: (s,x,x') \in {\mc S}{\times}{\mc X_1}{\times}{\mc X_2}\\
\label{FunctionalR1BoundsOnCardinalityForTheorem1}
r_1(\mu) &= \int_{u}d_{\mu}(u)[H_{\mu}(X_1|u)+H_{\mu}(Y_2,S,X_2|u)-H_{\mu}(X_1,X_2,Y_2,S|u)]\\
\label{FunctionalR2BoundsOnCardinalityForTheorem1}
r_2(\mu) &= \int_{u}d_{\mu}(u)[H_{\mu}(X_1,U_2|u)-H_{\mu}(X_1,U_2,Y_3|u)-H_{\mu}(U_2|u)+H_{\mu}(U_2,S|u)],
\end{align}
\label{FunctionalsForBoundOnU1AchievableRegionFullDecode-and-Forward}
\end{subequations} 
can be preserved with another measure $\mu'$ that has the form \eqref{MeasureForAchievabeRateNonCausalCaseDiscreteMemorylessChannel}. Observing that  there is a total of $|\mc S||\mc X_1||\mc X_2|+1$ functionals in \eqref{FunctionalsForBoundOnU1AchievableRegionFullDecode-and-Forward}, this is ensured by a standard application of the support lemma; and this shows that the cardinality of the alphabet of the auxiliary random variable $U_1$ can be limited as indicated in \eqref{BoundsOnCardinalityOfAuxiliaryRandonVariableU1ForAchievabeRateNonCausalCaseDiscreteMemorylessChannel} without altering the rate \eqref{AchievabeRateNonCausalCaseDiscreteMemorylessChannel}. We note that the inputs constraints for the source and the relay, which involve $\mu(S,U_1,U_2,X_1,X_2,Y_2,Y_3)$ only through its marginals over $(S,U_1,U_2,X_2,Y_2,Y_3)$ and $(S,U_1,U_2,X_1,Y_2,Y_3)$ respectively, are satisfied.

Once the alphabet of $U_1$ is fixed, we apply similar arguments to bound the alphabet of $U_2$, where this time $|\mc S||\mc X_1||\mc X_2|(|\mc S||\mc X_1||\mc X_2|+1)-1$ functionals must be satisfied in order to preserve the joint distribution of $S$, $U_1$, $X_1$, $X_2$, and one more functional to preserve 
\begin{align}
I_{\mu}(X_1,U_1,U_2;Y_3)-I_{\mu}(U_2;S|U_1)&=H_{\mu}(Y_3)-H_{\mu}(S)-H_{\mu}(U_1|U_2)+H_{\mu}(U_1,S|U_2)\nonumber\\
&+H_{\mu}(X_1,U_1|U_2)-H_{\mu}(X_1,U_1,Y_3|U_2),
\end{align}
yielding the bound indicated in \eqref{BoundsOnCardinalityOfAuxiliaryRandonVariableU2ForAchievabeRateNonCausalCaseDiscreteMemorylessChannel}.

\renewcommand{\theequation}{B-\arabic{equation}}
\subsection{Proof of Corollary \ref{CorollaryPartialDecodeAndForwardNonCausalCaseDiscreteMemorylessChannel}}\label{appendixCorollary1}
The proof combines rate-splitting \cite{H-MZ05} and the techniques used in the proof of Theorem \ref{TheoremAchievabeRateNonCausalCaseDiscreteMemorylessChannel}.

\noindent As we already mentioned in the discussion following Corollary \ref{CorollaryPartialDecodeAndForwardNonCausalCaseDiscreteMemorylessChannel}, we split the message $W$ to be transmitted from the source node into two independent parts $W_r$ and $W_d$; the relay forwards only the part $W_r$, at rate $R_r$, and the part $W_d$ is sent directly to the destination, at rate $R_d$. The total rate is $R=R_r+R_d$. We transmit in $B+1$ blocks, each of length $n$. During each of the first $B$ blocks, the source sends a message $w_i=(w_{r,i},w_{d,i})$, with $w_{r,i} \in [1,2^{nR_r}]$ and $w_{d,i} \in [1,2^{nR_d}]$ and $i=1,\hdots,B$ denotes the index of the block. For convenience, we let $w_{r,B+1}=w_{d,1}=1$ . For fixed $n$, the average rate $R\frac{B}{B+1}$ over $B+1$ blocks approaches $R$ as $B \longrightarrow +\infty$.

\textbf{Codebook generation:}
Fix a measure $P_{S,U_1,U_2,U,X_1,X_2,Y_2,Y_3}$ satisfying \eqref{MeasureForAchievabeRatePartialDecodeAndForwardNonCausalCaseDiscreteMemorylessChannel} and $\mathbb{E}[\varphi_i(X_i)] \leq \Gamma_i$, $i=1,2$. Fix $\epsilon > 0$ and let
\begin{subequations}
\begin{align}
J & =  2^{n(I(U_2;S|U_1)+2\epsilon)}\\
M_r & =  2^{n(R_r-2\epsilon)}\\
M_d & =  2^{n(R_d-4\epsilon)}.
\end{align}
\label{ValuesForBinningVariablesInCorollary1}
\end{subequations}
\begin{enumerate}
\item[1.] We generate $M_r$ i.i.d. codewords $\{\dv u_1(w'_r)\}$ indexed by $w'_r=1,\hdots,M_r$, each with i.i.d. components drawn according to $P_{U_1}$. For each $\dv u_1(w'_r)$, we generate $M_r$ i.i.d. codewords $\{\dv u(w'_r,w_r)\}$ at the source indexed by $w_r=1,\hdots,M_r$, and $J$ auxiliary codewords $\{\dv u_2(w'_r,j)\}$ at the relay indexed by $j=1,\hdots,J$. The codewords $\dv u(w'_r,w_r)$ and  $\dv u_2(w'_r,j)$ are with i.i.d. components  given $\dv u_1(w'_r)$ drawn according to $P_{U|U_1}$ and $P_{U_2|U_1}$, respectively.
\item[2.] For each $\dv u_1(w_r')$, for each $\dv u(w_r',w_r)$, we generate $M_d$ i.i.d. codewords $\{\dv x_1(w_r',w_r,w_d)\}$ indexed by $w_d=1,\hdots,M_d$, each with i.i.d. components given $(\dv u_1(w_r'),\dv u(w_r',w_r))$ drawn according to $P_{X_1|U_1,U}$.
\end{enumerate}

\textbf{Encoding:} 
At the beginning of block $i$, let $w_i=(w_{r,i},w_{d,i})$ be the new message to be sent from the source and $w_{i-1}=(w_{r,i-1},w_{d,i-1})$ be the message sent in the previous block $i-1$. 

 At the beginning of block $i$, the relay has decoded $w_{r,i-1}$ correctly, and the source transmits $\dv x_1(w_{r,i-1},w_{r,i},w_{d,i})$. The relay searches for the smallest $j \in \{1,\cdots,J\}$ such that ${\dv u_2}(w_{r,i-1},j)$ and $\dv s[i]$ are jointly typical given $\dv u_1(w_{r,i-1})$. Since the vectors ${\dv u_2}(w_{r,i-1},j)$ and $\dv s[i]$ are generated independently given $\dv u_1(w_{r,i-1})$ according to the memoryless distributions defined by the $n$-product of $P_{U_2|U_1}$ and the $n$-product of $Q_S$, respectively; and there are $J$ sequences in the bin indexed by $w_{r,i-1}$, the probability that there is no such sequence ${\dv u_2}$ goes to zero as $n \longrightarrow +\infty$. Denote the found $j$ by $j^{\star}=j(\dv s[i], w_{r,i-1})$. The relay then transmits a vector $\dv x_2(w_{r,i-1})$ with i.i.d. components conditionally given $(\dv u_1(w_{r,i-1}), \dv u_2(w_{r,i-1},j^{\star}), \dv s[i])$ drawn according to the marginal $P_{X_2|U_1,U_2,S}$ induced by \eqref{MeasureForAchievabeRatePartialDecodeAndForwardNonCausalCaseDiscreteMemorylessChannel}. Using arguments similar to those in the proof of Theorem~\ref{TheoremAchievabeRateNonCausalCaseDiscreteMemorylessChannel}, it can be shown that the inputs $\dv x_1(w_{r,i-1},w_{r,i},w_{d,i})$  and $\dv x_2(w_{r,i-1})$ satisfy the input constraints. 

\textbf{Decoding:} The decoding procedures at the source and the relay are as follows.
\begin{enumerate}
\item[1.] At the end of block $i$, the relay knows $w_{r,i-1}$ and declares that $\hat{w}_{r,i}$ is sent if there is a unique $\hat{w}_{r,i}$ such that $\dv u(w_{r,i-1},\hat{w}_{r,i})$, $\dv y_2[i]$ and $\dv s[i]$ are jointly typical given $\dv u_1(w_{r,i-1})$, $\dv u_2(w_{r,i-1},j^{\star})$ and $\dv x_2(w_{r,i-1})$. One can show that the decoding error in this step is small for sufficiently large $n$ if
\begin{equation}
R_r < I(U;Y_2|S,U_1,X_2).
\label{RateRrAchievabeRatePartialDecodeAndForwardNonCausalCaseDiscreteMemorylessChannel}
\end{equation}
\item[2.] At the end of the transmission, the destination has collected all the blocks of channel outputs $\dv y_3[1],\dv y_3[2],\hdots,\dv y_3[B+1]$, and can then perform backward-decoding by first decoding $(w_{r,B},w_{d,B+1})$ from $\dv y_3[B+1]$. 

First, it declares that the pair $(\hat{w}_{r,B},\hat{w}_{d,B+1})$ is sent if there is a unique pair $(\hat{w}_{r,B},\hat{w}_{d,B+1})$, with $\hat{w}_{r,B} \in \{1,\hdots,M_r\}$ and $\hat{w}_{d,B+1} \in \{1,\hdots,M_d\}$, there is $j_B \in \{1,\hdots,J\}$, such that $\dv u_1(\hat{w}_{r,B})$, $\dv u_2(\hat{w}_{r,B},j_B)$, $\dv u(\hat{w}_{r,B},1)$, $\dv x_1(\hat{w}_{r,B},1,\hat{w}_{d,B+1})$, $\dv y_3[B+1]$ are jointly typical. One can show that the decoding error in this step is small for sufficiently large $n$ if
\begin{align}
R_d &< I(X_1;Y_3|U,U_1,U_2)\nonumber\\
R_d &< I(X_1,U_2;Y_3|U,U_1)-I(U_2;S|U_1)\nonumber\\
R_r+R_d &< I(X_1,U,U_1,U_2;Y_3)-I(U_2;S|U_1)\nonumber\\
        &\stackrel{(a)}{=} I(X_1,U_1,U_2;Y_3)-I(U_2;S|U_1),
\label{RateRrRdAchievabeRatePartialDecodeAndForwardNonCausalCaseDiscreteMemorylessChannel}
\end{align}
where in $(a)$ we used the fact that $I(U;Y_3|U_1,U_2,X_1)=0$ under the distribution \eqref{MeasureForAchievabeRatePartialDecodeAndForwardNonCausalCaseDiscreteMemorylessChannel}. 

\noindent Next, for $b$ ranging from $B$ to $2$, the destination knows $w_{r,b}$ and decodes $(w_{r,b-1},w_{d,b})$ based on the information received in block $b$. It declares that the pair $(\hat{w}_{r,b-1},\hat{w}_{d,b})$ is sent if there is a unique pair $(\hat{w}_{r,b-1},\hat{w}_{d,b})$, with $\hat{w}_{r,b-1} \in \{1,\hdots,M_r\}$ and $\hat{w}_{d,b} \in \{1,\hdots,M_d\}$, there is $j_{b-1} \in \{1,\hdots,J\}$, such that $\dv u_1(\hat{w}_{r,b-1})$, $\dv u_2(\hat{w}_{r,b-1},j_{b-1})$, $\dv u(\hat{w}_{r,b-1},w_{r,b})$, $\dv x_1(\hat{w}_{r,b-1},w_{r,b},\hat{w}_{d,b})$, $\dv y_3[b]$ are jointly typical. One can show that the decoding error in this step is small for sufficiently large $n$ if \eqref{RateRrRdAchievabeRatePartialDecodeAndForwardNonCausalCaseDiscreteMemorylessChannel} is true.
\end{enumerate}

It remains to show that the rate \eqref{AchievabeRatePartialDecodeAndForwardNonCausalCaseDiscreteMemorylessChannel} is not altered if the sizes of the alphabets of the auxiliary random variables $U$, $U_1$ and $U_2$ are restricted as in \eqref{BoundsOnCardinalityOfAuxiliaryRandonVariablesForAchievabeRatePartialDecodeAndForwardNonCausalCaseDiscreteMemorylessChannel}. This can be easily done by following the steps in the proof of Theorem~\ref{TheoremAchievabeRateNonCausalCaseDiscreteMemorylessChannel}.

\renewcommand{\theequation}{C-\arabic{equation}}
\subsection{Proof of Theorem \ref{TheoremOuterBoundNonCausalCaseDiscreteMemorylessChannel}}\label{appendixTheorem2}

Consider a sequence of $(\epsilon_n,n,R,\dv \Gamma)-$codes with $\epsilon_n \rightarrow 0$ as $n \rightarrow +\infty$. We show that $R$ must be less than or equal $R^{\text{up}}(\dv \Gamma)$. By Fano's inequality, we have
\begin{equation}
H(W|Y_3^n) \leq nR\epsilon_n+1 \triangleq n\delta_n.
\end{equation}
Thus, 
\begin{align}
nR &= H(W) \leq I(W;Y_3^n)+n\delta_n\nonumber\\
\label{FanosInequality}
\end{align}

We upper bound $I(W;Y^n_3)$ as in the following lemma, the proof of which follows.

\begin{lemma}\label{LemmaProofTheoremOuterBound} 
\begin{subequations}
\begin{align}
\label{Equation1Lemma1}
& \text{i)} \:\:\: I(W;Y_3^n) \leq \sum_{i=1}^{n}I(X_{1,i},X_{2,i};Y_{3,i}|S_i)-I(S_i;X_{1,i}|Y_{3,i})\\
& \text{ii)} \:\:\: I(W;Y_3^n) \leq \sum_{i=1}^{n}I(X_{1,i};Y_{2,i},Y_{3,i}|S_i,X_{2,i}).
\label{Equation2Lemma1}
\end{align}
\label{EquationsLemma1}
\end{subequations}
\end{lemma}
\begin{proof}
To simplify the notation, we use $S^i=(S_1,S_2,\cdots,S_i)$, $Y_k^i=(Y_{k,1},Y_{k,2},\cdots,Y_{k,i})$, $k=2,3$, and $X_j^i=(X_{j,1},X_{j,2},\cdots,X_{j,i})$, $j=1,2$. 

We obtain the bound on $I(W;Y_3^n)$ given in (i) as follows.
{\allowdisplaybreaks
\begin{align}
I(W;Y_3^n) & = I(W,S^n;Y_3^n)-I(S^n;Y_3^n|W)\nonumber\\
	& = \sum_{i=1}^{n}I(W,S^n;Y_{3,i}|Y_3^{i-1})-H(S^n|W)+H(S^n|W,Y_3^n)\nonumber\\
	& = \sum_{i=1}^{n}[H(Y_{3,i}|Y_3^{i-1})-H(Y_{3,i}|W,S^n,Y_3^{i-1})\nonumber\\
	& \qquad - H(S_i)+H(S_i|W,Y_3^n,S^{i-1})]\nonumber\\
	& \stackrel{(a)}{\leq} \sum_{i=1}^{n}[H(Y_{3,i})-H(Y_{3,i}|X_{1,i},X_{2,i},S_i)\nonumber\\
	& \qquad - H(S_i)+H(S_i|W,Y_3^n,S^{i-1},X_{1,i})]\nonumber\\
	& \stackrel{(b)}{\leq} \sum_{i=1}^{n}[I(X_{1,i},X_{2,i},S_i;Y_{3,i})-H(S_i)+H(S_i|X_{1,i},Y_{3,i})]\nonumber\\
	& = \sum_{i=1}^{n}[I(X_{1,i},X_{2,i},S_i;Y_{3,i})-I(S_i;X_{1,i},Y_{3,i})]\nonumber\\
	& = \sum_{i=1}^{n}[I(X_{1,i},X_{2,i};Y_{3,i}|S_i)-I(S_i;X_{1,i}|Y_{3,i})],\nonumber
\end{align}}
where\\
$(a)$ follows from $(W,S^n,Y_3^{i-1}) \leftrightarrow (X_{1,i},X_{2,i},S_i) \leftrightarrow Y_{3,i}$ (a Markov chain); and the fact that $X_{1,i}$ is a deterministic function of $W$; and\\
$(b)$ follows from the fact that conditioning reduces entropy.\\

We  obtain the bound on $I(W;Y_3^n)$ given in (ii) as follows.
{\allowdisplaybreaks
\begin{align}
&I(W;Y_3^n)  \leq I(W;Y_2^n,Y_3^n)\nonumber\\
	& = H(W)-H(W|Y_2^n,Y_3^n)\nonumber\\
	& \stackrel{(c)}{\leq} H(W|S^n)-H(W|Y_2^n,Y_3^n,S^n)\nonumber\\
	& = \sum_{i=1}^{n}I(W;Y_{2,i},Y_{3,i}|Y_2^{i-1},Y_3^{i-1},S^n)\nonumber\\
	& \stackrel{(d)}{=} \sum_{i=1}^{n}I(W;Y_{2,i},Y_{3,i}|Y_2^{i-1},Y_3^{i-1},S^n,X_{2,i})\nonumber\\
	 & = \sum_{i=1}^{n}[H(Y_{2,i},Y_{3,i}|Y_2^{i-1},Y_3^{i-1},S^n,X_{2,i})\nonumber\\
& \qquad -H(Y_{2,i},Y_{3,i}|Y_2^{i-1},Y_3^{i-1},S^n,X_{2,i},W)]\nonumber\\
	& \stackrel{(e)}{=} \sum_{i=1}^{n}[H(Y_{2,i},Y_{3,i}|Y_2^{i-1},Y_3^{i-1},S^n,X_{2,i})\nonumber\\
	& \qquad -H(Y_{2,i},Y_{3,i}|Y_2^{i-1},Y_3^{i-1},S^n,X_{2,i},W,X_{1,i})]\nonumber\\
	& \leq \sum_{i=1}^{n}[H(Y_{2,i},Y_{3,i}|S_i,X_{2,i})\nonumber\\
        & \qquad -H(Y_{2,i},Y_{3,i}|Y_2^{i-1},Y_3^{i-1},S^n,X_{2,i},W,X_{1,i})]\nonumber\\
	& \stackrel{(f)}{=} \sum_{i=1}^{n}[H(Y_{2,i},Y_{3,i}|S_i,X_{2,i})-H(Y_{2,i},Y_{3,i}|S_i,X_{2,i},X_{1,i})]\nonumber\\
	 & = \sum_{i=1}^{n}I(X_{1,i};Y_{2,i},Y_{3,i}|S_i,X_{2,i}),\nonumber
\end{align}
}
where\\
$(c)$ follows from the fact that $W$ and $S^n$ are independent; and  $H(W|Y_2^n,Y_3^n) \geq H(W|Y_2^n,Y_3^n,S^n)$; \\
$(d)$ follows from the fact that $X_{2,i}$ is a deterministic function of $(S^n,Y_2^{i-1})$; \\
$(e)$ follows from  the fact that $X_{1,i}$ is a deterministic function of $W$; and\\
$(f)$ follows from the fact that the channel is discrete memoryless.\\
\end{proof}

% INPUT CONSTRAINTS

Consider now the input constraints. By definition the code satisfies
\begin{align}
\frac{1}{n}\sum_{i=1}^{n}\varphi_1(\phi^n_1(w)_i) &\leq \Gamma_1\nonumber\\
\frac{1}{n}\sum_{i=1}^{n}\varphi_2(\phi_{2,i}(y^{i-1}_2,s^n)) &\leq \Gamma_2.
\end{align}
for $w \in \{1,\cdots,M\}$.

\noindent We start with the input constraint of the source. Since each codeword satisfies the input constraint, their average over $w_1$ also satisfies the input constraint. Thus, we have
\begin{align}
\Gamma_1 & \geq \sum_{w=1}^{M} P(w)\varphi^n_1(x^n_1(w))\nonumber\\
         & = \sum_{w=1}^{M}P(w)\frac{1}{n}\sum_{i=1}^{n}\varphi_1(x_{1,i}(w))\nonumber\\
         & = \sum_{x^n_1} \sum_{w=1}^{M} P(w)P(x^n_1|w) \frac{1}{n}\sum_{i=1}^{n}\varphi_1(x_{1,i}(w))\nonumber\\
         & = \frac{1}{n}\sum_{i=1}^{n}\sum_{w=1}^{M}\sum_{x^n_1}P(w)P(x^n_1|w)\varphi_1(x_{1,i}(w))\nonumber\\
         & = \frac{1}{n}\sum_{i=1}^{n}\mathbb{E}_{X_{1,i}}[\varphi_1(X_{1,i})].
\label{InputConstraintOfSource__ProofOfUpperBound}
\end{align}

Similarly, for the input constraint of the relay, we have
\begin{align}
\Gamma_2 & \geq \sum_{w=1}^{M}\sum_{s^n} P(w)P(s^n) \frac{1}{n}\sum_{i=1}^{n}\varphi_2(\phi_{2,i}(y^{i-1}_2,s^n))\nonumber\\
         & = \sum_{w=1}^{M}\sum_{s^n,x^n_1,x^n_2,y^n_2} P(w)P(s^n)P(x^n_1|w)P(x^n_2|s^n,x^n_1)P(y^n_2|s^n,x^n_1,x^n_2)\nonumber\\
         & \qquad \qquad \qquad {\cdot}\frac{1}{n}\sum_{i=1}^{n}\varphi_2(\phi_{2,i}(y^{i-1}_2,s^n))\nonumber\\
         & = \frac{1}{n}\sum_{i=1}^{n}\mathbb{E}_{X_{2,i}}[\varphi_2(X_{2,i})].
\label{InputConstraintOfRelay__ProofOfUpperBound}
\end{align}

We introduce a random variable $T$ which is uniformly  distributed over $\{1,\cdots,n\}$. Set $S=S_T$, $X_1=X_{1,T}$, $X_2=X_{2,T}$, $Y_2=Y_{2,T}$, and $Y_3=Y_{3,T}$. We substitute $T$ into the above bounds on the message rate and the input constraints. Considering the bounds given in Lemma~\ref{LemmaProofTheoremOuterBound}, we obtain
\begin{align}
&\frac{1}{n}\sum_{i=1}^{n}I(X_{1,i},X_{2,i};Y_{3,i}|S_i)-I(S_i;X_{1,i}|Y_{3,i})\nonumber\\
&=I(X_1,X_2;Y_3|S,T)-I(S;X_1|Y_3,T)\nonumber\\
&=I(X_1,X_2,S;Y_3|T)-I(S;X_1,Y_3|T),
\label{OuterBoundWithTimeSharingRandomVariablePart1}
\end{align}
and
\begin{align}
\frac{1}{n}\sum_{i=1}^{n}I(X_{1,i};Y_{2,i},Y_{3,i}|S_i,X_{2,i}) &= I(X_1;Y_2,Y_3|S,X_2,T),
\label{OuterBoundWithTimeSharingRandomVariablePart2}
\end{align}
where the distribution on $(T,S,X_1,X_2,Y_2,Y_3)$  from a given code is of the form
\begin{align}
P_{T,S,X_1,X_2,Y_2,Y_3}=P_SP_TP_{X_1|T}P_{X_2|X_1,S,T}W_{Y_2,Y_3|S,X_1,X_2}.
\label{AllowedDistributionOuterBoundWithTimeSharingRandomVariable}
\end{align}
Similarly, substituting $T$ into the input constraints, we obtain
\begin{align}
\Gamma_k & \geq \frac{1}{n}\sum_{i=1}^{n}\sum_{x_{k,i}}P_{X_k}(x_{k,i})\varphi_k(x_{k,i})\nonumber\\
         & = \sum_{t=1}^{n} \frac{1}{n} \sum_{x_{k}}P_{X_k|T}(x_k|t)\varphi_k(x_k)\nonumber\\
         & = \mathbb{E}[\varphi_k(X_k)], \quad k=1,2.
\end{align}

We now eliminate the variable $T$ from \eqref{OuterBoundWithTimeSharingRandomVariablePart1} and \eqref{OuterBoundWithTimeSharingRandomVariablePart2} as follows. The RHS of \eqref{OuterBoundWithTimeSharingRandomVariablePart1} can be bounded as 
{\allowdisplaybreaks
\begin{align}
& I(X_1,X_2,S;Y_3|T)-I(S;X_1,Y_3|T)\nonumber\\
& \stackrel{(g)}{\leq} H(Y_3)-H(Y_3|X_1,X_2,S)-H(S|T)+H(S|X_1,Y_3,T)\nonumber\\
& = I(X_1,X_2,S;Y_3)-H(S|T)+H(S|X_1,Y_3,T)\nonumber \\
& \stackrel{(h)}{\leq} I(X_1,X_2,S;Y_3)-H(S)+H(S|X_1,Y_3)\nonumber \\
& = I(X_1,X_2,S;Y_3)-I(S;X_1,Y_3)\nonumber\\
& = I(X_1,X_2;Y_3|S)-I(S;X_1|Y_3),
\label{OuterBoundTheorem2Part1}
\end{align}}
where\\
$(g)$ holds since $H(Y_3|T) \leq H(Y_3)$ and $H(Y_3|X_1,X_2,S,T)=H(Y_3|X_1,X_2,S)$ (by the Markovian relation $T \leftrightarrow (X_1,X_2,S) \leftrightarrow Y_3$); and\\
$(h)$ holds  since $S$ is independent of $T$ and $H(S|X_1,Y_3,T) \leq H(S|X_1,Y_3)$.\\
Similarly, the RHS of \eqref{OuterBoundWithTimeSharingRandomVariablePart2} can be bounded as
\begin{align}
I(X_1;Y_2,Y_3|S,X_2,T) \leq I(X_1;Y_2,Y_3|S,X_2).
\label{OuterBoundTheorem2Part2}
\end{align} 

Finally, combining \eqref{FanosInequality}, \eqref{Equation1Lemma1}, \eqref{OuterBoundWithTimeSharingRandomVariablePart1}, \eqref{OuterBoundTheorem2Part1} at one hand, and \eqref{FanosInequality}, \eqref{Equation2Lemma1}, \eqref{OuterBoundWithTimeSharingRandomVariablePart2}, \eqref{OuterBoundTheorem2Part2} at the other hand, we get 
\begin{subequations}
\begin{align}
 R & \:\: \leq \:\: I(X_1,X_2;Y_3|S)-I(S;X_1|Y_3)\\
 R & \:\: \leq \:\: I(X_1;Y_2,Y_3|S,X_2),
\end{align}
\label{OuterBoundTheorem2}
\end{subequations}
where the distribution on $(S,X_1,X_2,Y_2,Y_3)$, obtained by  marginalizing \eqref{AllowedDistributionOuterBoundWithTimeSharingRandomVariable} over the variable $T$, has the form given in \eqref{MeasureForOuterBoundNonCausalCaseDiscreteMemorylessChannel} and satisfies $\mathbb{E}[\varphi_i(X_i)] \leq \Gamma_i$ for $i=1,2$.

We conclude that, for a given sequence of $(\epsilon_n,n,R,\dv \Gamma)-$codes with $\epsilon_n$ going to zero as $n$ goes to infinity, there exists a probability distribution of the form \eqref{MeasureForOuterBoundNonCausalCaseDiscreteMemorylessChannel} such that the rate $R$ satisfies \eqref{OuterBoundTheorem2} and the input constraints $\mathbb{E}[\varphi_i(X_i)] \leq \Gamma_i$, $i=1,2$, are satisfied. This completes the proof of Theorem \ref{TheoremOuterBoundNonCausalCaseDiscreteMemorylessChannel}.

\renewcommand{\theequation}{D-\arabic{equation}}
\subsection{Proof of Theorem \ref{TheoremAchievabeRateNonCausalCaseGaussianChannelFullDuplexRegime}}\label{appendixTheorem3}
In this proof, we compute the lower bound in Theorem~\ref{TheoremAchievabeRateNonCausalCaseDiscreteMemorylessChannel} using an appropriate jointly Gaussian distribution on $S$, $X_1$, $U_1$, $U_2$, $X_2$. The techniques used in this section rely strongly on those used in the proof of Theorem $6$ in \cite{SBSV07a}. 

\noindent We first evaluate the second term of the minimization in \eqref{AchievabeRateNonCausalCaseDiscreteMemorylessChannel} because this gives insights about the distribution that we should use to compute the lower bound. The second term of the minimization in \eqref{AchievabeRateNonCausalCaseDiscreteMemorylessChannel} can be written as 
\begin{align}
&I(X_1,U_1,U_2;Y_3)-I(U_2;S|U_1)=\nonumber\\
&\hspace{0cm}I(X_1,U_1;Y_3)+I(U_2;Y_3|X_1,U_1)-I(U_2;S|X_1,U_1),
\label{ExapandedSumRateAchievableRateFullDuplexGaussianRC}
\end{align}
which follows from the fact that $I(U_2;S|U_1)=I(U_2;S|U_1,X_1)$ for the considered distribution. 

We first focus on the evaluation of the term $[I(U_2;Y_3|X_1,U_1)-I(U_2;S|X_1,U_1)]$. To evaluate it, we assume that $X_1$ is zero mean Gaussian with variance $P_1$, $U_1$ is zero mean Gaussian with variance $\bar{\theta}P_2$, and $X_1$ and $U_1$ are jointly Gaussian with $\mathbb{E}[U_1X_1]=\rho'_{12}\sqrt{\bar{\theta}P_1P_2}$, for some $\theta \in [0,1]$, $\rho'_{12} \in [-1,1]$. The random variables $X_1$ and $U_1$ are independent of $S$ as shown by the distribution given in Theorem~\ref{TheoremAchievabeRateNonCausalCaseDiscreteMemorylessChannel}. We also consider
\begin{align}
X_2 &= U_1+\tilde{X}_2
\label{RelayInputDistributionProofAchievabeRateNonCausalCaseGaussianChannelFullDuplexRegime_temp}
\end{align}
where, $\tilde{X}_2$ is zero mean Gaussian with variance ${\theta}P_2$, is independent of both $X_1$ and $U_1$, and is jointly Gaussian with $S$ with $\mathbb{E}[\tilde{X}_2S]=\rho'_{2s}\sqrt{{\theta}P_2Q}$, for some $\rho'_{2s} \in [-1,1]$. Then, from \eqref{ChannelModelForFullDuplexRegimeGaussianRCWithState} and 
\eqref{RelayInputDistributionProofAchievabeRateNonCausalCaseGaussianChannelFullDuplexRegime_temp}, we can write $Y_3$ as 
\begin{equation}
Y_3=X_1+U_1+\tilde{X}_2+S+Z_3.
\label{ReceivedAtDestinationForFullDuplexRegimeGaussianRC}
\end{equation}
Let $\hat{\tilde{X}}_2=\mathbb{E}[\tilde{X}_2|S]$ be the optimal linear estimator of $\tilde{X}_2$ given $S$ under minimum mean square error criterion, and $X'_2$ be the resulting estimation error. The estimator $\hat{\tilde{X}}_2$ and the estimation error $X'_2$ are given by
\begin{align}
\hat{\tilde{X}}_2 &=\rho'_{2s}\sqrt{\frac{{\theta}P_2}{Q}}S\\
X'_2 &=\tilde{X}_2-\hat{\tilde{X}}_2.
\end{align}
We can alternatively write $Y_3$ in \eqref{ReceivedAtDestinationForFullDuplexRegimeGaussianRC} as
\begin{align}
Y_3 &=(\tilde{X}_2-\hat{\tilde{X}}_2)+\hat{\tilde{X}}_2+X_1+U_1+S+Z_3\nonumber\\
    &=X'_2+X_1+U_1+S'+Z_3,
\end{align}
where 
$$S' = \left (1+\rho'_{2s}\sqrt{\frac{{\theta}P_2}{Q}}\right)S.$$

We now consider the following new channel output $Y'_3$ given by 
\begin{equation}
Y'_3:=Y_3-\mathbb{E}[Y_3|X_1,U_1]=X'_2+S'+Z_3.
\label{EquivalentChannelForGDPCFullDuplexRegime}
\end{equation}
This new channel output $Y'_3$ is similar to the channel output considered in \cite{C83} because $X'_2$ is independent of the state $S'$. Hence, the capacity of this new channel is achieved if we use an auxiliary random variable 
\begin{align}
U_2 &=X'_2+{\alpha}S',
\label{DPCforEquivalentFictitiousChannelGaussianFullDuplex}
\end{align}
where $\alpha$ is Costa's parameter given by
\begin{align}
\alpha &= \frac{\mathbb{E}[X'^2_2]}{\mathbb{E}[X'^2_2]+\mathbb{E}[Z^2_3]} = \frac{{\theta}P_2(1-\rho'^2_{2s})}{{\theta}P_2(1-\rho'^2_{2s})+N_3}.
\label{CostasParameterIntermediaryStep}
\end{align}
Then we can easily show that
$$[I(U_2;Y_3|X_1,U_1)-I(U_2;S|X_1,U_1)] = [I(U_2;Y'_3)-I(U_2;S')]. $$ 
The term $[I(U_2;Y'_3)-I(U_2;S')]$ is maximized if $U_2$ is chosen as in \eqref{DPCforEquivalentFictitiousChannelGaussianFullDuplex}. Thus, we obtain 
\begin{align}
I(U_2;Y_3|X_1,U_1)-I(U_2;S|X_1,U_1)&= \frac{1}{2}\log\left(1+\frac{\mathbb{E}[X'^2_2]}{N_3}\right)\nonumber\\
&=  \frac{1}{2}\log\left(1+\frac{{\theta}P_2(1-\rho'^2_{2s})}{N_3}\right).
\label{SecondPartOfSecondTermInMinFullDuplexGaussianCase}
\end{align}
By substituting $X'_2$ and $S'$ in \eqref{DPCforEquivalentFictitiousChannelGaussianFullDuplex}, we get 
\begin{align}
U_2 &=\tilde{X}_2-\rho'_{2s}\sqrt{\frac{{\theta}P_2}{Q}}S+{\alpha}\Big(1+\rho'_{2s}\sqrt{\frac{{\theta}P_2}{Q}}\Big)S\nonumber\\
     &= \tilde{X}_2+{\alpha_{\text{opt}}}S,
\label{AuxiliaryRandomVariableIntermediaryStep}
\end{align} 
where
\begin{align}
\alpha_{\text{opt}} &= \left(1+\rho'_{2s}\sqrt{\frac{{\theta}P_2}{Q}}\right){\alpha}-\rho'_{2s}\sqrt{\frac{{\theta}P_2}{Q}}\nonumber\\
&=\frac{{\theta}P_2(1-\rho'^2_{2s})-\rho'_{2s}\sqrt{\frac{{\theta}P_2}{Q}}N_3}{{\theta}P_2(1-\rho'^2_{2s})+N_3}.
% &=\frac{{\theta}P_2(1-\rho'^2_{2s})}{{\theta}P_2(1-\rho'^2_{2s})+N_3}\left(1+\rho'_{2s}\sqrt{\frac{{\theta}P_2}{Q}}\right)-\rho'_{2s}\sqrt{\frac{{\theta}P_2}{Q}}.\nonumber
%\label{OptimalCostaParameterIntermediaryStepGeneralizedDPCatRelayFullDuplexGaussianRC}
\end{align}

The term $I(X_1,U_1;Y_3)$ on the RHS of \eqref{ExapandedSumRateAchievableRateFullDuplexGaussianRC} can be computed as 
\begin{align}
I(X_1,U_1;Y_3) &= h(Y_3)-h(Y_3|X_1,U_1)\nonumber\\
	       &= h(Y_3)-h(\tilde{X}_2+S+Z_3|X_1,U_1)\nonumber\\
               &\stackrel{(b)}{=}h(Y_3)-h(\tilde{X}_2+S+Z_3)\nonumber\\
               &=\frac{1}{2}\log\Big(\frac{\mathbb{E}[(X_1+X_2+S)^2]+\mathbb{E}[Z^2_3]}{\mathbb{E}[(\tilde{X}_2+S)^2]+\mathbb{E}[Z^2_3]}\Big)\nonumber\\
               &=\frac{1}{2}\log\Big(1+\frac{P_1+\bar{\theta}P_2+2\rho'_{12}\sqrt{\bar{\theta}P_1P_2}}{{\theta}P_2+Q+N_3+2\rho'_{2s}\sqrt{{\theta}P_2Q}}\Big),
\label{FirstPartOfSecondTermInMinFullDuplexGaussianCase}
\end{align}
where $(b)$ follows from the fact that $\tilde{X}_2$ and $S$ are independent of $(X_1,U_1)$. Then, by adding  \eqref{SecondPartOfSecondTermInMinFullDuplexGaussianCase} and \eqref{FirstPartOfSecondTermInMinFullDuplexGaussianCase} we get the second term of the minimization in \eqref{AchievabeRateNonCausalCaseGaussianChannelFullDuplexRegime}.

The first term of the minimization in \eqref{AchievabeRateNonCausalCaseDiscreteMemorylessChannel} can be written as 
\begin{align}
I(X_1;Y_2|S,U_1,X_2) &= h(Y_2|S,U_1,X_2)-h(Y_2|S,U_1,X_1,X_2)\nonumber\\
                     &= h(X_1+Z_2|S,U_1,X_2)-h(Z_2)\nonumber\\
                     &\stackrel{(a)}{=} h(X_1+Z_2|U_1)-h(Z_2)\nonumber\\
		     &=\frac{1}{2}\log(1+\frac{P_1(1-\rho'^2_{12})}{N_2}),
\label{FirstTermInMinFullDuplexGaussianCase}
\end{align}
where $(a)$ follows from the fact that $X_1$ and $(S,X_2)$ are independent conditionally on $U_1$. 

Finally, we obtain the rate on the RHS of \eqref{AchievabeRateNonCausalCaseGaussianChannelFullDuplexRegime} by maximization over all possible values of $\theta \in [0,1]$, $\rho'_{12} \in [-1,1]$ and $\rho'_{2s} \in [-1,1]$. Investigating the two terms of the minimization, we can easily see that it suffices to consider $\rho'_{12} \in [0,1]$ and $\rho'_{2s} \in [-1,0]$.

%-----------------------------------------------------

\renewcommand{\theequation}{E-\arabic{equation}}
\setcounter{equation}{0}  % reset counter
\subsection{Proof of Corollary~\ref{CorollaryAchievabeRatePartialDecodeAndForwardNonCausalCaseGaussianChannel}}\label{appendixCorollary3}

Recall the outline after Corollary~\ref{CorollaryAchievabeRatePartialDecodeAndForwardNonCausalCaseGaussianChannel}. We decompose the source input $X_1$ and the relay input $X_2$ as
\begin{align}
X_1 &= U+\tilde{X}_1\\
X_2 &= U_1+\tilde{X}_2,
\end{align}
where $U$ and $\tilde{X}_1$ are independent zero mean Gaussian random variables with variances $\bar{\gamma}P_1$ and ${\gamma}P_1$, respectively, for some $\gamma \in [0,1]$; and $U_1$ and $\tilde{X}_2$ are independent zero mean Gaussian random variables with variances $\bar{\theta}P_2$ and ${\theta}P_2$, respectively, for some $\theta \in [0,1]$. Furthermore, $\tilde{X}_1$ is independent of all other variables; $U$ and $U_1$ are correlated, with $\mathbb{E}[UU_1]=\rho'_{12}\sqrt{\bar{\theta}\bar{\gamma}P_1P_2}$ for some $\rho'_{12} \in [0,1]$, and are both independent of $S$; $\tilde{X}_2$ is independent of $U$, is correlated with $S$ with $\mathbb{E}[\tilde{X}_2S]=\rho'_{2s}\sqrt{{\theta}P_2Q}$ for some $\rho'_{2s} \in [-1,0]$, and is obtained using a GDPC the auxiliary random variable of which is given by
\begin{equation}
U_2 = \tilde{X}_2 +\Big[\alpha'(1+\rho'_{2s}\sqrt{\frac{{\theta}P_2}{Q}})-\rho'_{2s}\sqrt{\frac{{\theta}P_2}{Q}}\Big]S
\end{equation}
for some $\alpha' \in \mathbb{R}$.

Let
\begin{align}
T_0 &:=  I(U;Y_2|S,U_1,X_2)\\
T_1 &:= T_0+I(X_1;Y_3|U,U_1,U_2)\\
T_2 &:= T_1+I(U_2;Y_3|U,U_1)-I(U_2;S|U_1)\\
T_3 &:= I(X_1,U_1,U_2;Y_3)-I(U_2;S|U_1).
\end{align}
Also, define the following function and substitutions which we will use throughout the proof.
\begin{align}
P'_2 &:= {\theta}P_2(1-\rho'^2_{2s})\\
Q' &:=(\sqrt{Q}+\rho'_{2s}\sqrt{{\theta}P_2})^2\\
\Phi(\alpha',\theta,\rho'_{2s}) &:=\frac{P'_2Q'(1-\alpha')^2}{P'_2+\alpha'^2Q'},
\end{align}
and let $\tilde{Y}_3 :=\tilde{X}_2+S+Z_3$.

i) The computation of the quantities $T_0$ and $T_3$ can be done along the lines of those for the corresponding quantities in the proof of Theorem~\ref{TheoremAchievabeRateNonCausalCaseGaussianChannelFullDuplexRegime}. We obtain 
\begin{align}
\label{FirstPartOf__FirstTerm__AchievableRate__PartialDF}
T_0 &= \frac{1}{2}\log\Big(1+\frac{\bar{\gamma}P_1(1-\rho'^2_{12})}{N_2+{\gamma}P_1}\Big)
%T_3 &=\frac{1}{2}\log\Big(1+\frac{P_1+\bar{\theta}P_2+2\rho'_{12}\sqrt{\bar{\theta}\bar{\gamma}P_1P_2}}{{\theta}P_2+Q+N_3+2\rho'_{2s}\sqrt{{\theta}P_2Q}}\Big) + \frac{1}{2}\log\Big(\frac{P'_2(P'_2+Q'+N_3)}{P'_2Q'(1-\alpha')^2+N_3(P'_2+\alpha'^2Q')}\Big). 
\end{align}
and $T_3$ as given by \eqref{AchievabeRate__PartialDF__GaussianCase__FullDuplexRegime_T3}.
 
ii)  Now, we compute $T_1$.  
\begin{align}
I(X_1;Y_3|U,U_1,U_2) &= I(\tilde{X}_1;\tilde{X}_1+\tilde{Y}_3|U,U_1,U_2)\nonumber\\
&\stackrel{(a)}{=} I(\tilde{X}_1;\tilde{X}_1+\tilde{Y}_3|U_2)\nonumber\\
%&=h(\tilde{X}_1+\tilde{Y}_3|U_2)-h(\tilde{Y}_3|U_2,\tilde{X}_1)\nonumber\\
&\stackrel{(b)}{=} h(\tilde{X}_1+\tilde{Y}_3|U_2)-h(\tilde{Y}_3|U_2)\nonumber\\
%&= \frac{1}{2}\log\Big(\mathbb{E}[(\tilde{X}_1+\tilde{Y}_3)^2]-\mathbb{E}[(\tilde{X}_1+\tilde{Y}_3)\mathbb{E}[\tilde{X}_1+\tilde{Y}_3|U_2]]\Big)\nonumber\\
%&  -\frac{1}{2}\log\Big(\mathbb{E}[\tilde{Y}^2_3]-\mathbb{E}[\tilde{Y}_3\mathbb{E}[\tilde{Y}_3|U_2]]\Big)\nonumber\\
&\stackrel{(c)}{=}\frac{1}{2}\log\Big(1+\frac{\mathbb{E}[\tilde{X}^2_1]}{\mathbb{E}[\tilde{Y}^2_3]-\mathbb{E}[\tilde{Y}_3\mathbb{E}[\tilde{Y}_3|U_2]]}\Big)\nonumber\\
&\stackrel{(d)}{=}\frac{1}{2}\log\Big(1+\frac{{\gamma}P_1}{N_3+\Phi(\alpha',\theta,\rho'_{2s})}\Big)
\label{SecondPartOf__FirstTerm__AchievableRate__PartialDF}
\end{align}
where $(a)$ holds since $U$ and $U_1$ are independent of $\tilde{X}_1$, $\tilde{Y}_3$ and $U_2$; $(b)$ holds since $\tilde{X}_1$ is independent of $\tilde{Y}_3$ and $U_2$; $(c)$ holds since $\tilde{X}_1$, $U_2$ and $\tilde{Y}_3$ are jointly Gaussian, and $(d)$ follows by straightforward algebra using the fact that $\mathbb{E}[\tilde{Y}_3|U_2] = {\beta}U_2$, with
\begin{equation}
\beta = \frac{P'_2+\alpha'Q'}{P'_2+\alpha'^2Q'}.
\end{equation}

 \noindent Then, by adding  \eqref{FirstPartOf__FirstTerm__AchievableRate__PartialDF} and \eqref{SecondPartOf__FirstTerm__AchievableRate__PartialDF} we get $T_1$ as given by \eqref{AchievabeRate__PartialDF__GaussianCase__FullDuplexRegime_T1}.

iii) Finally, we compute $T_2$. It can be shown easily that
\begin{align}
I(U_2;S|U_1) &= \frac{1}{2}\log(\frac{P'_2+\alpha'^2Q'}{P'_2}).
\label{SecondPartOf__EquivalentForm__SecondTerm__AchievableRate__PartialDF}
\end{align}
Also, we have
\begin{align}
I(U_2;Y_3|U,U_1) &= I(U_2;\tilde{X}_1+\tilde{Y}_3|U,U_1) \nonumber\\
&\stackrel{(e)}{=} I(U_2;\tilde{X}_1+\tilde{Y}_3) \nonumber\\
&=h(\tilde{X}_1+\tilde{Y}_3)-h(\tilde{X}_1+\tilde{Y}_3|U_2)\nonumber\\
%&=\frac{1}{2}\log\Big(\mathbb{E}[(\tilde{X}_1+\tilde{Y}_3)^2]\Big)-\frac{1}{2}\log\Big(\mathbb{E}[(\tilde{X}_1+\tilde{Y}_3)^2]-\mathbb{E}[\tilde{Y}_3\mathbb{E}[\tilde{Y}_3|U_2]]\Big)\nonumber\\
&\stackrel{(f)}{=}\frac{1}{2}\log\Big(\frac{\mathbb{E}[\tilde{X}^2_1]+\mathbb{E}[\tilde{Y}^2_3]}{\mathbb{E}[\tilde{X}^2_1]+\mathbb{E}[\tilde{Y}^2_3]-\mathbb{E}[\tilde{Y}_3\mathbb{E}[\tilde{Y}_3|U_2]]}\Big)\nonumber\\
&\stackrel{(g)}{=}\frac{1}{2}\log\Big(\frac{P'_2+Q'+{\gamma}P_1+N_3}{N_3+{\gamma}P_1+\Phi(\alpha',\theta,\rho'_{2s})}\Big).
\label{FirstPartOf__EquivalentForm__SecondTerm__AchievableRate__PartialDF}
\end{align}
where $(e)$ holds since $U$ and $U_1$ are independent of $U_2$, $\tilde{X}_1$ and $\tilde{Y}_3$; $(f)$ holds since $\tilde{X}_1$, $U_2$ and $\tilde{Y}_3$ are jointly Gaussian, and $\tilde{X}_1$ is independent of $U_2$ and $\tilde{Y}_3$; and $(g)$ follows through straightforward algebra similar to in \eqref{SecondPartOf__FirstTerm__AchievableRate__PartialDF}.

\noindent Adding $T_1$ (given by \eqref{AchievabeRate__PartialDF__GaussianCase__FullDuplexRegime_T1}) and \eqref{FirstPartOf__EquivalentForm__SecondTerm__AchievableRate__PartialDF} and subtracting \eqref{SecondPartOf__EquivalentForm__SecondTerm__AchievableRate__PartialDF}, we get $T_2$ as given by \eqref{AchievabeRate__PartialDF__GaussianCase__FullDuplexRegime_T2}.

%\begin{align}
%T_2 &= \frac{1}{2}\log\Big(1+\frac{\bar{\gamma}P_1(1-\rho'^2_{12})}{N_2+{\gamma}P_1}\Big)+\frac{1}{2}\log\Big(\frac{P'_2(P'_2+Q'+{\gamma}P_1+N_3)}{P'_2Q'(1-\alpha')^2+N_3(P'_2+\alpha'^2Q')}\Big).
%\label{ResultComputation__SecondTerm__AchievableRate__PartialDF}
%\end{align}

%-----------------------------------------------------
\renewcommand{\theequation}{E-\arabic{equation}}
\subsection{Proof of Theorem \ref{TheoremOuterBoundNonCausalCaseGaussianChannelFullDuplexRegime}}\label{appendixTheorem4}
In this section we use the upper bound for the DM case in Theorem \ref{TheoremOuterBoundNonCausalCaseDiscreteMemorylessChannel} to compute the upper bound on the capacity of the state-dependent full-duplex Gaussian RC with informed relay. 

Fix a joint distribution of $X_1,X_2,S,Y_2,Y_3$ of the form \eqref{MeasureForOuterBoundNonCausalCaseDiscreteMemorylessChannel} satisfying 
\begin{align}
&\mathbb{E}[X^2_1]=\tilde{P}_1 \leq P_1, \quad \mathbb{E}[X^2_2]=\tilde{P}_2 \leq P_2,\nonumber\\
&\mathbb{E}[X_1X_2]=\sigma_{12},\quad \mathbb{E}[X_2S]=\sigma_{2s},\quad \mathbb{E}[X_1S]=0.
\label{FixedSecondMomentsOuterBoundFullDuplexGaussianCase}
\end{align}
We shall also use the correlation coefficients $\rho_{12}$ and $\rho_{2s}$ defined as
\begin{equation}
\rho_{12}=\frac{\sigma_{12}}{\sqrt{\tilde{P}_1\tilde{P}_2}},\quad \rho_{2s}=\frac{\sigma_{2s}}{\sqrt{\tilde{P}_2Q}}.
\end{equation}

We first compute the first term in the minimization on the RHS of \eqref{OuterBoundNonCausalCaseDiscreteMemorylessChannel}. Let $\dv Y=(X_1+Z_2,X_1+Z_3)^T$. We have
{\allowdisplaybreaks
\begin{align}
I(X_1;Y_2,Y_3|S,X_2) &= h(Y_2,Y_3|S,X_2)-h(Y_2,Y_3|S,X_1,X_2)\nonumber\\
&= h(X_1+Z_2,X_1+Z_3|S,X_2)-h(Z_2,Z_3)\nonumber\\
&\stackrel{(a)}{\leq} \frac{1}{2}\log\Bigg|\mathbb{E}\Big[\Big(\dv Y-\mathbb{E}[\dv Y|S,X_2]\Big)\Big(\dv Y-\mathbb{E}[\dv Y|S,X_2]\Big)^T\Big]\Bigg|\nonumber\\
&\hspace{6cm}-\frac{1}{2}\log(N_2N_3)\nonumber\\
&= \frac{1}{2}\log\frac{\Big|\mathbb{E}[\dv Y\dv Y^T]-\mathbb{E}[\mathbb{E}[\dv Y|S,X_2]\mathbb{E}[\dv Y|S,X_2]^T]\Big|}{N_2N_3}\nonumber\\
& \stackrel{(b)}{=}\frac{1}{2}\log\Big(1+\tilde{P}_1(1-\frac{\rho^2_{12}}{1-\rho^2_{2s}})(\frac{1}{N_2}+\frac{1}{N_3})\Big),
\label{FirstTermUpperBoundGaussianFullDuplex}
\end{align}}
where, $|\cdot|$ denotes the determinant operator,  \\ 
$(a)$ follows from the fact that the conditional differential entropy $h(X_1+Z_2,X_1+Z_3|S,X_2)$ is maximized if  $(S,X_1,X_2,Z_2,Z_3)$ are jointly Gaussian, and \\
$(b)$ follows from the fact the vector $(S,X_1,X_2,Z_2,Z_3)$ is a jointly Gaussian vector and  the MMSE estimator of $\dv Y$ given $(S,X_2)$ is 
\begin{equation}
\mathbb{E}[\dv Y|S,X_2]=(-\frac{\sigma_{12}\sigma_{2s}}{\tilde{P}_2Q-\sigma^2_{2s}}S+\frac{\sigma_{12}Q}{\tilde{P}_2Q-\sigma^2_{2s}}X_2){\cdot}(1,1)^T.
\end{equation}

\iffalse
Furthermore, if the channel is physically degraded, i.e., the distribution on  $(X_1,X_2,S,Y_2,Y_3)$ can be written as in \eqref{MeasureForOuterBoundDegradedChannelNonCausalCaseDiscreteMemorylessChannel}, it can be easily observed that the bound \eqref{FirstTermUpperBoundGaussianFullDuplex} can be tightened as 
\begin{align}
I(X_1;Y_2,Y_3|S,X_2) &=I(X_1;Y_2|S,X_2)\nonumber\\
&= h(X_1+Z_2|S,X_2)-h(Z_2)\nonumber\\
&\leq \frac{1}{2}\log\mathbb{E}\left[\left(X_1+Z_2-\mathbb{E}[X_1+Z_2|S,X_2]\right)^2\right]-\frac{1}{2}\log(N_2)\nonumber\\
&=\frac{1}{2}\log\left(1+\frac{\tilde{P}_1(1-\rho^2_{12}-\rho^2_{2s})}{N_2(1-\rho^2_{2s})}\right).
\label{FirstTermUpperBoundPhysicallyDegradedGaussianFullDuplex}
\end{align}
\fi

We now compute the term $[I(X_1,X_2;Y_3|S)-I(X_1;S|Y_3)]$. We have
\begin{align}
I(X_1,X_2;Y_3|S)-I(X_1;S|Y_3) &= h(Y_3|S)-h(Y_3|X_1,X_2,S)-h(S|Y_3)+h(S|X_1,Y_3)\nonumber\\
&= h(Y_3)-h(S)+h(S|X_1,Y_3)-h(Z_3).
\label{SecondTermUpperBoundFullDuplex}
\end{align}
For fixed second moments \eqref{FixedSecondMomentsOuterBoundFullDuplexGaussianCase}, we have
\begin{align}
h(Y_3) & \leq \frac{1}{2}\log(2\pi{e})(\tilde{P}_1+\tilde{P}_2+2\sigma_{12}+2\sigma_{2s}+Q+N_3),
\label{EntropyOfReceivedForSecondTermUpperBoundFullDuplex}
\end{align}
where equality is attained if $Y_3$ is Gaussian. Similarly, the term $h(S|X_1,Y_3)$ is maximized if $(S,X_1,Y_3)$ are jointly Gaussian.  Let $\hat{S}(X_1,Y_3)=\mathbb{E}[S|X_1,Y_3]$ be the MMSE estimator of $S$ given $(X_1,Y_3)$, i.e., 
\begin{align}
\hat{S}(X_1,Y_3) &= \mathbb{E}[S|X_1,X_2+S+Z_3]\nonumber\\
&= \gamma_1X_1+\gamma_2(X_2+S+Z_3)
\end{align}
with
\begin{align}
\gamma_1 &= -\frac{\sigma_{12}(Q+\sigma_{2s})}{\tilde{P}_1(\tilde{P}_2+2\sigma_{2s}+Q+N_3)-\sigma^2_{12}}\nonumber\\
\gamma_2 &= \frac{\tilde{P}_1(Q+\sigma_{2s})}{\tilde{P}_1(\tilde{P}_2+2\sigma_{2s}+Q+N_3)-\sigma^2_{12}}.
\end{align}
Then we have
\begin{align}
 h(S|X_1,Y_3) & = h(S-\hat{S}(X_1,Y_3)|X_1,Y_3)\nonumber\\
             &\leq h(S-\gamma_1X_1-\gamma_2(X_2+S+Z_3))\nonumber\\
             &= \frac{1}{2}\log(2\pi{e})\mathbb{E}\Big[\Big(S-\gamma_1X_1-\gamma_2(X_2+S+Z_3)\Big)^2\Big]\nonumber\\
             &= \frac{1}{2}\log\Big((2\pi{e})\frac{Q\tilde{P}_1\tilde{P}_2+\tilde{P}_1N_3Q-\sigma^2_{2s}\tilde{P}_1-\sigma^2_{12}Q}{\tilde{P}_1(\tilde{P}_2+2\sigma_{2s}+Q+N_3)-\sigma^2_{12}}\Big),
\label{ConditionalEntropyForSecondTermUpperBoundFullDuplex}
\end{align}
where the inequality is attained with equality if $S,X_1,X_2,Y_3$ are jointly Gaussian. From \eqref{SecondTermUpperBoundFullDuplex}, \eqref{EntropyOfReceivedForSecondTermUpperBoundFullDuplex} and \eqref{ConditionalEntropyForSecondTermUpperBoundFullDuplex}, we obtain
\begin{align}
 I(X_1,X_2;Y_3|S)-I(X_1;S|Y_3) &= \frac{1}{2}\log\Bigg(\frac{(\tilde{P}_1+\tilde{P}_2+2\sigma_{12}+2\sigma_{2s}+Q+N_3)}{(\tilde{P}_1\tilde{P}_2+2\tilde{P}_1\sigma_{2s}+\tilde{P}_1Q+\tilde{P}_1N_3-\sigma^2_{12})}\nonumber\\
&\hspace{2.5cm}{\cdot}\frac{(Q\tilde{P}_1\tilde{P}_2+\tilde{P}_1N_3Q-\sigma^2_{2s}\tilde{P}_1-\sigma^2_{12}Q)}{QN_3}\Bigg)\nonumber\\
&=\frac{1}{2}\log\Big(1+\frac{(\sqrt{\tilde{P}_1}+\rho_{12}\sqrt{\tilde{P}_2})^2}{\tilde{P}_2(1-\rho^2_{12}-\rho^2_{2s})+(\sqrt{Q}+\rho_{2s}\sqrt{\tilde{P}_2})^2+N_3}\Big)\nonumber\\
&\hspace{2.5cm}+\frac{1}{2}\log\Big(1+\frac{\tilde{P}_2(1-\rho^2_{12}-\rho^2_{2s})}{N_3}\Big).
\label{SecondTermUpperBoundGaussianFullDuplex}
\end{align}

For convenience, let us define the function $\Theta_1(\tilde{P}_1,\rho_{12},\rho_{2s})$ as the RHS of \eqref{FirstTermUpperBoundGaussianFullDuplex} and the function $\Theta_2(\tilde{P}_1,\tilde{P}_2,\rho_{12},\rho_{2s})$ as the RHS of \eqref{SecondTermUpperBoundGaussianFullDuplex}. From the above analysis, the capacity of the channel is upper-bounded as
\begin{align}
C \leq \max\:\min \{\Theta_1(\tilde{P}_1,\rho_{12},\rho_{2s}),\Theta_2(\tilde{P}_1,\tilde{P}_2,\rho_{12},\rho_{2s})\}
\label{IntermediaryStepOuterBoundGaussianFullDuplex}
\end{align}
where the maximization is over all covariance matrices $\Lambda_{X_1,X_2,S,Z_2,Z_3}$ of $(X_1,X_2,S,Z_2,Z_3)$,
\begin{align}
&\Lambda_{X_1,X_2,S,Z_2,Z_3}=\nonumber\\
&{\left(
\begin{array}{ccccc}
\tilde{P}_1 & \rho_{12}\sqrt{\tilde{P}_1\tilde{P}_2} & 0 & 0 & 0\\
\rho_{12}\sqrt{\tilde{P}_1\tilde{P}_2} & \tilde{P}_2 & \rho_{2s}\sqrt{\tilde{P}_2Q} & 0 & 0\\
0 & \rho_{2s}\sqrt{\tilde{P}_2Q} & Q & 0 & 0\\
0 & 0 & 0 & N_2 & 0\\
0 & 0 & 0 & 0 & N_3
\end{array}
\right),}
\label{AllowableCovarianceMatrixOuterBoundFullDuplex}
\end{align}
that satisfy
\begin{equation}
\tilde{P}_1 \leq P_1, \quad \tilde{P}_2 \leq P_2
\end{equation}
and have non-negative discriminant, 
\begin{equation}
Q\tilde{P}_1\tilde{P}_2N_2N_3(1-\rho^2_{12}-\rho^2_{2s})\geq 0,
\end{equation}
i.e., for $Q > 0$,
\begin{equation}
\rho^2_{12}+\rho^2_{2s} \leq 1.
\label{ConditionOnRhosOuterBoundGaussianFullDuplex}
\end{equation}
Furthermore, investigating $\Theta_1(\tilde{P}_1,\rho_{12},\rho_{2s})$ and $\Theta_2(\tilde{P}_1,\tilde{P}_2,\rho_{12},\rho_{2s})$, it can be seen that it suffices to consider $\rho_{12} \in [0,1]$ and $\rho_{2s} \in [-1,0]$ for the maximization in \eqref{IntermediaryStepOuterBoundGaussianFullDuplex}.

To complete the proof, we should show that $\Theta_1(\tilde{P}_1,\rho_{12},\rho_{2s})$ and $\Theta_2(\tilde{P}_1,\tilde{P}_2,\rho_{12},\rho_{2s})$ are maximized at $\tilde{P}_1=P_1$ and $\tilde{P}_2=P_2.$ 
It is easy to show that $\Theta_1(\tilde{P}_1,\rho_{12},\rho_{2s})$ and $\Theta_2(\tilde{P}_1,\tilde{P}_2,\rho_{12},\rho_{2s})$ increase monotonically with $\tilde{P}_1$ for fixed $\rho_{12}$, $\rho_{2s}$, $\tilde{P}_2$. Then we can replace $\tilde{P}_1$ with $P_1$ in both $\Theta_1(\tilde{P}_1,\rho_{12},\rho_{2s})$ and $\Theta_2(\tilde{P}_1,\tilde{P}_2,\rho_{12},\rho_{2s})$. To show that $\tilde{P}_2$ can be replaced by $P_2$, we use the following intuitive argument. Since the term $\Theta_1(P_1,\rho_{12},\rho_{2s})$ does not depend on $\tilde{P}_2$ for given $\rho_{12}$ and $\rho_{2s}$,  it remains to show that  $\tilde{P}_2$ can be replaced with $P_2$ in only the term $\Theta_2(\tilde{P}_1,\tilde{P}_2,\rho_{12},\rho_{2s})$. The term $\Theta_2(\tilde{P}_1,\tilde{P}_2,\rho_{12},\rho_{2s})$ is the sum rate of a two-user MAC with asymmetric CSI in which the informed encoder knows the message of the uninformed encoder \cite[Theorem 6]{SBSV07a}. Then, considering this MAC, it can be argued \cite{SBSV07a} that for the sum-rate to be maximized the informed encoder should use the entire power available, i.e., $P_2$. This concludes the proof of Theorem \ref{TheoremOuterBoundNonCausalCaseGaussianChannelFullDuplexRegime}.

%{\color{red} May be there is a problem with the last argument--- regarding the achievability of the mutual info. difference in the upper bound when viewed as a sum rate for the MAC !}

\renewcommand{\theequation}{F-\arabic{equation}}
\subsection{Proof of Observation \ref{CapacityForSpecialCases}}\label{appendixProposition1}

We first prove the first statement in Observation~\ref{CapacityForSpecialCases}. Let us denote $N_2^{\star}$ as the RHS of \eqref{SnrRangeForChannelCapacityLowSnrAtRelay}. We have
\begin{align}
 R_{\text{G}}^{\text{lo}} & \stackrel{(a)}{\geq} \min \Big\{\frac{1}{2}\log(1+\frac{P_1}{N_2}),\nonumber\\
&\qquad \max_{-1 \leq \rho'_{2s} \leq 0}\;\frac{1}{2}\log\Big(1+\frac{P_1}{P_2+Q+N_3+2\rho'_{2s}\sqrt{P_2Q}}\Big)+\frac{1}{2}\log(1+\frac{P_2(1-\rho'^2_{2s})}{N_3})\Big\}\nonumber\\
&\stackrel{(b)}{=} \frac{1}{2}\log(1+\frac{P_1}{N_2}) \nonumber \\
&:=  R_{\text{DG}},
\label{FirstStepProofCapacityVeryStrongSideInformationDegradedGaussianChannel}
\end{align}
where $(a)$ follows by putting $\rho'_{12}=0$ and $\theta=1$ in \eqref{AchievabeRateNonCausalCaseGaussianChannelFullDuplexRegime}, and $(b)$ follows if $N_2 \geq N_2^{\star}$.

\noindent Then, it is easy to observe that
\begin{equation}
 R_{\text{DG}}^{\text{up}} \leq R_{\text{DG}}.
 \label{FirstStepProofCapacityVeryStrongSideInformationDegradedGaussianChannel_temp1}
 \end{equation}
From \eqref{FirstStepProofCapacityVeryStrongSideInformationDegradedGaussianChannel} and
\eqref{FirstStepProofCapacityVeryStrongSideInformationDegradedGaussianChannel_temp1}, we get that
\begin{equation}
R_{\text{DG}} \leq R_{\text{G}}^{\text{lo}} \leq C_{\text{DG}} \leq R_{\text{DG}}^{\text{up}} \leq R_{\text{DG}}.
\end{equation}
Then we can conclude that the lower bound and upper bound meet if $N_2 \geq N_2^{\star}$.

Let us now prove the second statement in Observation~\ref{CapacityForSpecialCases}. 
If the pair $(\rho_{12},\rho_{2s})$ that maximizes the upper bound in Corollary \ref{CorollaryUpperBoundDegradedGaussianChannelFullDuplexRegime} satisfies the condition in \eqref{MaximizationRangeOuterBoundNonCausalCaseGaussianChannelFullDuplexRegime} with equality, i.e., $\rho^2_{12}+\rho^2_{2s}=1$, then we choose $\varrho_{2s}=\rho_{2s}$, $\varrho_{12}=\rho_{12}$, and $\theta=\varrho^2_{2s}$ ( i.e., $\bar{\theta}=\varrho^2_{12}$) in the lower bound \eqref{EquivalentFormForAchievabeRateNonCausalCaseGaussianChannelFullDuplexRegime} to achieve the upper bound, and thus obtain channel capacity in this case. 

\renewcommand{\theequation}{G-\arabic{equation}}
\subsection{Proofs for Time Division Relaying}\label{appendixTimeDivisionRelaying}
\subsubsection{Proof of Proposition \ref{PropositionOuterBoundNonCausalCaseGaussianChannelTimeDivisionRelaying}}
Let $(X_{1,1}^{(1)},X_{1,2}^{(1)},\ldots,X_{1,\lfloor \nu n \rfloor}^{(1)})$ and $(X_{1,\lfloor \nu n \rfloor+1}^{(2)},X_{1,\lfloor \nu n \rfloor+2}^{(2)},\ldots,X_{1,n}^{(2)})$ be the transmitted sequences from the source during the relay-receive period and the relay-transmit period, respectively. The relay receives $Y_{2,1},Y_{2,2},\ldots,Y_{2,{\lfloor}{\nu}n{\rfloor}}$ during the relay-receive period and transmits a sequence $X_{2,{\lfloor}{\nu}n{\rfloor}+1},X_{2,{\lfloor}{\nu}n{\rfloor}+2},\cdots,X_{2,n}$ during the relay-transmit period. From Fano's inequality~\eqref{FanosInequality} and Lemma~\ref{LemmaProofTheoremOuterBound}, we have the following 
\begin{align}
nR  \leq \:\min\: \bigg\{& \sum_{i=1}^{n}I(X_{1,i};Y_{2,i},Y_{3,i}|S_i,X_{2,i}),\nonumber\\
& \sum_{i=1}^{n} I(X_{1,i},X_{2,i};Y_{3,i}|S_i)-I(X_{1,i};S_i|Y_{3,i})\bigg\} +n\delta_n.
\end{align}
We now specialize this bound to the TD mode for which we have $X_{2,i}=0$ for $i \leq {\lfloor}{\nu}n{\rfloor}$ (as the relay does not transmit during the relay-receive period) and $Y_{2,i}=0$ for $i \geq {\lfloor}{\nu}n{\rfloor}+1$ (as the relay does not receive during the relay-transmit period). This gives
\begin{align}
 nR  \leq \:\min\: \bigg\{& \sum_{i=1}^{{\lfloor}{\nu}n{\rfloor}}I(X_{1,i}^{(1)};Y_{2,i},Y_{3,i}^{(1)}|S_i^{(1)},X_{2,i}=0)+\sum_{i={\lfloor}{\nu}n{\rfloor}+1}^{n}I(X_{1,i}^{(2)};Y_{3,i}^{(2)}|S_i^{(2)},X_{2,i}),\nonumber\\ 
&\sum_{i=1}^{{\lfloor}{\nu}n{\rfloor}}I(X_{1,i}^{(1)};Y_{3,i}^{(1)}|S_i^{(1)},X_{2,i}=0)-I(X_{1,i}^{(1)};S_i^{(1)}|Y_{3,i}^{(1)})\nonumber\\
&+\sum_{i={\lfloor}{\nu}n{\rfloor}+1}^{n} I(X_{1,i}^{(2)},X_{2,i};Y_{3,i}^{(2)}|S_i^{(2)})-I(X_{1,i}^{(2)};S_i^{(2)}|Y_{3,i}^{(2)})\bigg\} +n\delta_n.
\end{align}
By letting $n \rightarrow \infty$ and using standard arguments \cite{CT91}, we get the single letter upper bound on capacity
\begin{align}
C  \leq \max\:\min\:\bigg\{& {\nu}I(X_1^{(1)};Y_2,Y_3^{(1)}|S^{(1)},X_2=0)+\bar{\nu}I(X_1^{(2)};Y_3^{(2)}|S^{(2)},X_2),\nonumber\\
&{\nu}I(X_1^{(1)};Y_3^{(1)}|S^{(1)},X_2=0)-{\nu}I(X_1^{(1)};S^{(1)}|Y_3^{(1)})\nonumber\\
&+\bar{\nu}I(X_1^{(2)},X_2;Y_3^{(2)}|S^{(2)})-\bar{\nu}I(X_1^{(2)};S^{(2)}|Y_3^{(2)})\bigg\},
\label{SingleLetterOuterBoundGaussianChannelTimeDivisionRelaying}
\end{align}
where the maximization is over all joint distributions of the form
\begin{align}
&Q_{S^{(1)}}P_{X_1^{(1)}}W_{Y_2,Y_3^{(1)}|X_1^{(1)},S^{(1)}}Q_{S^{(2)}}P_{X_1^{(2)}}P_{X_2|X_1^{(2)},S^{(2)}}W_{Y_3^{(2)}|X_1^{(2)},X_2,S^{(2)}}.
\end{align}
The bound in \eqref{SingleLetterOuterBoundGaussianChannelTimeDivisionRelaying} is the counterpart, to the TD mode, of the upper bound \eqref{OuterBoundNonCausalCaseDiscreteMemorylessChannel} for the full-duplex case. By  closely following the arguments and the algebra used in the proof of Theorem \ref{TheoremOuterBoundNonCausalCaseGaussianChannelFullDuplexRegime}, it can be shown that this bound is maximized by choosing $S^{(1)},S^{(2)},X_1^{(1)},X_1^{(2)},X_2,Y_2,Y_3^{(1)},Y_3^{(2)}$ that are jointly Gaussian, with $X_1^{(1)}$ with power $P_1^{(1)}$ is independent of $S^{(1)}$, and $X_1^{(2)}$ and $X_2$ with power $P_1^{(2)}$ and $P_2$, respectively, are such that
\begin{equation*}
\mathbb{E}[X_1^{(2)}X_2]=\rho_{12}\sqrt{P_1^{(2)}P_2},\quad \mathbb{E}[X_1^{(2)}S^{(2)}]=0,\quad \mathbb{E}[X_2S^{(2)}]=\rho_{2s}\sqrt{P_2Q^{(2)}}.
\end{equation*} 
Using this distribution, the evaluation of the RHS of \eqref{SingleLetterOuterBoundGaussianChannelTimeDivisionRelaying} gives the RHS of \eqref{OuterBoundNonCausalCaseGaussianChannelTimeDivisionRelaying}. 

\subsubsection{Proof of Proposition \ref{PropositionAchievabeRateNonCausalCaseGaussianChannelTimeDivisionRelaying}}
The proof follows by combining the technique of rate-splitting \cite{H-MZ05} and the Generalized DPC described in Section \ref{secIV_subsecA} for the full-duplex mode. Rate splitting has the message $W$ to be transmitted from the source node split into two independent parts: $W_d$ transmitted directly to the destination at rate $R_d$, and $W_r$ transmitted through the relay at rate $R_r$, with a total rate $R=R_r+R_d$. 

The encoding and transmission scheme is as follows. During the relay-receive period, the source sends a Gaussian signal $X_{1,i}^{(1)}$ which carries $W_r$ only and is independently drawn with a random variable $X_1^{(1)}\sim \calN(0,P_1^{(1)})$ which is independent of the channel state $S^{(1)}$. During the relay-transmit period, the source transmits a Gaussian signal $X_{1,i}^{(2)}$ which carries both $W_r$ and $W_d$ and is independently drawn with $X_1^{(2)}\sim \calN(0,P_1^{(2)})$. During the relay-transmit period, the relay sends a Gaussian signal $X_{2,i}$ which carries $W_r$ only and is given by
\begin{align}
& X_{2,i}=U_{1,i}+\tilde{X}_{2,i},
\end{align} 
where $U_{1,i}$ is drawn with $ U_1 \sim \calN(0,\bar{\theta}P_2)$ and $\tilde{X}_{2,i}$ is obtained via a GDPC considering $S^{(2)}$ as non-causal channel state information during this period. 

\noindent The random variables $U_1$ and $X_1^{(2)}$ are jointly Gaussian with $\mathbb{E}[X_1^{(2)}X_2]=\mathbb{E}[X_1^{(2)}U_1]=\rho'_{12}\sqrt{\bar{\theta}P_1^{(2)}P_2}$, and are both independent of the state $S^{(2)}$. For the GDPC, we use the following auxiliary random variable to generate the auxiliary codewords $U_{2,i}$,
\begin{align}
U_2=\tilde{X}_2+ \Big[\alpha'(1+\rho'_{2s}\sqrt{\frac{{\theta}P_2}{Q^{(2)}}})-\rho'_{2s}\sqrt{\frac{{\theta}P_2}{Q^{(2)}}}\Big]S^{(2)},
\label{GDPC__TimeDivisionRelaying}
\end{align}
where $\tilde{X}_2 \sim \calN(0, \theta P_2)$ is jointly Gaussian with $S^{(2)}$, with $\mathbb{E}[X_2S^{(2)}]=\mathbb{E}[\tilde{X}_2S_2^{(2)}]=\rho'_{2s}\sqrt{{\theta}P_2Q^{(2)}}$; and ${\alpha}'$ is a scale parameter. Thus, using the GDPC given by \eqref{GDPC__TimeDivisionRelaying}, $\tilde{X}_{2,i}$ is generated as
\begin{align}
\tilde{X}_{2,i}= U_{2,i}-\Big[\alpha'(1+\rho'_{2s}\sqrt{\frac{{\theta}P_2}{Q^{(2)}}})-\rho'_{2s}\sqrt{\frac{{\theta}P_2}{Q^{(2)}}}\Big]S^{(2)}_i
\end{align}
where $U_{2,i}$ is independently drawn with $U_{2}$. 

\noindent Furthermore, we let $X_{1,i}^{(2)}=\rho'_{12}\sqrt{P_1^{(2)}/\bar{\theta}P_2}U_{1,i}+\tilde{X}_{1,i}^{(2)}$, where $\tilde{X}_{1,i}^{(2)}$ is independently drawn with $\tilde{X}_{1}^{(2)} \sim \calN(0,(1-\rho'^2_{12})P_1^{(2)})$, is independent of $U_1, X_2, S^{(2)}$, and carries $W_d$ only.

For the decoding procedures at the source and the relay, we give simple arguments based on intuition (the rigorous decoding uses joint typicality arguments). Also, since all the random variables are i.i.d., we sometimes omit the time index. The relay subtracts out $S^{(1)}$ from the received $Y_2$ and then decodes $W_r$.  Message $W_r$ can be decoded correctly at the relay as long as
\begin{equation}
R_r < \frac{\nu}{2}\log\Big(1+\frac{P_1^{(1)}}{N_2}\Big).
\label{Condition1OnRrTimeDivisionRelaying}
\end{equation}

% END OF SEQUENTIAL DECODING AT DESTINATION.

%----------------------- JOINT DECODING OF (Wr,Wd) AT DESTINATION -----------------%
\noindent The destination jointly decodes $W_r$ and $W_d$ from $(Y_3^{(1)},Y_3^{(2)})$. One can show that this can be done reliable as long as
\begin{align}
\label{Condition1OnRdTimeDivisionRelaying}
R_d &<  \bar{\nu}I(X_1^{(2)};Y_3^{(2)}|U_1,U_2)\\
\label{Condition2OnRdTimeDivisionRelaying}
R_d &< \bar{\nu}[I(X_1^{(2)},U_2;Y_3^{(2)}|U_1)-I(U_2;S^{(2)}|U_1)]\\
R_r+R_d &< {\nu}I(X_1^{(1)};Y_3^{(1)})+\bar{\nu}[I(X_1^{(2)},U_1,U_2;Y_3^{(2)})-I(U_2;S^{(2)}|U_1)].
\label{ConditionOnRrRdTimeDivisionRelaying}
\end{align}

\noindent Adding \eqref{Condition1OnRrTimeDivisionRelaying} and \eqref{Condition1OnRdTimeDivisionRelaying} at one hand, and \eqref{Condition1OnRrTimeDivisionRelaying} and \eqref{Condition2OnRdTimeDivisionRelaying} at the other hand, and using \eqref{ConditionOnRrRdTimeDivisionRelaying}, we obtain 
\begin{align}
\label{Condition1OnRrRdTimeDivisionRelaying}
R &< \frac{\nu}{2}\log(1+\frac{P_1^{(1)}}{N_2}) +\bar{\nu}I(X_1^{(2)};Y_3^{(2)}|U_1,U_2)\\
\label{Condition3OnRrRdTimeDivisionRelaying}
R &< \frac{\nu}{2}\log(1+\frac{P_1^{(1)}}{N_2}) +\bar{\nu}[I(X_1^{(2)},U_2;Y_3^{(2)}|U_1)-I(U_2;S^{(2)}|U_1)]\\
R &< {\nu}I(X_1^{(1)};Y_3^{(1)})+\bar{\nu}[I(X_1^{(2)},U_1,U_2;Y_3^{(2)})-I(U_2;S^{(2)}|U_1)].
\label{Condition2OnRrRdTimeDivisionRelaying}
\end{align}

\noindent The computation of the mutual information terms in \eqref{Condition1OnRrRdTimeDivisionRelaying}, \eqref{Condition3OnRrRdTimeDivisionRelaying} and \eqref{Condition2OnRrRdTimeDivisionRelaying} involves straightforward algebra which is very similar to that in the proofs of Theorem~\ref{TheoremAchievabeRateNonCausalCaseGaussianChannelFullDuplexRegime} in Appendix~\ref{appendixTheorem3} and of Corollary~\ref{CorollaryPartialDecodeAndForwardNonCausalCaseDiscreteMemorylessChannel} in Appendix ~\ref{appendixCorollary3}; and, so, we omit the details for brevity. More specifically, define
$$P'_2:={\theta}P_2(1-\rho'^2_{2s}), \qquad Q'^{(2)}:=\big(\sqrt{Q^{(2)}}+\rho'_{2s}\sqrt{{\theta}P_2}\big)^2.$$
Also, recall $\Phi(\alpha',\theta,\rho'_{2s})$ as defined in \eqref{IntermediateFunction1ForAchievableRate1NonCausalCaseGaussianChannelTimeDivisionRelaying}. Then, we have the following.

\noindent The mutual information on the RHS of \eqref{Condition1OnRrRdTimeDivisionRelaying} can be computed as in \eqref{SecondPartOf__FirstTerm__AchievableRate__PartialDF} to obtain\begin{align}
 I(X_1^{(2)};Y_3^{(2)}|U_1,U_2) &= \frac{1}{2}\log\Big(1+\frac{(1-\rho'^2_{12})P_1^{(2)}}{N_3+\Phi(\alpha',\theta,\rho'_{2s})}\Big).
\label{MutualInformationEvaluationForCondition1OnRrRdTimeDivisionRelaying}
\end{align}

\noindent The conditional mutual information difference on the RHS of \eqref{Condition3OnRrRdTimeDivisionRelaying} is similar to $T_2$ in Appendix ~\ref{appendixCorollary3} and it gives 
\begin{equation}
I(X_1^{(2)},U_2;Y_3^{(2)}|U_1)-I(U_2;S^{(2)}|U_1) = \frac{1}{2}\log\Big(\frac{P'_2(P'_2+Q'^{(2)}+N_3+(1-\rho'^2_{12})P_1^{(2)})}{P'_2Q'^{(2)}(1-\alpha')^2+N_3(P'_2+\alpha'^2Q'^{(2)})}\Big).
\label{MutualInformationEvaluationForCondition3OnRrRdTimeDivisionRelaying}
\end{equation}

\noindent The evaluation of the term $[I(X_1^{(2)},U_1,U_2;Y_3^{(2)})-I(U_2;S^{(2)}|U_1)]$ is similar to that of \eqref{ExapandedSumRateAchievableRateFullDuplexGaussianRC} in Appendix \ref{appendixTheorem3}, and we obtain
\begin{align}
&I(X_1^{(2)},U_1,U_2;Y_3^{(2)})-I(U_2;S^{(2)}|U_1)\nonumber\\
&=\frac{1}{2}\log\Big(1+\frac{P_1^{(2)}+\bar{\theta}P_2+2\rho'_{12}\sqrt{\bar{\theta}P_1^{(2)}P_2}}{{\theta}P_2+Q^{(2)}+2\rho'_{2s}\sqrt{{\theta}P_2Q^{(2)}}+N_3}\Big)+\frac{1}{2}\log\Big(\frac{P'_2(P'_2+Q'^{(2)}+N_3)}{P'_2Q'^{(2)}(1-\alpha')^2+N_3(P'_2+\alpha'^2Q'^{(2)})}\Big).
\label{MutualInformationEvaluationForCondition2OnRrRdTimeDivisionRelaying__SecondPart}
\end{align}
%with $$P'_2:={\theta}P_2(1-\rho'^2_{2s}), \qquad Q'^{(2)}:=\big(\sqrt{Q^{(2)}}+\rho'_{2s}\sqrt{{\theta}P_2}\big)^2.$$
Also, it is easy to show that
\begin{align}
I(X_1^{(1)};Y_3^{(1)}) &=\frac{1}{2}\log\Big(1+\frac{P_1^{(1)}}{N_3+Q^{(1)}}\Big).
\label{MutualInformationEvaluationForCondition2OnRrRdTimeDivisionRelaying__FirstPart}
\end{align}

\noindent Finally, we obtain \eqref{AchievableRate1NonCausalCaseGaussianChannelTimeDivisionRelaying} using \eqref{Condition1OnRrRdTimeDivisionRelaying} and \eqref{MutualInformationEvaluationForCondition1OnRrRdTimeDivisionRelaying}; we obtain \eqref{AchievableRate2NonCausalCaseGaussianChannelTimeDivisionRelaying} using \eqref{Condition3OnRrRdTimeDivisionRelaying} and \eqref{MutualInformationEvaluationForCondition3OnRrRdTimeDivisionRelaying}; and we obtain \eqref{AchievableRate3NonCausalCaseGaussianChannelTimeDivisionRelaying} using \eqref{Condition2OnRrRdTimeDivisionRelaying}, \eqref{MutualInformationEvaluationForCondition2OnRrRdTimeDivisionRelaying__SecondPart} and \eqref{MutualInformationEvaluationForCondition2OnRrRdTimeDivisionRelaying__FirstPart}. This completes the proof.

\bibliographystyle{IEEEtran}
\bibliography{Draft__RevisedVersion}

% that's all folks
\end{document}